\documentclass[journal=jpccck,manuscript=article]{achemso}

\usepackage{chemformula} 
\usepackage[T1]{fontenc} 
\usepackage{amsmath}
\usepackage{caption}
\usepackage{subcaption}
\usepackage{appendix}
\usepackage{longtable}
\usepackage{wrapfig}
\usepackage{xr}
\usepackage{array}
\usepackage{ragged2e}
\usepackage{pdfpages}
\SectionNumbersOn

\makeatletter
\newcommand*{\addFileDependency}[1]{
  \typeout{(#1)}
  \@addtofilelist{#1}
  \IfFileExists{#1}{}{\typeout{No file #1.}}
}
\makeatother
 
\newcommand*{\myexternaldocument}[1]{%
    \externaldocument{#1}%
    \addFileDependency{#1.tex}%
    \addFileDependency{#1.aux}%
}

\myexternaldocument{supporting_info}

\newcommand{\ang}{\mbox{\normalfont\AA}}

\author{Holden L. Parks}
\author{Alan. J. H. McGaughey}
\author{Venkatasubramanian Viswanathan}
\email{venkvis@cmu.edu}
\affiliation[CMU]{Department of Mechanical Engineering, Carnegie Mellon University, Pittsburgh, Pennsylvania 15213, USA}

\title[Harmonic vibrations and ZPE]
  {Uncertainty quantification in first-principles predictions of harmonic vibrational frequencies of molecules and molecular complexes}

\abbreviations{IR,NMR,UV}

\begin{document}

\begin{abstract}
Accurate prediction of molecular vibrational frequencies is important to identify spectroscopic signatures and reaction thermodynamics.  In this work, we develop a method to quantify uncertainty associated with density functional theory predicted harmonic vibration frequencies utilizing the built-in error estimation capabilities of the BEEF-vdW exchange-correlation functional.  The method is computationally efficiency by estimating the uncertainty at nearly the same computational cost as a single vibrational frequency calculation.  We demonstrate the utility and robustness of the method by showing that the uncertainty estimates bounds the self-consistent calculations of six exchange correlation functionals for small molecules, rare gas dimers, and molecular complexes from the S22 dataset. Ten rare-gas dimers and the S22 dataset of molecular complexes provide a rigorous test as they are systems with complicated vibrational motion and non-covalent interactions. Using coefficient of variation as a uncertainty metric, we find that modes involving bending or torsional motion and those dominated by non-covalent interactions are found to have higher uncertainty in their predicted frequencies than covalent stretching modes.  Given the simplicity of the method, we believe that this method can be easily adopted and should form a routine part of DFT-predicted harmonic frequency analysis.
\end{abstract}


\section{Introduction}

Accurate prediction of molecular vibrational frequencies by ab initio methods is important in many areas of chemistry and physics\cite{cramer,mcquarrie}. Calculating the enthalpy and free energy of a reaction, for example, requires zero-point energy (ZPE) and finite temperature contributions, both of which depend on vibrational frequencies. The accuracy of the frequencies depends on the level of theory used to model the system \cite{sinha}, convergence criteria \cite{pople}, and if anharmonic corrections have been included\cite{barone}.   A widely-used method to predict vibrational frequencies is density functional theory (DFT) \cite{dft}. The accuracy of a DFT calculation depends strongly on the choice of the exchange-correlation (XC) functional. There has been much work devoted to identifying the best XC functional and DFT calculation parameters for accurately predicting the frequencies of small-to medium-sized molecules. \cite{sinha, laury, scott, wong, neugebauer, patton2} Effort has also been extended to predicting frequencies of weakly-bonded systems.\cite{patton1,tao} Patton and Pederson \cite{patton1} and Tao and Perdew \cite{tao} showed that DFT calculations of rare-gas dimers at the generalized-gradient (GGA) \cite{gga} level correct the overbinding introduced by the local density approximation (LDA) functional. This body of results has shown that DFT-predicted frequencies at the GGA level may be accurate to within tens of meV of experimental frequencies for many small molecules. However, for complex vibration modes, the sensitivity of the predictions to the choice of XC functionals remains unclear.

One na\"ive way to estimate the uncertainty associated with DFT-predicted frequencies is to perform the calculation with multiple XC functionals. The selection of functionals is somewhat arbitrary, however, and the calculation must be performed multiple times making it computationally inefficient. The Bayesian error estimation functional with van der Waals correlation (BEEF-vdW) \cite{beef} is a GGA-level XC functional that can systematically estimate uncertainty in DFT predictions. It possesses built-in uncertainty estimation capabilities in the form of an ensemble of GGA XC functionals that are calibrated to reproduce the error observed between experimental measurements and DFT predictions. Using the BEEF-vdW ensemble is computationally efficient compared to performing many calculations using different XC functionals, as results for thousands of XC functionals are obtained non-self-consistently by using one self-consistent calculation. BEEF-vdW has been applied to quantify uncertainty in magnetic ground states \cite{houchins} heterogeneous catalysis \cite{medford,sumaria, rune1}, electrocatalysis \cite{deshpande,krishnamurthy, rune2}, and mechanical properties of solid electrolytes \cite{zeeshan}. The BEEF-vdW model space includes contributions from the non-local vdW-DF2 XC functional \cite{df2}. It therefore offers potential improvements compared to other GGA-level functionals in describing van der Waals forces, which are traditionally not well described by DFT \cite{zhao}.

In this work, we present a computationally efficient method to estimate the uncertainty of harmonic vibrational frequencies of non-periodic systems such as molecules and molecular complexes. The harmonic approximation is valid at temperatures well below the dissociation limit \cite{atkins2}, which are on the order of $1,000$ K for most systems considered herein but can be as low as $370$ K for the rare-gas dimers \cite{luo, blanksby}. Frequency ensembles are calculated by solving an eigenvalue problem for an ensemble of Hessian matrices that depend on the second derivatives of the energy of the system with respect to its nuclear coordinates. The formulation of the eigenvalue problem is presented in Section \ref{sect: freq}. The ensemble of Hessians is calculated from an ensemble of energies determined with BEEF-vdW, which is described in Section \ref{sect: beef}. Computational details are provided in Section \ref{sect: comp details}.

We first apply our method to a set of small benchmark molecules in Section \ref{subsect: benchmarks} to determine if it can capture the uncertainty associated with choosing an XC functional. We also make comparisons to experimental measurements. We then consider a case study of the ten rare-gas dimers studied by Patton and Pederson \cite{patton1} and Tao and Perdew \cite{tao} in Section \ref{subsect: rare gas dimers}. Rare-gas dimers offer a simple test of the ability of the BEEF-vdW XC functional and the BEEF-vdW ensemble to describe weakly-bonded systems. In Section \ref{subsect: S22}, we study the non-covalently bonded complexes in the S22 dataset \cite{s22}. Because some XC functionals are known to perform well on rare-gas dimers but poorly on larger complexes (and vice versa) \cite{cerny,s22}, the S22 dataset offers an extensive test of the ability of BEEF-vdW to describe complicated vibrations in non-covalently bonded systems. We compare the predictions for all the molecules and molecular complexes in Section \ref{subsect: comparison}. In considering all the results, we utilize coefficient of variation as a metric to quantify uncertainty and should that the coefficient of variation is largest for modes that involve bending or torsion or whose bonds are non-covalent. Localized stretching modes tend to have lower uncertainty in their predicted frequencies. The spread of the frequencies can also be used to quantify the uncertainty associated with quantities derived from the frequencies (e.g., Gibbs free energy or ZPE).  Given the simplicity of the developed method, we believe that this should form a routine part of DFT-predicted harmonic frequency analysis.

\section{Methods}\label{sec: methods}

\subsection{Harmonic vibrational frequencies}\label{sect: freq}

In this section, we discuss a method for predicting harmonic vibrational frequencies by computing energies using DFT. Let $N$ be the number of atoms in a molecule, and let $i$ and $j$ index the atoms so that $1\leq i,j\leq N$. $\alpha$ and $\beta$ denote the Cartesian directions [i.e., $\alpha,\beta=x(1),y(2),z(3)$]. We make the harmonic approximation, in which a Taylor series expansion of the system's potential energy, $E$, is truncated after the second-order term \cite{atkins,quong,mcgaughey}. In this approximation, the force on atom $i$ in the $\alpha$-direction, $F_i^\alpha$, is proportional to the displacement of every other atom in the system from its equilibrium position. Newton's 2nd law for atom $i$ in the $\alpha$-direction is thus

\begin{equation}\label{eqn: sho}
\begin{split}
m_i \ddot{x}_i^\alpha&=F_i^\alpha\\ &=-\sum_{\beta}\sum_{j}\left(\frac{\partial^2E}{\partial x_i^\alpha \partial x_j^\beta}\right) x_j^\beta\\
&=-\sum_{\beta}\sum_{j}\Phi_{ij}^{\alpha\beta}x_j^\beta,
\end{split}
\end{equation}

\noindent where $m_i$ is the mass of atom $i$. The second derivatives in eq \ref{eqn: sho} are called the harmonic force constants, $\Phi_{ij}^{\alpha\beta}$. By assuming temporally periodic atomic motions, the set of equations represented by eq \ref{eqn: sho} for all atoms in all directions can be converted into the eigenvalue problem

\begin{equation}\label{eqn: eig}
\omega^2 \mathbf{e} = \mathbf{He}.
\end{equation}

\noindent Here, $\omega$ is a vibrational frequency, $\mathbf{e}$ is a vector of length $3N$ describing the motion of the atoms, i.e. the mode shape, and $\mathbf{H}$ is the 3$N\times3N$ Hessian matrix. The elements of the Hessian are related to the harmonic force constants by

\begin{equation}\label{hess}
\text{H}_{3*(i-1)+\alpha,\text{ }3*(j-1)+\beta} =\frac{1}{\sqrt{m_i m_j}}\Phi_{ij}^{\alpha\beta} .
\end{equation}

 The harmonic force constants are calculated by numerically approximating the second derivative in eq \ref{eqn: sho} using a central finite difference of the energies with respect to perturbations of the equilibrium structure. Using the shorthand $E(x_i^\alpha \pm h,x_j^\beta \pm h) = E_{i\pm h, j\pm h}$, where $h$ is the perturbation magnitude of the atomic displacement, the central difference formulas are \cite{fd}

\begin{equation}\label{eqn: fd}
\Phi_{ij}^{\alpha\beta}\approx
\begin{cases} 
\frac{1}{4h^2}\bigg(E_{i+h,j+h}+E_{i-h,j-h}-E_{i+h,j-h}-E_{i-h,j+h}\bigg),&\text{if}\ i\neq j\\
-\sum_{k\neq i} \Phi_{ik}^{\alpha\beta},  &\text{if}\ i=j.\\
\end{cases}
\end{equation}

\noindent Note that we do not directly calculate the harmonic force constants where $i=j$. This practice reduces numerical error associated with the eggbox effect by enforcing conservation of momentum (i.e., translational invariance) \cite{frederiksen}. The harmonic force constants are symmetrical with respect to permutation of the atomic and direction indices \cite{esfarjani} (i.e., $\Phi_{ij}^{\alpha\beta} =\Phi_{ji}^{\beta\alpha}$), so that we only need to calculate the upper triangular portion of the Hessian. Other molecule-specific symmetries can further reduce the number of force constants to be calculated, but we do not consider these here. 

The harmonic force constants can also be calculated with a finite difference of the force on atom $i$ \cite{quong}. We use the energies because, as will be explained in Section \ref{sect: beef}, BEEF-vdW provides uncertainty estimation in the system energy and not in the atomic forces.

\subsection{Bayesian error estimation}\label{sect: beef}

The energy of a system can be predicted using DFT and takes the form \cite{dft}

\begin{equation}
    E = E_{KE}+E_{ions}+E_{e-e}+E_{xc},
\end{equation}

\noindent where $E$ is the electronic energy of the system used in eq \ref{eqn: fd}. $E_{KE}$ is the kinetic energy, $E_{ions}$ is the potential energy of the electrons due to Coulombic interactions with the ions, and $E_{e-e}$ is the potential energy of the electrons due to Coulombic interactions with other electrons, all of which can be calculated exactly. $E_{xc}$ is the exchange-correlation energy, whose value is approximated by the chosen XC functional. 

BEEF-vdW is an XC functional at the GGA level\cite{beef}. The BEEF-vdW model space takes the form 
 
\begin{equation}\label{eqn: beef}
E_{xc}  =\sum_{m=0}^{29}a_m E_{m}^{\text{GGA}-\text{x}}+\alpha_cE^{\text{LDA}-\text{c}}+(1-\alpha_c)E^{\text{PBE}-\text{c}}+E^{\text{nl}-\text{c}},
\end{equation}
 
 \noindent where $a_m$ and $\alpha_c$ are multiplicative factors, $E^{\text{LDA}-\text{c}}$ is a correlation contribution from the local Perdew-Wang LDA correlation \cite{lda}, $E^{\text{PBE}-\text{c}}$ is a correlation contribution from the PBE semi-local correlation \cite{pbe}, $E^{\text{nl}-\text{c}}$ is a correlation contribution from the vdW-DF2 non-local correlation \cite{df2}, and $E_m^{\text{GGA}-\text{x}}$ is the contribution to the exchange energy given by

\begin{equation}\label{eqn: legendre}
 E_m^{\text{GGA}-\text{x}}=\int \epsilon_\text{x}^{\text{UEG}}[n(\mathbf{r})]B_m\left\{s[n(\mathbf{r}), \nabla n(\mathbf{r})]\right\}d\mathbf{r}.
\end{equation}

\noindent In eq \ref{eqn: legendre}, $n(\mathbf{r})$ and $\nabla n(\mathbf{r})$ are the electron density and its gradient, $s$ is a function that parameterizes $n$ and $\nabla n$, $\epsilon_\text{x}^{\text{UEG}}$ is the exchange energy density of the uniform electron gas, and $B_m$ is the $m$th Legendre polynomial. To determine the optimal BEEF-vdW XC functional, Wellendorff et al. fit $a_m$ and $\alpha_c$ to energetic and structural data describing bonding in a variety of chemical and condensed matter systems. These parameters were regularized to prevent overfitting to the training data. 

BEEF-vdW provides a systematic approach to estimating uncertainty in a DFT energy calculation by using an ensemble of XC functionals around the optimal BEEF-vdW XC functional. A self-consistent DFT calculation is first performed using the optimal parameters, yielding a converged electron density. This density is then used with distributions of $a_m$ and $\alpha_c$ to non-self-consistently generate an ensemble of energies using eq \ref{eqn: beef}. The distributions of $a_m$ and $\alpha_c$ are tuned such that the spread of the ensemble energies reproduces the errors observed when comparing the experimental training data to BEEF-vdW self-consistent predictions using the optimal XC functional.


BEEF-vdW provides an ensemble of energies rather than a single energy for a DFT calculation. This ensemble can be propagated in eq \ref{eqn: fd} to obtain an ensemble of numerical derivatives $\partial^2 E/\partial x_i^\alpha x_j^\beta$ and Hessians. The eigenvalue problem can be solved for each Hessian in the ensemble to determine the ensemble of frequencies.

\subsection{Computational details}\label{sect: comp details}

Self-consistent DFT calculations were performed with the real-space projector-augmented wave method \cite{paw1,paw2} as implemented in GPAW \cite{gpaw1,gpaw2}. The BEEF-vdW XC functional was used with 2000 ensemble functionals for each calculation. Using more than 2000 functionals has been found to have little effect on the standard deviation of the ensemble energy values \cite{beef, zeeshan}. We used a real-space grid spacing of $0.12$ $\ang$ for the rare-gas dimers and $0.18$ $\ang$ otherwise. Molecules were surrounded by vacuum in cubic boxes. Box lengths were determined so that adding eight additional real-space grid points along each axis changed the energy of the relaxed structure by no more than $0.01$ eV. This criterion resulted in a box length of at least $10$ $\ang$ for each system. Equilibrium geometries were determined by relaxing the structure so that each atom experienced a force of less than $0.01$ eV$/\ang$. Starting geometries for the S22 dataset were obtained from the Benchmark Energy and Geometry DataBase \cite{begdb}. All single-point calculations were converged so that the energy variation between the final three iterations was less than $10^{-8}$ eV. 

\begin{figure}
    \centering
    \begin{subfigure}{3.5 in}
        \centering
        \includegraphics[width=3.25 in]{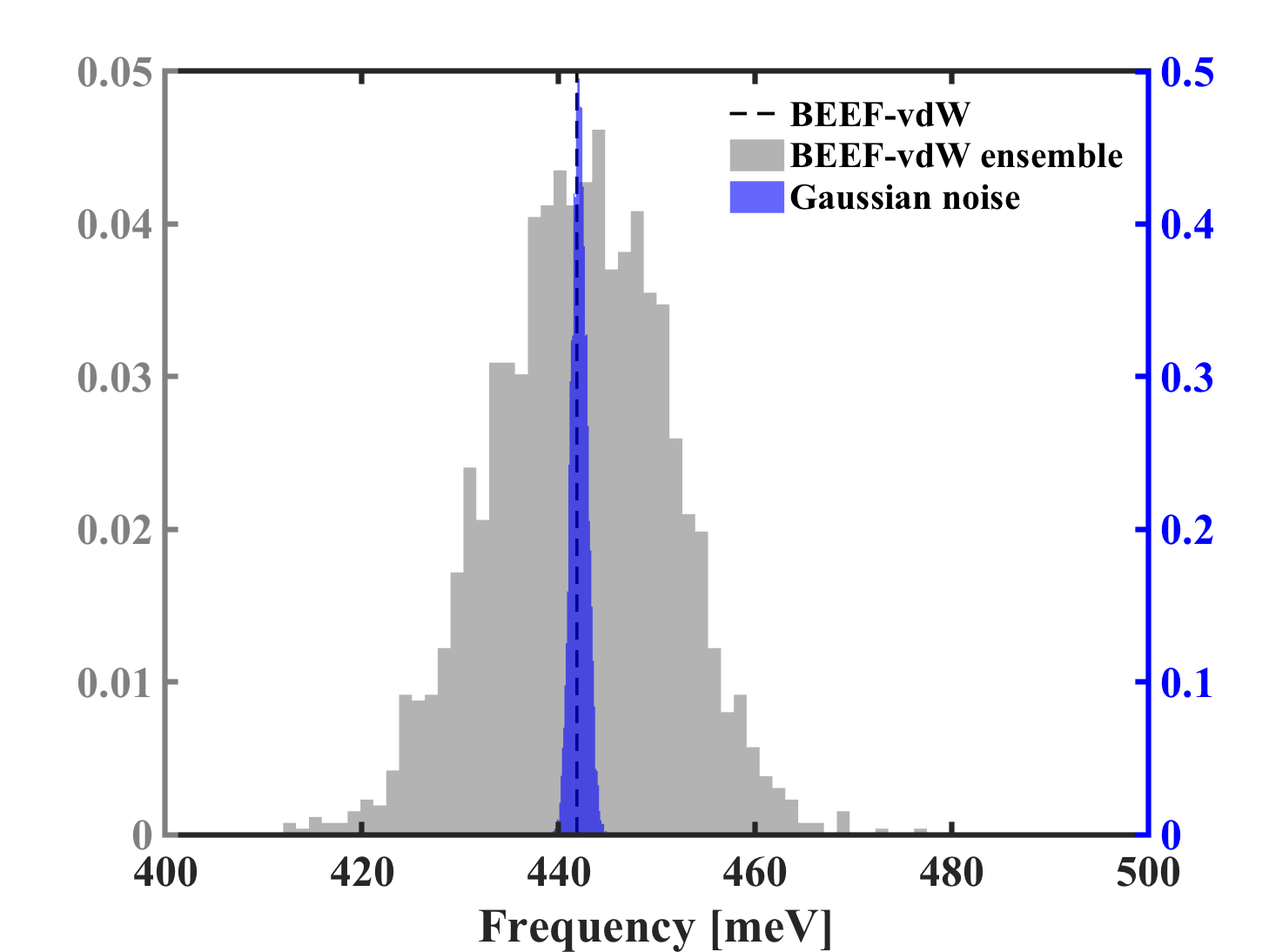}
        \caption{}\label{fig: uncertainty comp high}
    \end{subfigure}\hfill
    \begin{subfigure}{3.5 in}
        \centering
        \includegraphics[width=3.5 in]{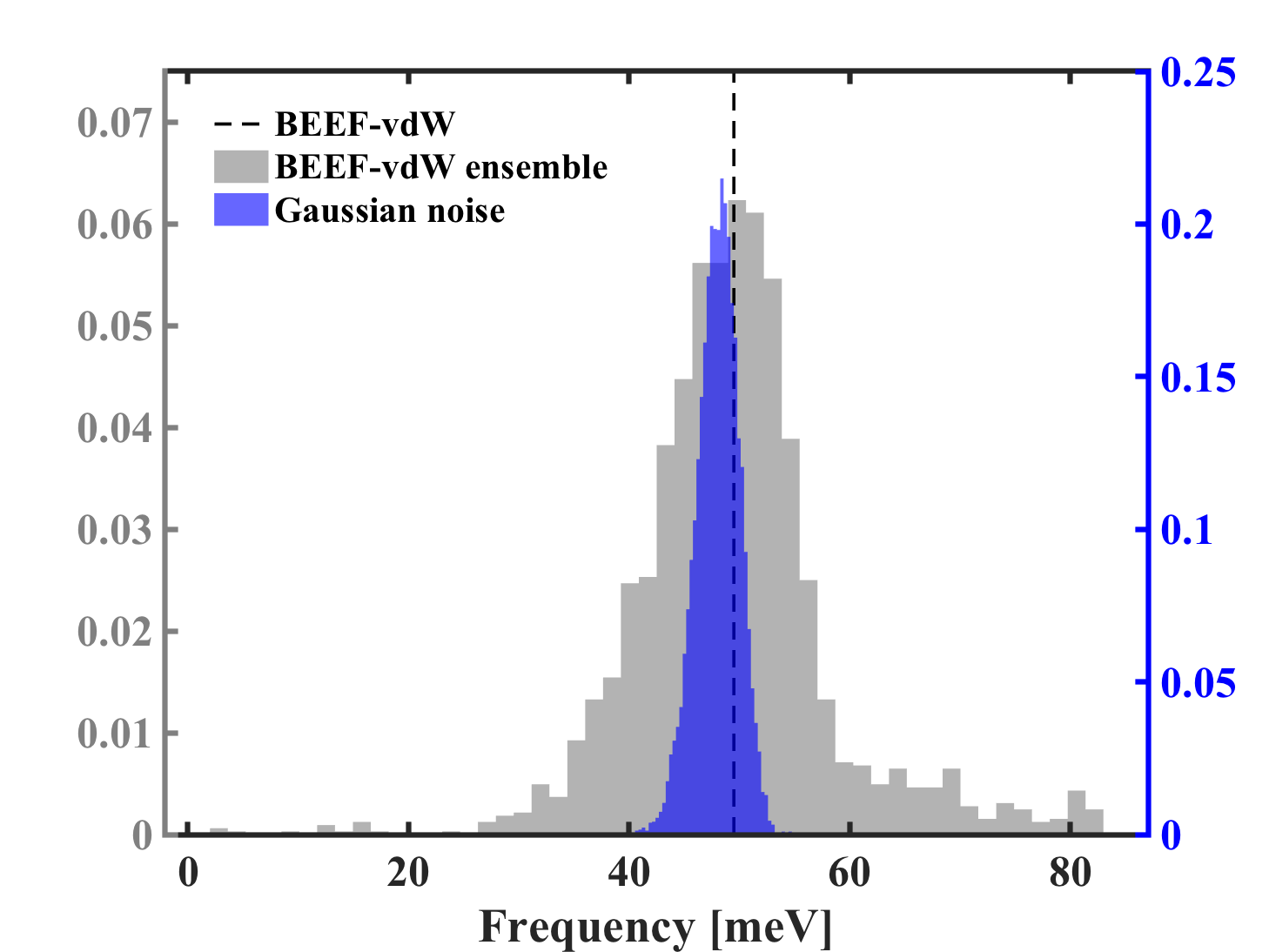}
        \caption{}\label{fig: uncertainty comp low}
    \end{subfigure}
    \caption{Numerical uncertainty (blue histogram, right axis scale), BEEF-vdW ensemble (gray histogram, left axis scale), and BEEF-vdW prediction (dotted black line) for (a) high and (b) low frequency vibration of the benzene-ammonia complex. Histograms in both figures are normalized to have a total area of unity. The spread of frequencies due to numerical uncertainty in both cases is smaller than the BEEF-vdW ensemble spread.}
	\label{fig: uncertainty comp}
\end{figure}

To obtain perturbed energies for numerical estimation of the second derivative in eq \ref{eqn: fd}, atomic displacements ($h$) of at most $\pm 0.01\text{ }\ang$ were applied. The $0.01\text{ }\ang$ displacement size led to variations in the harmonic force constants on the order of $0.001$ eV$/$\ang$^2$ and at most $0.0625$ eV$/$\ang$^2$ with respect to force constants calculated with displacements as small as $0.001$ $\ang$. A smaller displacement leads to smaller energy variations, which require more stringent convergence criteria and longer computation time, but can yield more accurate numerical derivatives. Numerical variation in the force constants will impact the calculated frequencies, but this effect is suppressed by two factors: (i) the requirement that the force constants satisfy conservation of momentum (eq \ref{eqn: fd}), and (ii) the low sensitivity of the Hessian to the eigenvalue problem because it is Hermitian\cite{franklin}. A comparison of the effect of numerical uncertainty in the force constants to the BEEF-vdW uncertainty for two vibrational frequencies (one high and one low) of the benzene-ammonia complex, a member of the S22 dataset, is shown in Figure \ref{fig: uncertainty comp}. The numerical uncertainty histograms were generated by adding draws from a Gaussian distribution $N(0,0.0625)$ eV$/$\ang$^2$ to the force constants calculated using the BEEF-vdW XC functional. This process was repeated $10,000$ times to ensure converged error estimates. For both frequencies, the spread due to the numerical uncertainty in the force constants is smaller than the ensemble spread, indicating that in considering the latter we may ignore the effects of the former.

For larger molecules and molecular complexes, the atomic motions described by the eigenvectors in eq \ref{eqn: eig} are often delocalized. To analyze the atomic motions in terms of the movement of local groups of atoms, they were transformed from Cartesian to internal coordinates. The internal coordinates were determined using the MolMod Python package \cite{molmod}. Bond length, bend angle, and dihedral angle internal coordinates, corresponding to stretching, bending, and torsional modes, were considered. The atomic motions were then mapped onto the internal coordinates by following the decomposition method outlined by Boatz and Gordon\cite{boatz}. The components of stretch, bend, and torsion internal coordinate motion for each mode sums to unity within a numerical error of at most $0.01$. As such, the relative contribution of each type of internal coordinate motion can be compared across different modes. Intermolecular modes are excluded from this mapping procedure, as their motion does not correspond to motion of the internal coordinates.

\section{Results}

\subsection{Benchmark molecules}\label{subsect: benchmarks}

To test the effectiveness of the proposed method, we analyze the estimated uncertainty due to choice of XC functional in predictions of vibrational frequencies of small molecules.   We first examine a set of eight small molecules: H$_2$, N$_2$, CO, CO$_2$, H$_2$O, NH$_3$, H$_2$CO, each of which has only stretching and bending modes. We then compare the results to those for two larger molecules, HCOOH and C$_2$H$_6$, that have more complicated modes (e.g., a torsional mode in the case of C$_2$H$_6$). We designate these ten molecules as our ``benchmark'' set. Frequencies were predicted using the PBE \cite{pbe}, RPBE \cite{rpbe}, PBEsol \cite{pbesol}, PW91 \cite{pw91}, and optPBE-vdW \cite{optpbe}, and BEEF-vdW functionals self-consistently. 

Results for the eight small molecules are presented in Table \ref{tbl: small molecules} and for HCOOH and C$_2$H$_6$ in Table \ref{tbl: small molecules c2h6 and hcooh}. Experimental values and BEEF-vdW ensemble statistics are also shown. To test for normality in the BEEF-vdW ensemble, we use the skew and kurtosis to calculate the Jarque-Bera (JB) statistic \cite{jarque, bowman}, given by JB $= \frac{n}{6}\left[S^2+\frac{K^2}{4} \right]$, where $n=2000$ is the sample size of the ensemble, $S$ is the sample skew, and $K$ is the sample kurtosis. Under the null hypothesis that the ensemble is Gaussian, JB is approximately described by a $\chi^2(2)$ distribution for large $n$. We choose to reject the null hypothesis at a $95\%$ confidence level ($\alpha=0.05$), corresponding to a $\chi^2$ value of approximately 6 (i.e., we label the ensemble as non-Gaussian if JB $>6$). We use the standard deviation ($\sigma$) and coefficient of variation, $\text{COV} =\sigma/\mu$, where $\mu$ is the mean of the distribution, of the ensembles as measures of uncertainty due to choice of XC functional. A low COV indicates that the functionals tend to agree in their frequency predictions, while a high COV indicates disagreement. In Section \ref{subsect: comparison} we find that a COV of $0.25$ reasonably separates ensembles with low and high uncertainty.

\begin{table*}
\resizebox{7in}{!}{
  \begin{tabular}{l|lll | >{\RaggedLeft\arraybackslash}p{1.2cm}>{\RaggedLeft\arraybackslash}p{1.2cm}>{\RaggedLeft\arraybackslash}p{1.2cm}>{\RaggedLeft\arraybackslash}p{1.2cm}>{\RaggedLeft\arraybackslash}p{1.2cm}>{\RaggedLeft\arraybackslash}p{1.2cm}>{\RaggedLeft\arraybackslash}p{1.2cm}|>{\RaggedLeft\arraybackslash}p{1.2cm}>{\RaggedLeft\arraybackslash}p{1.2cm}>{\RaggedLeft\arraybackslash}p{1.2cm}>{\RaggedLeft\arraybackslash}p{1.2cm}>{\RaggedLeft\arraybackslash}p{1.2cm}>{\RaggedLeft\arraybackslash}p{1.2cm}}
    \hline
     & Stretch & Bend & Tors. & Expt. & PBE & RPBE & PBEsol & PW91 & optPBE & BEEF & $\mu$ & $\sigma$ & COV & Skew & Kurt. & JB \\
     &  &  &  &  &  &  &  &  & -vdW & -vdW  &  &  &  &  &    \\
    \hline
    H$_2$   &1 &0 &0 & 545.5\cite{huber} & 536.1 & 540.4 & 527.7 & 538.6 & 543.5 & 555.0  & 555.3 & 7.5 & 0.01 & -0.03 & -0.07 & 0.7 \\
    CH$_4$ &0 &1 &0 & 168.2 \cite{johnson} & 159.0 & 160.1 & 156.2 & 159.6 & 164.6 & 164.5  & 163.7  & 10.1 & 0.06 & -0.17 & 0.20 & 13.0 \\
           &0 &1 &0 &  194.3 \cite{johnson}& 186.8 & 187.5 & 184.5 & 187.4 & 191.6 & 191.6 & 191.1 & 7.6 & 0.04 & -0.10  & 0.10 & 4.2 \\
           &1 &0 &0 &  376.5 \cite{johnson}& 368.5 & 367.4 & 367.0 & 369.4 & 372.0 & 372.0 & 372.2  & 6.2 & 0.02 & -0.06  & 0.10 & 2.0 \\
           &1 &0 &0 & 391.5 \cite{johnson} & 382.6 & 381.1 & 382.2 & 383.2 & 385.1 & 385.1 & 385.4 & 7.9 & 0.02 & -0.09  & 0.08 & 3.2 \\
    NH$_3$ &0 &1 &0 & 126.7 \cite{johnson}& 125.8 & 128.7 & 122.4 & 125.4 & 127.9 & 130.6   & 129.5 &  13.2 & 0.10 & -0.38  & 0.72 & 91.3 \\
           &0 &1 &0 & 209.6 \cite{johnson}& 200.3 & 201.5 & 197.7 & 200.8 & 202.7 & 205.8  & 205.0 & 8.7  & 0.04 &  -0.13 & 0.17 & 8.0 \\
           &1 &0 &0 &  434.7 \cite{johnson} & 420.5 & 418.9 & 420.4 & 421.5 & 418.6 & 425.2 & 425.4 & 7.4 & 0.02 & -0.06 & 0.04 & 1.3 \\
           &1 &0 &0 & 443.5 \cite{johnson}& 435.7 & 433.8 & 436.3 & 436.6 & 433.0 & 440.1 & 440.4 & 8.5 & 0.02 & -0.08 & 0.06 & 2.4 \\
    H$_2$O & 0&1 &0 & 204.3 \cite{johnson}& 197.1 & 198.7 & 194.5 & 197.1 & 199.1 & 202.3 & 201.8 & 9.4 & 0.09 & -0.32 & 0.47 & 52.5 \\
           &1 &0 &0 &  475.1 \cite{johnson}& 461.1 & 460.0 & 461.7 & 462.2 & 458.5 & 467.1 & 467.4 & 8.2 & 0.02 & -0.08  & 0 & 2.1\\
           &1 &0 &0 &  488.8 \cite{johnson} & 473.7 & 472.5 & 474.7 & 474.9 & 470.9 & 479.7 & 480.0 &  8.6 & 0.03 & -0.11 & 0.04 & 4.2 \\
    CO  &1 &0 &0 &  269.0\cite{huber}& 265.3 & 262.3 & 267.2 & 265.8 & 264.3 & 265.4 & 265.5 & 5.7 & 0.02 & -0.08 & 0.05 & 2.3 \\
    N$_2$   &1 &0 &0 &  292.3\cite{huber} & 292.4 & 289.4 & 294.1 & 292.9 & 291.0 & 293.6 & 293.8 & 6.0 & 0.02 & -0.08 & 0.01 & 2.1 \\
	H$_2$CO &0.04 &0.96 &0 & 147.7 \cite{johnson}& 142.4 & 142.3 & 141.4 & 142.8 & 142.8 & 144.8 & 144.1 & 9.4 & 0.07 & -0.17 & 0.22 & 13.7 \\
            &0 &1 &0 & 159.6 \cite{johnson}& 151.3 & 151.2 & 150.3 & 151.8 & 152.3 & 154.1 & 153.7 & 7.3 & 0.05 & -0.09  & 0.09 & 3.4 \\
            &0.10 &0.90 &0 & 193.8 \cite{johnson}& 183.5 & 183.8 & 181.7 & 184.1 & 185.0 & 187.3 & 186.3 & 5.5 & 0.03 & -0.70  & 0.53 & 186.7 \\
            &0.90 &0.10 &0& 218.7 \cite{johnson}& 219.0 & 217.0 & 221.1 & 219.2 & 217.5 & 219.4 & 220.0 & 3.4 & 0.02 & 0.55  & 0.28 & 107.4 \\
            &1 &0 &0 &  365.0 \cite{johnson}& 346.5 & 345.9 & 344.6 & 347.5 & 346.1 & 351.6 & 351.8 & 7.3 & 0.02 & -0.08  & 0.07 & 2.5 \\
           &1 &0 &0 &  373.0 \cite{johnson}& 352.3 & 351.4 & 350.6 & 353.3 & 351.7 & 357.5 & 357.8 & 7.5 & 0.02 & -0.09  & 0.05 & 2.9 \\
    CO$_2$  & 0 &1 &0 & 82.7 \cite{shimanouchi}& 79.7 & 79.1 & 80.2 & 79.7 & 78.7 & 79.6  & 79.0 & 7.3  & 0.09 & -0.32 & 0.47 & 52.4 \\
            & 1 &0 &0 & 165.3 \cite{shimanouchi}& 164.4 & 162.6 & 165.9 & 164.5 & 163.3 & 164.5 & 164.6 & 3.6 & 0.02 & -0.08 & 0 & 2.1 \\
            &1 &0 &0 & 291.2\cite{shimanouchi}& 292.7 & 289.5 & 296.0 & 292.8 & 290.0 & 292.4 & 292.6 & 7.7 & 0.03 & -0.11  & 0.04 & 4.2 \\
            \hline
          ME  &  &  &  &  & 7.8 & 8.3 & 8.6 & 7.2 & 7.3 & 3.8  &  &  &  &  &  \\
          MAE &  &  &  &  & 8.0 & 8.5 & 9.5 & 7.4 & 7.4 & 5.3  &  &  &  &  &  \\
   \hline
  \end{tabular}}
  \caption{Harmonic vibrational frequencies for eight small molecules in meV as predicted with the PBE, RPBE, PBEsol, PW91, optPBE-vdW, and BEEF-vdW XC functional. Experimental values and the BEEF-vdW ensemble statistics are also listed. Standard deviation values are in meV. COV, skew, and kurtosis are dimensionless.}
  \label{tbl: small molecules}
\end{table*}

\begin{table*}
\resizebox{7in}{!}{
  \begin{tabular}{l|lll | >{\RaggedLeft\arraybackslash}p{1.2cm}>{\RaggedLeft\arraybackslash}p{1.2cm}>{\RaggedLeft\arraybackslash}p{1.2cm}>{\RaggedLeft\arraybackslash}p{1.2cm}>{\RaggedLeft\arraybackslash}p{1.2cm}>{\RaggedLeft\arraybackslash}p{1.2cm}>{\RaggedLeft\arraybackslash}p{1.2cm}|>{\RaggedLeft\arraybackslash}p{1.2cm}>{\RaggedLeft\arraybackslash}p{1.2cm}>{\RaggedLeft\arraybackslash}p{1.2cm}>{\RaggedLeft\arraybackslash}p{1.2cm}>{\RaggedLeft\arraybackslash}p{1.2cm}>{\RaggedLeft\arraybackslash}p{1.2cm}}
    \hline
     & Stretch & Bend & Tors. & Expt. & PBE & RPBE & PBEsol & PW91 & optPBE & BEEF & $\mu$ & $\sigma$ & COV & Skew & Kurt. & JB \\
     &  &  &  &  &  &  &  &  & -vdW & -vdW  &  &  &  & &   &  \\
    \hline
    C$_2$H$_6$ & 0& 0& 1& 37.6 \cite{johnson}& 36.7 & 36.5 & 37.1 & 36.7 & 37.6 & 37.8 & 34.6 & 23.8 & 0.69 & -0.14  & -1.12 & 111.1 \\ 
               &0 &0.73 &0.27 &  101.9 \cite{johnson}& 99.1 & 99.4 & 98.0 & 99.3 & 100.1 & 101.3 & 99.5  & 11.5 & 0.12 & -0.72  & 1.03 & 261.2 \\
               &0.97 &0.03 &0 &  126.0  \cite{johnson}& 122.8 & 120.6 & 125.0 & 122.6 & 120.7 & 122.0 & 122.3 & 2.7 & 0.02 & 0.16 & -0.45 & 25.4 \\
               &0 &0.64 &0 &  154.5 \cite{johnson}& 145.9 & 146.1 & 144.3 & 146.3 & 147.6 & 149.4 & 148.8 & 8.1 & 0.05 & -0.11 & 0.12 & 5.2 \\
               &0 &1 &0 &  178.3  \cite{johnson}& 168.2 & 168.8 & 165.9 & 168.9 & 170.8 & 172.9 & 172.3 & 9.3 & 0.05 & -0.15  & 0.16 & 9.6 \\
               &0.03 &0.97 &0 &  179.6 \cite{johnson}& 168.7 & 170.1 & 167.9 & 170.1 & 171.6 & 173.9 & 173.4 & 8.8  & 0.05 & -0.07 & 0.01 & 1.6 \\
               &0 &0.71 &0.29 &  189.2 \cite{johnson}& 180.0 & 180.7 & 177.6 & 180.6 & 182.2 & 184.8  & 183.6 & 7.9 &  0.04  & -0.12 & 0.17 & 7.2 \\
               &0 &0.62 &0.38 &  192.4 \cite{johnson}& 180.0 & 180.8 & 177.6 & 180.7 & 182.3 & 184.8 & 184.2 & 8.1 & 0.04 & -0.09  & 0.07 & 3.1 \\
               &1 &0 &0 &  377.3  \cite{johnson}& 367.7 & 366.8 & 366.4 & 368.6 & 366.1 & 371.2 & 371.4 & 6.7 & 0.02 & -0.08  & 0.09 & 2.8 \\
               &1 &0 &0 &  379.5 \cite{johnson}& 368.0 & 367.0 & 366.7 & 368.9 & 366.4 & 371.4 & 371.6 & 6.6 & 0.02 & -0.07  & 0.10 & 2.5 \\
               &1 &0 &0 &  389.3 \cite{johnson}& 377.1 & 375.9 & 376.5 & 377.8 & 374.6 & 380.3 & 377.6 & 8.0 & 0.02 & -0.09  & 0.07 & 2.5 \\
               &1 &0 &0 &  393.6 \cite{johnson}& 377.5 & 376.3 & 376.8 & 378.1 & 374.9 & 380.4 & 380.5 & 7.9 & 0.02 & -0.09  & 0.08 & 3.2 \\
    HCOOH &0.10 &0.90 & 0&  77.5 \cite{shimanouchi}& 74.8 & 74.6 & 74.8 & 75.0 & 75.1 & 76.1 & 73.0 & 9.1 & 0.12 & -2.74 & 10.91 & 1.2e4 \\
          &0 &0 &1 &  79.1 \cite{shimanouchi}& 83.5 & 82.4 & 84.6 & 83.4 & 82.2 & 83.2 & 84.0 & 12.2 & 0.14 & 0.48 & 0.41 & 90.8 \\
          &0 & 0 &1 & 128.1  \cite{shimanouchi}& 124.2 & 123.9 & 123.9 & 124.5 & 123.9 & 125.7 & 124.0 & 8.0 & 0.06 & -0.83 & 0.47 & 248.0 \\
          &0.73 &0.27 & 0&  137.0 \cite{shimanouchi}& 133.4 & 131.6 & 135.4 & 133.6 & 132.0 & 133.6 & 133.8 & 3.5 & 0.03 & 0.99 & 5.63 & 2968.1 \\
          & 0.20& 0.80& 0& 152.4  \cite{shimanouchi}& 155.5 & 155.3 & 155.8 & 155.8 & 155.9 & 158.2 & 158.7 & 5.6 & 0.04 & 0.67 & 0.17 & 152.0 \\
          &0.07 &0.93 &0 & 172.0  \cite{shimanouchi}& 167.4 & 167.9 & 165.6 & 167.9 & 168.7 & 171.3 & 170.8 & 6.5 & 0.04 & -0.20  & -0.23 & 17.7 \\
          & 0.91& 0.09&0  &  219.4 \cite{shimanouchi}& 218.2 & 216.2 & 220.5 & 218.4 & 216.6 & 218.9 & 219.3 & 4.7 & 0.02 & 0.21  & -0.08 & 15.2 \\
          &1 &0 &0 & 364.9  \cite{shimanouchi}& 367.6 & 366.7 & 365.7 & 368.6 & 367.3 & 373.0 & 373.2 & 7.2 & 0.02 & -0.09  & 0.08 & 3.2 \\
          & 1& 0& 0&  442.6 \cite{shimanouchi}& 448.1 & 448.6 & 447.3 & 449.1 & 446.4 & 455.4 & 455.7 & 8.4 & 0.02 & -0.07  & 0.06 & 1.9 \\
               \hline          
          ME  &  &  &  &  & 5.1 & 5.5 & 5.7 & 4.6 & 5.2 & 2.2  &  &  &  &  &  \\
          MAE &  &  &  &  & 6.6 & 6.9 & 7.1 & 6.3 & 6.4 & 5.2  &  &  &  &  &  \\
   \hline
  \end{tabular}}
    \caption{Harmonic vibrational frequencies in meV as predicted with the PBE, RPBE, PBEsol, PW91, optPBE-vdW, and BEEF-vdW XC functionals for C$_2$H$_6$ and HCOOH. Experimental values and the BEEF-vdW ensemble statistics are also listed. Standard deviation values are in meV. COV, skew, and kurtosis are dimensionless.}
  \label{tbl: small molecules c2h6 and hcooh}
\end{table*}

\begin{figure}
  \includegraphics[width=3.25 in]{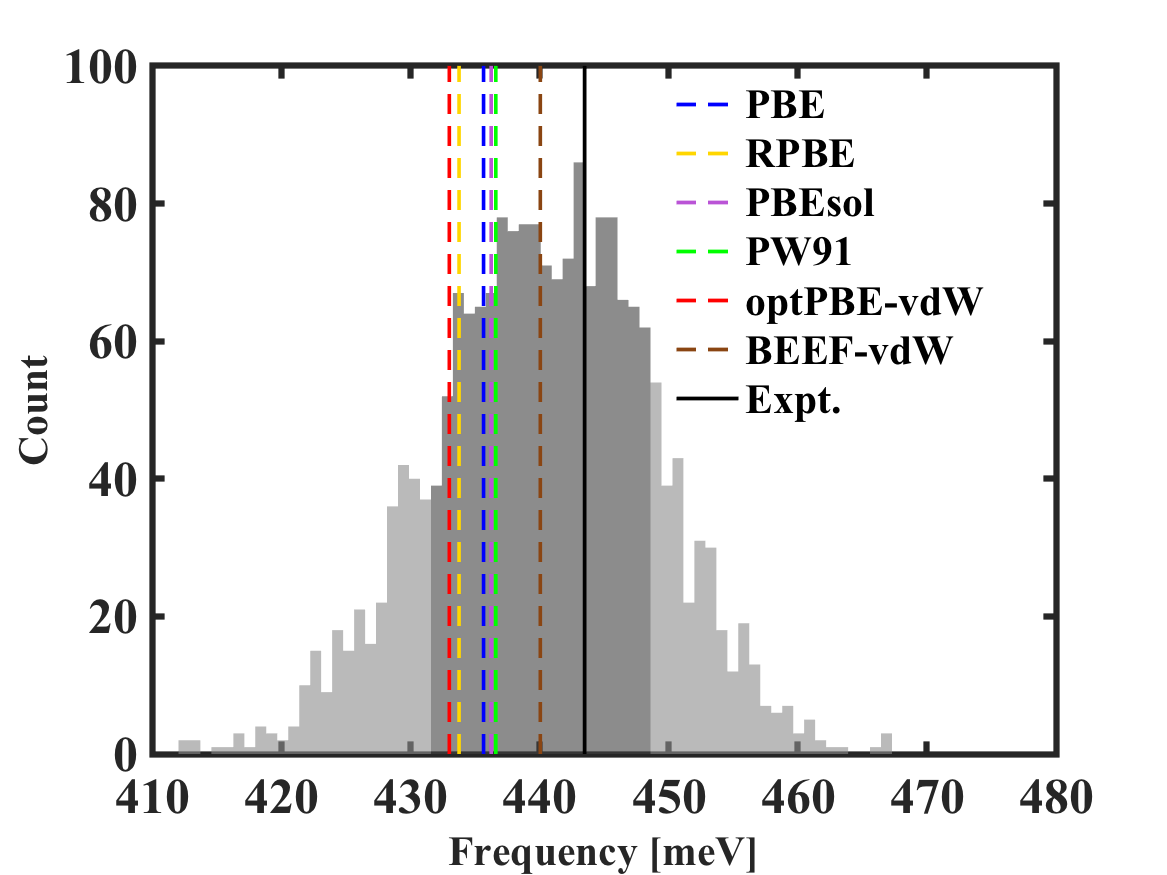}
  \caption{Experimental and DFT predictions of the harmonic vibrational frequency of the asymmetric stretch mode of NH$_3$, with the BEEF-vdW ensemble shown in gray. The darker gray indicates a range of $\pm \sigma$ around the mean of the BEEF-vdW ensemble mean.}
  \label{fig: nh3_highfreq_vib}
\end{figure}

As a representative example, consider the asymmetric stretching mode of NH$_3$. The BEEF-vdW ensemble of frequencies, predictions using other functionals, and the experimental value ($443.5$ meV \cite{johnson}) are plotted in Figure \ref{fig: nh3_highfreq_vib}. The internal coordinate decomposition of the mode in Table \ref{tbl: small molecules} indicates that it is a pure stretching mode. We report a BEEF-vdW value of $440.1$ meV and an ensemble standard deviation of $8.5$ meV. The ensemble bounds the experimental frequency to within one standard deviation, demonstrating that DFT can accurately predict this frequency. In addition, the ensemble bounds the frequencies predicted by the other XC functionals to within one standard deviation, which indicates that the ensemble can reproduce the predictions of other XC functionals. Based on the low COV of the BEEF-vdW ensemble ($0.02$), there is also little disagreement in ensemble functionals in predictin this frequency. The ensemble has low skew ($-0.08$) and kurtosis ($0.06$). These values yield a JB of $2.4$, so that the ensemble is Gaussian. 

In the Supporting Information, we propagate the uncertainty estimates provided by the NH$_3$ frequency ensembles for all NH$_3$ modes to quantify uncertainty in NH$_3$ ZPE. We also show how ensembles for N$_2$, H$_2$, and NH$_3$ can be used to quantify uncertainty in predicting the vibrational entropy for the Haber process, 3H$_2$+N$_2\rightarrow$ NH$_3$.

The data in Table \ref{tbl: small molecules} indicate that most of the 23 modes of the small molecules are pure stretching or pure bending. There are only a few modes with internal coordinate mixing, e.g., the third mode for H$_2$CO, which is $0.90$ stretching and $0.10$ bending. In terms of mean absolute error (MAE) in comparison to the experimental measurements, BEEF-vdW is the best XC functional for predicting vibrational frequencies. The experimental frequencies for 16 of the 23 frequencies are bounded to within one standard deviation of the BEEF-vdW ensemble, with the largest deviation ($2.07\sigma$) coming from the sixth mode of H$_2$CO. On average, the predictions via BEEF-vdW deviate $0.70\sigma$ from the experimental values.

The predictions of the other XC functionals are also generally bounded by the BEEF-vdW ensemble to within one standard deviation, with each XC functional predicting at most three frequencies outside of $\pm\sigma$. The average deviation is smallest for PW91 ($0.44\sigma$), followed by optPBE-vdW ($0.46\sigma$), PBE ($0.52\sigma$), RPBE ($0.62\sigma$), and PBEsol ($0.74\sigma$).

The largest COV for an ensemble is $0.10$, for the lowest frequency mode of NH$_3$. The low COV values indicate agreement among the ensemble functionals in predicting the frequencies. Each ensemble has relatively low skew and kurtosis values, the highest values being a skew of $-0.70$ for the third H$_2$CO mode and a kurtosis of $0.72$ for the first NH$_3$ mode. Eight modes have JB $>6$ and are non-Gaussian. Each of these eight modes is a pure bend mode or a mix of bending and stretching.

The results presented in Table \ref{tbl: small molecules} indicate that for small molecules with simple vibrational motion, the BEEF-vdW ensemble can bound both the spread of predictions of other XC functionals and experimental frequencies. To test the robustness of the BEEF-vdW ensemble for molecules containing torsional vibrations, we next apply the method to HCOOH and C$_2$H$_6$ and the results are presented in Table \ref{tbl: small molecules c2h6 and hcooh}. By the JB test, 12 of the 21 modes have non-Gaussian ensembles, none of which correspond to pure stretch modes. 

Based on the MAE, BEEF-vdW yields the frequencies closest to the experimental measurements. For 15 of the 21 modes, the experimental frequencies are bounded to within one standard deviation of the BEEF-vdW prediction, with an average deviation of $0.74\sigma$. Amongst the XC functionals considered, PW91 again has the smallest deviation ($0.32\sigma$) and PBEsol again has the largest ($0.58\sigma$). These deviations are smaller in comparison to those for the eight smaller molecules and seem to indicate better agreement among XC functionals for predicting of these frequencies. The ensembles for several modes, however, indicate this is not always the case. The two lowest frequency modes for both HCOOH and C$_2$H$_6$ have ensemble COV higher than the highest value reported in Table \ref{tbl: small molecules}, with a maximum of $0.69$ for the C$_2$H$_6$ torsional mode. This torsional mode also has the largest standard deviation, $23.8$ meV, of the modes presented in both tables. While the ensemble for this mode easily bounds the experimental and other DFT-predicted frequencies, its high COV indicates disagreement among GGA-level functionals in predicting this frequency. This mode is also noteworthy because some 5\% of the ensemble functionals predict a imaginary frequencies. These these imaginary frequencies correspond to negative solutions to the eigenvalue problem in eq \ref{eqn: eig} and indicate that some ensemble XC functionals predict that this mode is not stable. These values are not included in calculation of any ensemble statistics. There is also a significant fraction of ensemble functionals, over 15\%, which predict very small (nonzero but less than 2 meV) frequencies. These findings are in agreement with previous analysis showing the difficulty in describing the torsional energy landscape of C$_2$H$_6$ within GGA-level DFT\cite{werpetinski}.  


\subsection{Rare-gas dimers}\label{subsect: rare gas dimers}

In Section \ref{subsect: benchmarks}, we determined that the proposed methodology can reasonably bound frequencies both from experiment and predictions using other XC functionals for a benchmark set of ten molecules. We next consider ten rare-gas dimers built from He, Ne, Ar, and Kr. These systems are diatomic molecules and therefore have only one stretching mode. Rare-gas dimers are bonded by dispersion interactions and have been studied to determine the applicability of DFT to simple van der Waals systems \cite{zhang}. Patton and Pederson \cite{patton1} and Tao and Perdew \cite{tao} found that while GGA functionals corrected the overbinding tendency of the LDA functional, they overestimated the interaction strength when the outer electron shell consisted of $s$ electrons (as in He$_2$) and underestimated interaction strength when the outer shell consisted of $p$ electrons (as in Ne$_2$) \cite{zhao}. The rare-gas dimers offer an interesting case study for both the BEEF-vdW XC functional, which explicitly models non-local interactions, and the BEEF-vdW ensemble, as the vibrational frequencies are small enough that the uncertainty in the prediction approaches the magnitude of the frequency itself ($\sim 10$ meV).

\begin{table*}
\resizebox{7in}{!}{
  \begin{tabular}{l>{\RaggedLeft\arraybackslash}p{1.2cm}>{\RaggedLeft\arraybackslash}p{1.2cm}>{\RaggedLeft\arraybackslash}p{1.2cm}>{\RaggedLeft\arraybackslash}p{1.2cm}>{\RaggedLeft\arraybackslash}p{1.2cm}>{\RaggedLeft\arraybackslash}p{1.2cm}>{\RaggedLeft\arraybackslash}p{1.2cm}>{\RaggedLeft\arraybackslash}p{1.2cm}>{\RaggedLeft\arraybackslash}p{1.2cm}|>{\RaggedLeft\arraybackslash}p{1.2cm}>{\RaggedLeft\arraybackslash}p{1.2cm}>{\RaggedLeft\arraybackslash}p{1.2cm}>{\RaggedLeft\arraybackslash}p{1.2cm}>{\RaggedLeft\arraybackslash}p{1.2cm}>{\RaggedLeft\arraybackslash}p{1.2cm}}
    \hline
     & Expt. & PBE\cite{tao} & PW91\cite{patton1} &  TPSSh\cite{tao}  &  PBE & RPBE & optPBE & vdW & BEEF & $\mu$ & $\sigma$ & COV & Skew & Kurt. & JB \\
     &  &  &  &  &  & & -vdW & -DF2 &-vdW & & & & & & \\
    \hline
    He$_2$   & 4.116 & 8.936 & 12.4 & 4.342 & 8.4 & 10.0 & 10.3 & 9.1  & 15.6 & 15.1 & 4.7 & 0.31 & 0.76 & 0.20 & 195.9 \\
    HeNe     & 4.334 & 6.539 &  10.0 & 6.609  & 6.9 & 8.3 & 9.2 & 9.5  & 13.1 & 13.1 & 5.9 & 0.45 & 0 & -0.68  & 38.5\\
    HeAr     & 4.317  & 6.220& 8.1  & 4.896 & 7.6 & 6.7 & 8.5 & 7.5 & 11.0 & 10.9 & 5.6 & 0.52 & -0.03 & -0.86 & 61.9 \\
    HeKr     & 3.972 & 4.842 & 7.6 &  5.606  & 7.0 & 6.1 & 8.1 & 6.9 & 11.2 & 10.9 & 5.5 & 0.51 & -0.13 & -0.86 & 67.3 \\
    Ne$_2$   & 3.536 & 3.942 & 6.2  &  4.066 & 5.6 & 7.0 & 6.3 & 5.1 & 8.7 & 8.5 & 4.4 & 0.52 & -0.07 & -0.86 & 63.3 \\
    NeAr     & 3.494 & 3.681& 4.6 & 2.311 & 3.9 & 5.2 & 4.5 & 3.6   & 4.2 & 5.0 & 3.1 & 0.61 & 0.32 & -0.77 & 83.5 \\
    NeKr     & 3.243  & 3.335& 4.1  & 2.834 & 4.9 & 4.1 & 5.5 & 4.9  & 5.5 & 5.4 & 3.0 & 0.55 & 0 & -0.89 & 66.0 \\
    Ar$_2$   & 3.837 & 3.086& 3.2 &  2.534 & 2.8 & 4.5 & 5.8 & 5.3  & 6.4 & 6.0 & 2.3 & 0.38 & -0.73 & 0.16 & 179.8 \\
    ArKr     & 3.465 & 2.420& 2.9 &  2.289 & 3.3 & 2.7 & 3.9 & 3.8  & 4.6 & 4.3 & 1.9 & 0.43 & -0.40 & -0.47 & 71.7 \\
    Kr$_2$   & 2.923 & 1.808& 2.1&  2.120 & 2.2 & 2.2 & 3.7 & 3.5  & 4.2 & 4.0 & 1.4 & 0.36 & -0.71 & 0.29 & 175.0 \\
    \hline
    ME &   & -0.757& -2.4 & -0.037 &  -1.5 & -2.0 & -2.9 & -2.2  & -4.7 &  & &  &  &  & \\
    MAE &   & 1.339& 2.8 & 1.012 &  1.9 & 2.3 & 2.9 & 2.2  & 4.7 &  &  &  &  & & \\
   \hline
  \end{tabular}}
  \caption{Harmonic vibrational frequencies for rare-gas dimers in meV as predicted with the PBE, RPBE, optPBE-vdW, vdW-DF2, and BEEF-vdW XC functionals. BEEF-vdW ensemble statistics are also presented. Standard deviation values are in meV. COV, skew, and kurtosis are dimensionless. Experimental values are from Ogilvie and Wang \cite{ogilvie1,ogilvie2}. The PBE and TPSSh values from Tao and Perdew \cite{tao} include BSSE corrections.}
  \label{tbl: rare gas dimers}
\end{table*}

We compare the BEEF-vdW ensemble of frequencies for each molecule to those predicted by the PBE, RPBE, optPBE-vdW \cite{optpbe}, vdW-DF2 \cite{df2}, and BEEF-vdW XC functionals in Table \ref{tbl: rare gas dimers}. Experimental values are from Ogilvie and Wang \cite{ogilvie1, ogilvie2}. We also include DFT values from Patton and Pedersen \cite{patton1} and Tao and Perdew \cite{tao}. The tendency of GGA functionals to correct the severe overestimation of frequencies by LDA is consistent with our PBE and RPBE frequencies. For example, Tao and Perdew reported a frequency of $122.10$ meV for He$_2$ using LDA, while our PBE value of $8.4$ meV for the same quantity is much closer to the experimental value. 

The three vdW functionals tested, optPBE-vdW, vdW-DF2, and BEEF-vdW, overestimate the frequency for each dimer. BEEF-vdW in particular performs poorly. For example, it predicts a frequency of $13.1$ meV for HeNe, which has an experimental value of $4.334$ meV, and overpredicts the experimental frequencies by an average of $4.7$ meV. The previously-observed tendency of GGA functionals to overestimate $s$-shell electron interaction energy and underestimate $p$-shell electron interaction energy \cite{tao,ruzsinszky,furche} is also reproduced in our results. The exceptions are the vdW functionals, which consistently overestimate the vibrational frequency for every dimer. This tendency can also be observed in considering the binding energy for each dimer, which is calculated from the electronic energy difference between the dimer and its constituent monomers. A table of binding energies can be found in the Supporting Information. In considering the MAE in comparison to the experimental frequencies, the most accurate predictions come via Tao and Perdew using the meta-GGA level TPSSh functional ($1.012$ meV MAE). The best GGA-level functional is PBE ($1.339$ meV MAE for Tao and Perdew's values, $1.9$ meV MAE in this work).

Statistics for the BEEF-vdW ensembles are presented on the right side of Table \ref{tbl: rare gas dimers}. By the JB test, most notably, none of the ensembles are Gaussian. With the exception of He$_2$, the experimental value for each frequency is bounded to within two standard deviations of the BEEF-vdW result, with an average deviation of $1.12\sigma$. For each dimer, the majority of the frequencies in the ensemble overestimate the experimental value. The frequency predictions from the other tested XC functionals are also bounded by the ensemble to within two standard deviations, but agreement between BEEF-vdW and the other functionals is worse than for the benchmark molecules in Section \ref{subsect: benchmarks}. The smallest average deviation from the BEEF-vdW predictions came from optPBE-vdW ($0.44\sigma$), followed by vdW-DF2 ($0.60\sigma$), PW91 via Patton and Pederson ($0.73\sigma$), RPBE ($0.81\sigma$), PBE ($0.86\sigma$), PBE via Tao and Perdew ($1.08\sigma$), and TPSSh via Tao and Perdew ($1.25\sigma$). This higher spread is also present in the BEEF-vdW ensembles. The COV of every ensemble is higher than the COV of any frequency ensemble in Section \ref{subsect: benchmarks}, with the exception of the C$_2$H$_6$ torsional mode. The high COV values and consistent overprediction of the experimental frequencies reflect the difficulty that GGA-level XC functionals have in consistently and correctly describing the van der Waals interactions in rare-gas dimers.

\subsection{S22}\label{subsect: S22}

The rare-gas dimers from Section \ref{subsect: rare gas dimers} may not be representative of non-covalent interactions in larger systems \cite{cerny, s22}. As such, we next apply our methodology to S22, a set of 22 non-covalently bonded molecular complexes \cite{s22}. The complexes in S22 are bonded by hydrogen bonds, dispersion bonds, or a mix of the two. The size of each complex varies between 6 and 30 atoms. The complexes in S22 contain both intermolecular and intramolecular vibrational degrees of freedom. Note that there are always six intermolecular vibrational modes for each complex. By predicting their vibrational frequencies, it will be possible to examine a given XC functional's ability to describe both covalent and non-covalent interactions in the same system.

The results for the vibrational frequencies of the water dimer are presented in Table \ref{tbl: water dimer} with internal coordinate decomposition for each mode. We include frequencies from Xu and Goddard \cite{xu} predicted with the GGA functional BLYP \cite{blyp1, blyp2} and the hybrid functional B3LYP \cite{blyp1, blyp2, b3lyp1, b3lyp2}. From Dunn et al. \cite{dunn}, we include frequencies predicted with MP2 theory \cite{mp2} using the aug-cc-pVDZ basis set. The results for the entire S22 dataset are available in the Supporting Information. 

BEEF-vdW predicts the highest intramolecular frequencies (modes 7 to 12), with values comparable to the B3LYP frequencies reported by Xu and Goddard. With respect to the experimental intramolecular frequencies reported by Fredin et al.\cite{fredin}, BEEF-vdW is the best functional out of the six tested in this work, with an MAE of 3.8 meV. The best results come via Xu and Goddard \cite{xu} using B3LYP. A comparison of intermolecular frequencies is not possible because, with the exception of mode 4, the experimental measurements have not been corrected for anharmonic effects, and therefore the reported frequencies are not harmonic. We do note, however, that BEEF-vdW strongly overestimates the intermolecular frequencies in comparison to the other XC functionals. All six experimental intramolecular frequencies are bounded to within one standard deviation of the BEEF-vdW value, with an average deviation of $0.49\sigma$. Ensembles for the intramolecular modes all have COV lower than $0.05$, which is consistent with ensembles for H$_2$O reported in Table \ref{tbl: small molecules}. Ensembles for intermolecular modes have relatively high COV, comparable to the COV values of the C$_2$H$_6$ torsional mode and the rare-gas dimer modes, with the highest (1.22) coming from the lowest frequency mode. Intermolecular modes of other S22 complexes also tend to have higher ensemble COV values. All six intermolecular ensembles for the water dimer are non-Gaussian, while this is true of only two of the six intramolecular modes. As with the C$_2$H$_6$ torsional mode, some BEEF-vdW ensemble functionals predict unstable imaginary frequencies for mode 1 (458 functionals) and mode 2 (23 functionals).  Challenges in describing interaction of small molecule clusters with different XC approximations has been previously noted \cite{gillan}.

\begin{table*}
\resizebox{7in}{!}{
  \begin{tabular}{l | lll | >{\RaggedLeft\arraybackslash}p{1.2cm}>{\RaggedLeft\arraybackslash}p{1.2cm}>{\RaggedLeft\arraybackslash}p{1.2cm}>{\RaggedLeft\arraybackslash}p{1.2cm}>{\RaggedLeft\arraybackslash}p{1.2cm}>{\RaggedLeft\arraybackslash}p{1.2cm}>{\RaggedLeft\arraybackslash}p{1.2cm}>{\RaggedLeft\arraybackslash}p{1.2cm}>{\RaggedLeft\arraybackslash}p{1.2cm}  >{\RaggedLeft\arraybackslash}p{1.2cm} | >{\RaggedLeft\arraybackslash}p{1.2cm}>{\RaggedLeft\arraybackslash}p{1.2cm}>{\RaggedLeft\arraybackslash}p{1.2cm}>{\RaggedLeft\arraybackslash}p{1.2cm}>{\RaggedLeft\arraybackslash}p{1.2cm}>{\RaggedLeft\arraybackslash}p{1.2cm}}
    \hline
    Mode & Stretch & Bend & Tors & Expt. & PBE & RPBE & PBEsol & PW91 & optPBE & BLYP\cite{xu} & B3LYP\cite{xu} & MP2\cite{dunn}  & BEEF  & $\mu$ & $\sigma$ & COV& Skew & Kurt. & JB\\
     &  &  &  &  &  &  &  &  & -vdW &  & &  & -vdW & &  & &  \\
    \hline
    1 &  &  &  & 10.9\cite{braly} & 17.0 & 13.2  & 17.2 & 14.2 & 13.4  & 16.1 & 16.5 & 16.0  & 13.5 & 26.3 & 25.2 & 1.22 &  0.49 & -1 & 164.7 \\
    2 &  &  &  & 12.8\cite{braly} & 19.7 & 15.3   & 21.3 & 18.8 & 17.4   & 19.3 & 19.7 & 18.5& 25.3 & 23.3 & 25.3 & 1.10 & 0.73 & -0.77 & 228.9 \\
    3 &  &  &  & 13.4\cite{braly} & 21.1 & 18.5   & 22.3 & 20.0 & 17.8   & 20.2 & 19.8 & 18.8 & 35.7 & 26.1 & 25.6 & 0.98 & 0.63 & -0.90 & 199.7 \\
    4 &  &  &  & 18.6\cite{dyke} & 23.4 & 19.2   & 27.0 & 23.4 & 20.9   & 23.8 & 23.9 & 22.9 & 39.8 & 31.1 & 23.6 & 0.76 & 0.64 & -0.79 & 188.6 \\
    5 &  &  &  & 38.6\cite{bouteiller} & 45.9 & 38.4   & 49.0 & 44.0 & 42.1  & 46.1 & 46.5 & 44.6& 47.7 & 40.3  & 22.6 & 0.56 & 0.51 & -0.90 & 154.6 \\
    6 &  &  &  & 64.8\cite{bouteiller} & 79.5 & 69.1   & 88.2 & 80.2 & 72.2   & 77.0 & 78.6 & 79.8 & 74.6& 64.8 & 21.1 & 0.33 & -0.32 & -0.47 & 52.5 \\
    7 & 0 & 1 & 0 & 204.9\cite{fredin} & 196.9 & 198.7  & 193.9 & 196.8 & 199.0  & 198.9 & 203.0 & 201.5 & 204.6 & 201.6 &  9.2 & 0.05 & -0.15 & 0.19 & 10.5 \\
    8 & 0 & 1 &0  & 206.9\cite{fredin} & 200.0 & 201.1  & 198.2 & 200.4 & 201.0  & 201.0 & 205.3 & 203.8& 206.9 & 203.7 & 9.1 & 0.04 & -0.16 & 0.17 & 10.9 \\
    9 & 1 & 0 & 0 & 460.9\cite{fredin} & 440.4 & 447.8  & 431.6 & 440.3 & 444.6  & 437.1 & 456.0 & 459.3& 456.5 & 457.0 & 9.2 & 0.02 & -0.06 & -0.03 & 1.4 \\
    10 & 1 &0  &  0 & 470.7\cite{fredin} & 460.3 & 459.9  & 460.7 & 461.7 & 457.9  & 452.9 & 470.4 & 470.9& 465.6 & 466.6 & 8.2 & 0.02 & -0.06 & 0.03 & 1.2 \\
    11 & 1 & 0 & 0 & 481.2\cite{fredin} & 470.0 & 469.6  & 470.6 & 471.3 & 467.8  & 462.3 & 480.2 & 484.3 & 475.3 & 476.9 & 8.8 & 0.02 & -0.06 & 0.03 & 1.3 \\
    12 & 1 & 0 &0  & 483.4\cite{fredin} & 472.5 & 471.9  & 473.2 & 473.8 & 469.8  & 464.7 & 482.1 & 486.8 & 476.6 & 478.7 & 8.6 & 0.02 & -0.07 & 0.04 & 1.6 \\
    \hline
    ME &  &  & &  & 11.3 & 9.8 & 13.3 & 10.6 &  11.3 & 15.2 & 1.8 & 0.2 & 3.8 &  &  &  &  &  &  \\
    MAE &  &  & &  & 11.3 & 9.8  & 13.3 & 10.6 & 11.3  & 15.2 & 1.8 & 2.5 & 3.8 &  &  &  &  &  &  \\
   \hline
  \end{tabular}}
  \caption{Water dimer vibrational frequencies (meV) predicted with various XC functionals. Modes 1 through 6 are intermolecular, while modes 7 through 12 are intramolecular. BLYP and B3LYP values were predicted with aug-cc-pVTZ(-f) basis sets\cite{xu}, and MP2 values with the aug-cc-pVDZ basis set\cite{dunn}. MP2 values were multiplied by a scale factor by Dunn et al. and are rescaled here. All experimental intermolecular modes except mode 4 are not corrected for anharmonic effects. ME and MAE are calculated for intramolecular frequencies only.}
  \label{tbl: water dimer}
\end{table*}

\subsection{Comparison}\label{subsect: comparison}

\begin{figure}
    \centering
    \begin{subfigure}{3.5 in}
        \centering
        \includegraphics[width=3.25 in]{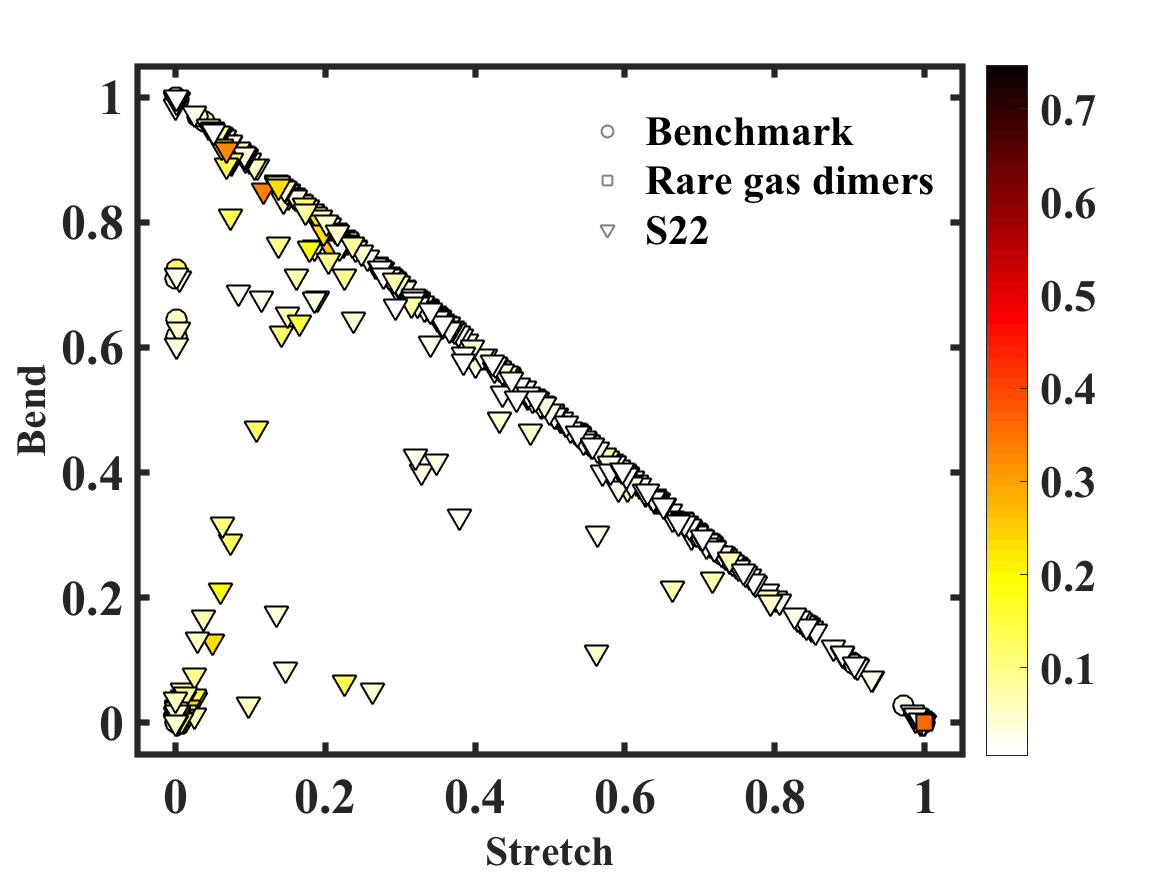}
        \caption{}\label{fig: bend stretch}
    \end{subfigure}\hfill
    \begin{subfigure}{3.5 in}
        \centering
        \includegraphics[width=3.25 in]{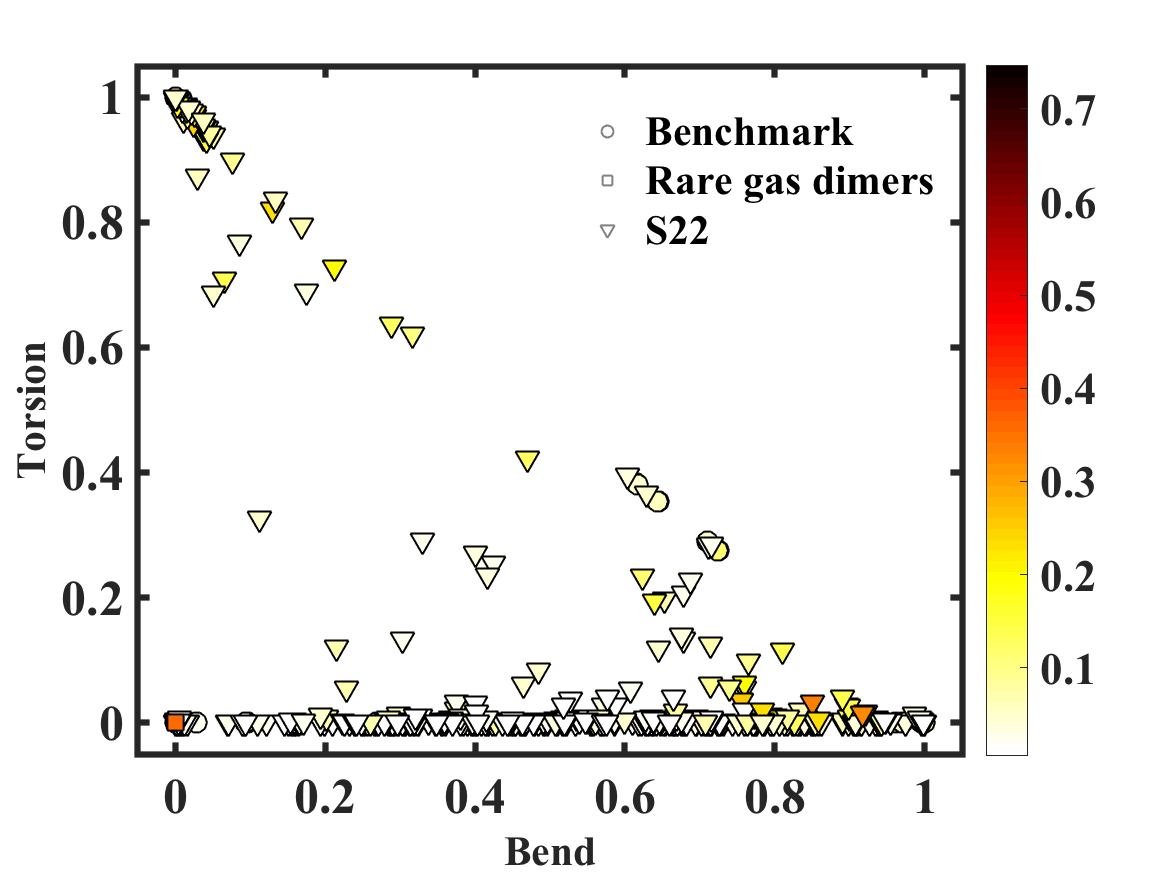}
        \caption{}\label{fig: bend torsion}
    \end{subfigure}
    \caption{Coefficient of variation of the BEEF-vdW ensemble for each mode as functions of (a) stretching and bending character and (b) bending and torsional character. Modes from the benchmark, rare-gas dimer, and S22 datasets are all included.}
	\label{fig: bend stretch torsion 2d plots}
\end{figure}

Based on the decomposition of the normal modes in terms of stretching, bending, and torsional motion for every molecule and complex considered, we now examine the relationship between the physical motion of the mode and the spread of its BEEF-vdW ensemble, as measured by COV. In Figure \ref{fig: bend stretch}, the bending component is plotted against the stretching component with marker color determined by the magnitude of the COV. Figure \ref{fig: bend torsion} is similar, with the bending component plotted versus the torsional component. The data plotted in Figures \ref{fig: bend stretch} and \ref{fig: bend torsion} indicate that COV tends to (i) increase as the component of bending or torsion increases and (ii) decrease as the stretching component increases. The rare-gas dimers, which have high COV stretch modes, are an exception. 

\begin{figure}
  \includegraphics[width=3.25 in]{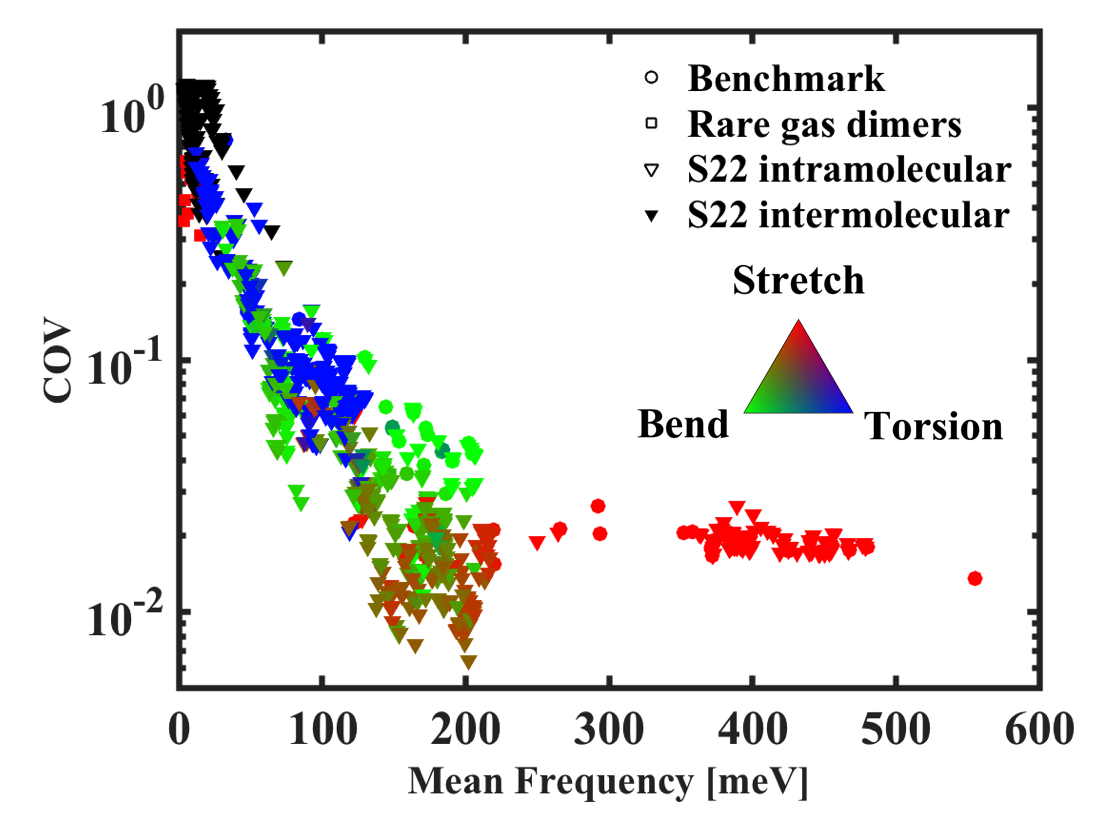}
  \caption{The COV as a function of BEEF-vdW ensemble mean frequency, with the physical character of the mode indicated by the marker color. Pure stretch, bend, and torsion modes are red, green, and blue. S22 intermolecular modes are black.}
  \label{fig: cov vs mean}
\end{figure}

The COV is plotted as a function of the BEEF-vdW ensemble mean frequency in Figure \ref{fig: cov vs mean}. The physical character of the mode is indicated by the marker color, with red, green, and blue representing pure stretch, bend, and torsion. The intermolecular modes for the S22 complexes are marked in black. The COV decreases as the mean frequency increases, with the S22 intermolecular modes having the largest spreads. Modes with frequencies higher than 200 meV are almost entirely stretch modes with low COV. Torsional and bending modes, rare-gas dimer stretching modes, and intermolecular modes tend to have frequencies lower than 200 meV and higher COV.

\begin{figure*}
  \includegraphics[width=7 in]{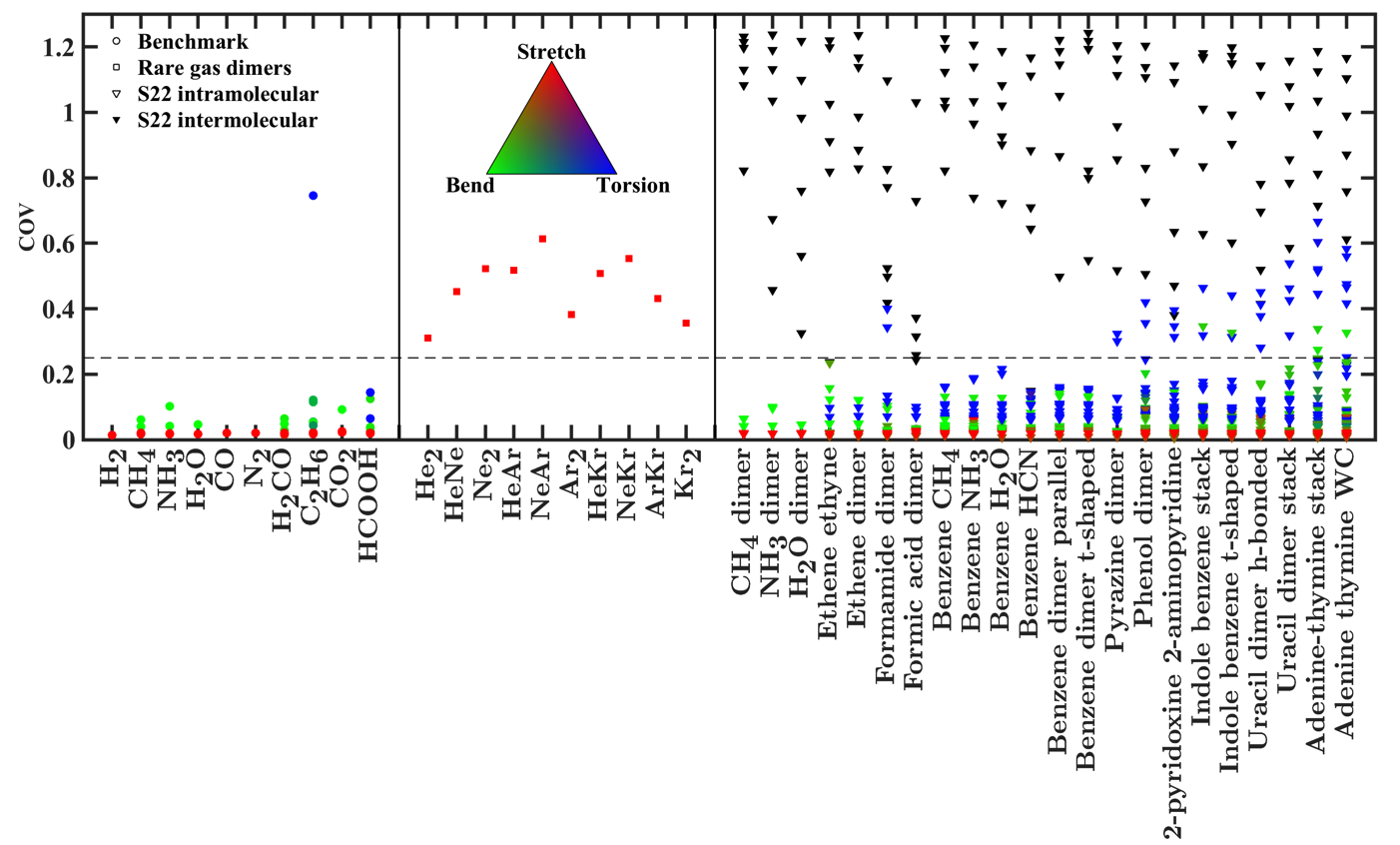}
  \caption{The COV for all modes in each of the three datasets considered in this work. The physical character is indicated by the color of the marker. Molecules within each dataset (benchmarks, rare-gas dimers, and S22) are listed in order of increasing mass.}
  \label{fig: cov allmodes}
\end{figure*}

The COV for all modes, organized by dataset and molecule, are provided in Figure \ref{fig: cov allmodes}. Within each dataset, the molecules and complexes are listed in order of increasing mass. An analogous plot with a logarithmic vertical axis is provided in the Supporting Information. From Figures \ref{fig: cov vs mean} and \ref{fig: cov allmodes}, a general ordering of mode types based on their COV can be observed. Stretching modes tend to have the lowest COV, followed by bending modes, torsional modes, and intermolecular modes. For every molecule in S22, nearly all intermolecular modes have higher COV than any other mode. The relatively high COV values for bending and torsional modes, in contrast with the stretching modes, indicates disagreement among XC functionals in describing these vibrations. As stretching modes are typically localized while bending and torsional modes are not\cite{halls}, our results suggest that XC functionals at the GGA level have difficulty consistently describing a molecule's non-local vibrational behavior.  The stretching modes of the rare-gas dimers have high COV and are an exception to this rule. These systems are bonded by van der Waals interactions, however, and the high COV values correctly reflect disagreement in XC functionals in describing non-covalent interactions. This inconsistency also causes the high COV values of the intermolecular S22 modes.

Based on Figure \ref{fig: cov allmodes} there is a separation of modes at a COV value of  $0.25$, which is marked by a horizontal dashed line. Modes above this boundary include the high COV stretch modes of the rare-gas dimers, most of the S22 intermolecular modes, and some bending and torsional modes from C$_2$H$_6$ and several S22 complexes. Based on this separation, we propose a criterion of a COV equal to $0.25$ for disagreement among XC functionals in predicting the frequency. For ensembles with COV larger than $0.25$, extra care should be taken in choosing an XC functional for the frequency prediction.

\section{Summary}

We presented a computationally efficient method to quantify the uncertainty due to the choice of the XC functional at the GGA level in DFT predictions of harmonic vibrational frequencies. To test the robustness of this method, we considered three sets of molecules of varying size and complexity. The first set consisted of small benchmark molecules (Tables \ref{tbl: small molecules} and \ref{tbl: small molecules c2h6 and hcooh}), none of which were larger than 8 atoms. We found that the BEEF-vdW ensemble bounds most experimental frequencies and the frequency predictions of six other XC functionals to within one standard deviation (e.g., Figure \ref{fig: nh3_highfreq_vib}). We then applied the method to ten rare-gas dimers (Table \ref{tbl: rare gas dimers}) and the S22 dataset (Table \ref{tbl: water dimer}) of molecular complexes, which offered a variety of systems that differed in their bonding environments (i.e., covalent, hydrogen, van der Waals) and physical nature of the their vibrational modes (i.e., stretch, bend, and torsion, and intermolecular). Our results show that frequency predictions for modes with delocalized motion, such as torsion and bending, and modes involving non-covalent bonds are more sensitive to choice of XC functional in comparison to predictions for localized and covalent stretch modes (Figures \ref{fig: bend stretch torsion 2d plots}, \ref{fig: cov vs mean}, and \ref{fig: cov allmodes}). Our proposed method can therefore be used to link DFT uncertainty to the physical behavior of the system.

\begin{acknowledgement}
H. L. P. acknowledges support from a Presidential Fellowship at Carnegie Mellon University and helpful feedback from Dilip Krishnamurthy.  V. V. acknowledges support from the National Science Foundation under award CBET-1554273.

\end{acknowledgement}

\newpage

\bibliography{bibliography}



\section*{Supplementary material}

# Supporting Information: Uncertainty quantification in first-principles predictions of harmonic vibrational frequencies of molecules and molecular complexes

Holden L. Parks, Alan. J. H. McGaughey, and Venkatasubramanian Viswanathan*

*Department of Mechanical Engineering, Carnegie Mellon University, Pittsburgh, Pennsylvania 15213, USA*

# S1 Rare-gas dimers binding energy

In Table S1 we present binding energies in meV for the rare-gas dimers, which is calculated from the electronic energy difference between the dimer and its constituent monomers

$$E_{bind} = E_{dimer} - (E_{monomer,\ 1} + E_{monomer,\ 2}). \tag{S.1}$$

Mean error (ME) and mean absolute error (MAE) are also reported in the table. Negative ME values mean that, on average, the functional tends to overestimate the binding energy. The vdW functionals (optPBE-vdW, vdW-DF2, and BEEF-vdW) all overestimate binding energy, with BEEF-vdW performing particularly poorly. The best results by MAE are from Tao and Perdew[1] using the TPSSH functional.

Table S1: Binding energies for rare-gas dimers in meV as calculated with PBE, RPBE, optPBE-vdW, vdW-DF2, and BEEF-vdW. Experimental values are from Ogilvie and Wang.[3,4] The PBE and TPSSh binding energies from Tao and Perdew[1] include BSSE corrections.

| Molecule | Expt. | PBE | PBE[1] | RPBE | optPBE-vdW | vdW-DF2 | PW91[2] | BEEF-vdW | TPSSh[1] |
|---|---|---|---|---|---|---|---|---|---|
| He$_2$ | 0.9 | 3.2 | 2.7 | 4.9 | 8.7 | 4.2 | 10.3 | 10.6 | 1.4 |
| HeNe | 1.8 | 4.8 | 2.7 | 6.5 | 12.6 | 7.2 | 12.3 | 16.5 | 1.4 |
| HeAr | 2.5 | 7.8 | 3.5 | 5.8 | 13.4 | 7.5 | 11.4 | 15.9 | 1.9 |
| HeKr | 2.5 | 6.7 | 3.5 | 5.5 | 13.1 | 7.3 | 11.1 | 15.6 | 1.9 |
| Ne$_2$ | 3.6 | 6.4 | 4.9 | 7.4 | 19.1 | 12.3 | 14.3 | 22.6 | 2.4 |
| NeAr | 5.8 | 6.7 | 5.4 | 7.3 | 21.4 | 13.8 | 14.4 | 24.0 | 2.4 |
| NeKr | 6.1 | 5.8 | 5.7 | 7.6 | 22.3 | 15.7 | 14.5 | 25.6 | 2.7 |
| Ar$_2$ | 12.3 | 3.8 | 6.0 | 5.4 | 25.2 | 18.5 | 14.2 | 26.7 | 2.7 |
| ArKr | 15.6 | 6.5 | 6.3 | 7.3 | 28.3 | 20.3 | 14.2 | 28.8 | 3.0 |
| Kr$_2$ | 17.3 | 7.0 | 6.5 | 6.9 | 29.7 | 23.1 | 14.3 | 29.3 | 3.3 |
| ME | | 0.1 | 2.1 | 0.4 | -12.5 | -6.2 | -6.3 | -14.7 | 4.5 |
| MA | | 4.7 | 3.3 | 4.7 | 12.5 | 6.2 | 7.1 | 14.7 | 4.6 |

## S2 COV scatter plot (log scale)

In Figure S1 we plot the COV on a logarthmic vertical scale for all modes considered in this work, organized by dataset and molecule. There is an analogous plot without logarithmic scale in the main text.

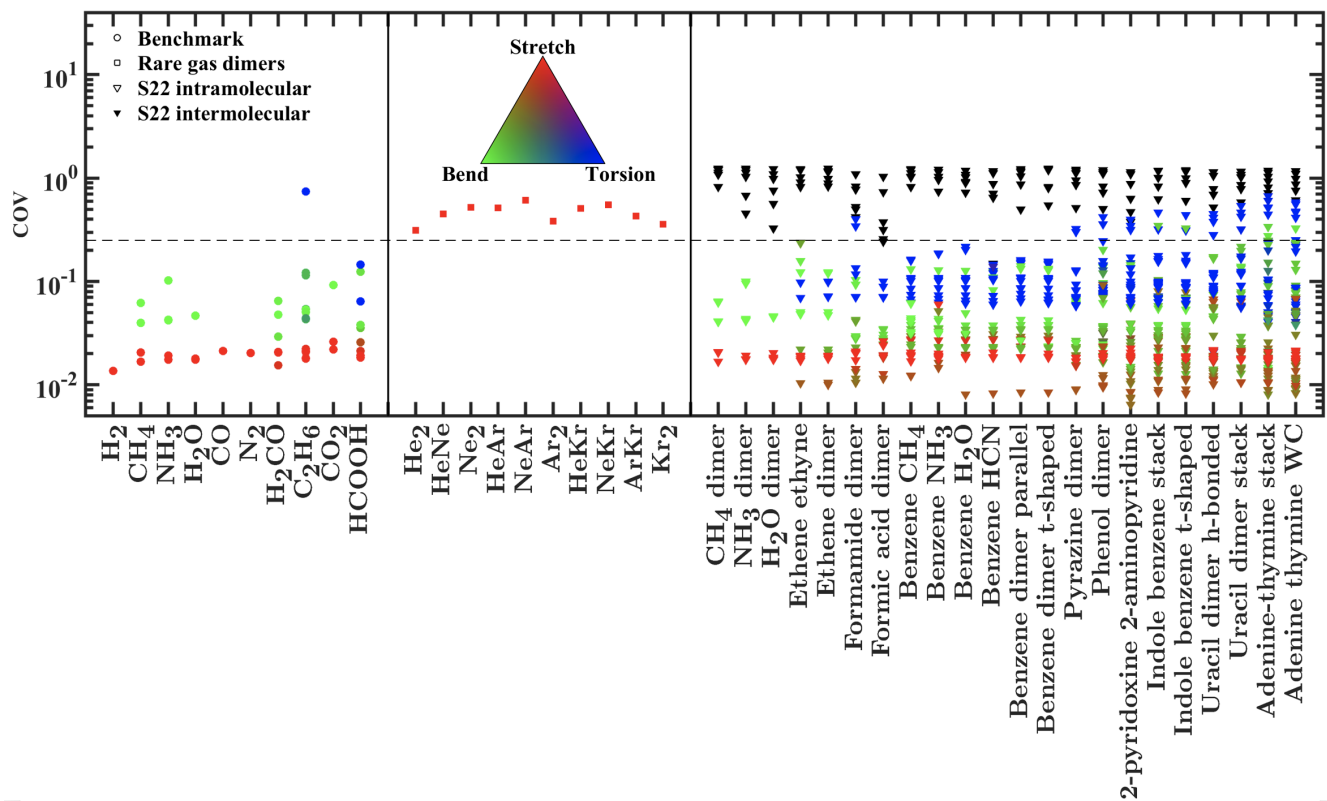

Figure S1: The coefficient of variation, plotted on a log scale, for all modes in each of the three datasets considered in this work, with modes separated by molecule. The physical character is indicated by the color of the marker. Molecules within each dataset (benchmarks, rare gas dimers, and S22) are listed in order of increasing mass.

# S3 Vibrational thermodynamic quantities

The vibrational degrees of freedom in a non-periodic system contribute to its free energy, enthalpy, ZPE, and entropy. If the system considered is at a temperature well below its dissociation limit, then the use of the harmonic approximation is justified[5] and the vibrational components of these thermodynamic quantities may be determined using harmonic mode frequencies obtained from the method outlined in the main text. We use $\omega_k$ to denote the frequency associated with the $k$th vibrational mode and $\epsilon_k$ to denote the energy associated with this mode, so that $\epsilon_k = \hbar\omega_k$, where $\hbar$ is the reduced Planck constant.

The Gibbs free energy of a system is given by

$$G = H - TS, \tag{S.2}$$

where $H$ is the enthalpy, $T$ is the temperature, and $S$ is the entropy. We assume that the molecules of interest are ideal gases. The enthalpy of an idea gas is[6]

$$H = E_{elec} + E_{ZPE} + \int_0^T C_p dT, \tag{S.3}$$

where $E_{elec}$ is the electronic energy (the direct output of a DFT calculation), $E_{ZPE}$ is the ZPE, and $C_p$ is the constant pressure specific heat. The ZPE is[7,8]

$$E_{ZPE} = \frac{1}{2} \sum_k \epsilon_k. \tag{S.4}$$

For an ideal gas, $C_p$ is given by

$$C_p = k_B + C_{v,rot} + C_{v,trans} + C_{v,vib}, \tag{S.5}$$

where $k_B$ is the Boltzmann constant and comes from converting the constant volume specific heat to the constant pressure specific heat, $C_{v,rot}$ is the constant-volume rotational specific heat, $C_{v,trans}$ is the constant-volume translational specific heat, and $C_{v,vib}$ is the constant-

volume vibrational specific heat. For an ideal gas, $C_{v,trans} = 3k_B/2$ and $C_{v,rot}$ is $k_B$ if the system is linear and $3k_B/2$ otherwise. The vibrational specific heat is given by

$$C_{v,vib} = \sum_k \frac{\exp\left(\frac{\epsilon_k}{k_B T}\right)}{k_B} \left[\frac{\epsilon_k/T}{\exp\left(\frac{\epsilon_k}{k_B T}\right) - 1}\right]^2. \quad (S.6)$$

Thus the integrals of $C_{v,trans}$, $C_{v,rot}$, and $k_B$ are constants scaled by $T$. The integral of $C_{v,vib}$ becomes[7,8]

$$\int_0^T C_{v,vib} dT = \sum_k \frac{\epsilon_k}{\exp\left(\frac{\epsilon_k}{k_B T}\right) - 1}. \quad (S.7)$$

The entropy is given by[6]

$$S = -k_B \ln\frac{P}{P_0} + S_{rot} + S_{trans} + S_{elec} + S_{vib}, \quad (S.8)$$

where $P$ is the pressure and $P_0$ is standard pressure, 1 atm. Using the ideal gas approximation, the rotational ($S_{rot}$), translational ($S_{trans}$), and electrical ($S_{elec}$) contributions to entropy depend only on some combination of the temperature, symmetry number, and total spin state of the molecule (i.e., they do not depend on the $\omega_k$). The vibrational entropy component $S_{vib}$ is[7,8]

$$S_{vib} = \sum_k \frac{\epsilon_k}{\exp\left(\frac{\epsilon_i}{k_B T}\right)} - \ln\left[1 - \exp\left(-\frac{\epsilon_k}{k_B T}\right)\right]. \quad (S.9)$$

By calculating the ensemble of vibrational frequencies detailed in the main text, it is possible to determine ensembles of thermodynamic quantities. For example, in Figure S2a, we plot the ensemble for the ZPE of $NH_3$, which is calculated with eq S.4 using the frequencies for $NH_3$ from the main text. The distribution bounds the predictions of the other tested XC functionals and has a small standard deviation of 0.006 eV. In Figure S2b, we plot the ensemble for the entropy of reaction at standard temperature and pressure (273.15 K and 1

bar) for the Haber process, $3H_2+N_2 \rightarrow 2NH_3$, a crucial process for the industrial production of ammonia.[9] The entropy of reaction is given by

$$\Delta S = S_{NH_3} - \left(\frac{1}{2}S_{N_2} + \frac{3}{2}S_{H_2}\right), \tag{S.10}$$

where $S_{NH_3}$, $S_{N_2}$, and $S_{H_2}$ are calculated using eq S.8. The entropy ensemble has standard deviation of $1.6 \times 10^{-3}$ meV/K. The small standard deviations for both distributions indicate that there is little variation in predicted ZPE and entropy values between different XC functionals. However, here we have only considered thermodynamic ensembles calculated from frequencies of $N_2$, $H_2$, and $NH_3$, all of which are simple molecules. Larger molecules, such as the S22 molecular complexes, could yield thermodynamic ensembles with varied behavior.

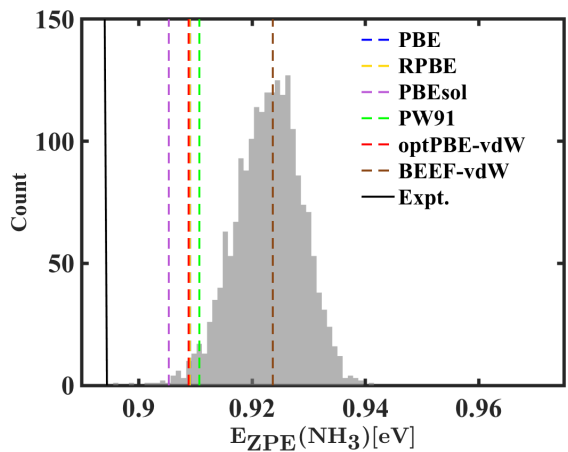

(a)

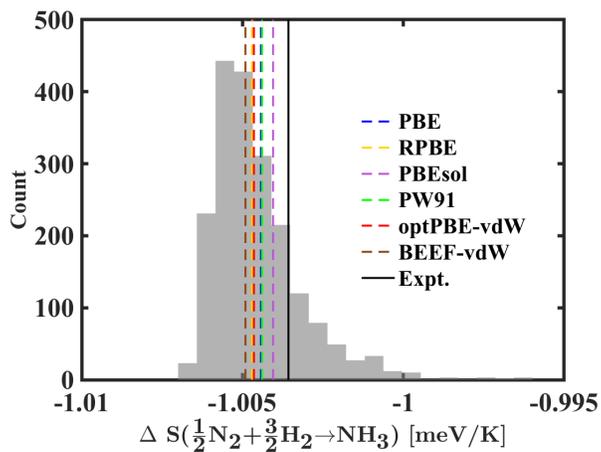

(b)

Figure S2: Distributions for (a) the zero-point energy of $NH_3$, with standard deviation 0.006 eV and (b) entropy of reaction at standard temperature and pressure (273.15 K and 1 bar) for the Haber process, with standard deviation $1.6 \times 10^{-3}$ meV/K. Experimental values are calculated from experimental frequencies from Shimanouchi.[10]

# S4  S22 BEEF-vdW ensemble statistics

This section contains results for the S22 dataset of molecular complexes. For each vibrational mode, the tables contain internal coordinate decompositions, with S, B, and T standing for stretch, bend, and torsion component; predictions using the PBE, RPBE, PBEsol, PW91, optPBE-vdW, and BEEF-vdW XC functionals; and BEEF-vdW ensemble statistics $\mu$ (mean), $\sigma$ (standard deviation), COV (coefficient of variation), skew, Kurt. (kurtosis), JB (Jarque-Bera statistic), and Imag. (number of ensemble functionals predicting imaginary frequencies for this mode). Imaginary frequencies have zero real part, but these values are not included in calculations of ensemble statistics.

Table S2: $CH_4$ dimer harmonic vibrational frequencies.

|  | S | B | T | PBE | RPBE | PBEsol | PW91 | optPBE-vdW | BEEF-vdW | $\mu$ | $\sigma$ | COV | Skew | Kurt. | JB | Imag. |
|---|---|---|---|---|---|---|---|---|---|---|---|---|---|---|---|---|
| 1 | | | | 1.5 | 0.0 | 0.0 | 1.7 | 2.2 | 37.2 | 36.3 | 17.5 | 0.48 | 0.0 | -0.56 | 26.1 | 1021 |
| 2 | | | | 2.4 | 0.0 | 2.3 | 2.3 | 4.3 | 37.8 | 32.0 | 20.1 | 0.63 | -0.02 | -0.85 | 60.8 | 874 |
| 3 | | | | 4.5 | 1.6 | 3.1 | 3.1 | 5.0 | 38.4 | 26.3 | 22.2 | 0.84 | 0.24 | -1.14 | 127.0 | 596 |
| 4 | | | | 6.0 | 2.5 | 4.0 | 5.1 | 7.6 | 38.7 | 23.0 | 21.9 | 0.95 | 0.53 | -1.01 | 176.8 | 326 |
| 5 | | | | 6.8 | 4.6 | 4.9 | 6.3 | 7.7 | 39.2 | 21.3 | 21.7 | 1.02 | 0.67 | -0.87 | 213.4 | 117 |
| 6 | | | | 8.7 | 7.1 | 6.8 | 8.9 | 9.5 | 39.4 | 23.7 | 19.3 | 0.81 | 0.76 | -0.47 | 213.0 | 16 |
| 7 | 0.0 | 1.0 | 0.0 | 158.4 | 159.7 | 155.4 | 159.1 | 161.6 | 171.2 | 162.9 | 10.5 | 0.06 | -0.2 | 0.23 | 17.5 | 0 |
| 8 | 0.0 | 1.0 | 0.0 | 158.8 | 160.1 | 156.0 | 159.6 | 161.9 | 171.2 | 163.4 | 10.3 | 0.06 | -0.18 | 0.22 | 15.1 | 0 |
| 9 | 0.0 | 1.0 | 0.0 | 158.9 | 160.1 | 156.0 | 159.6 | 162.0 | 171.4 | 163.7 | 10.3 | 0.06 | -0.18 | 0.21 | 14.8 | 0 |
| 10 | 0.0 | 1.0 | 0.0 | 159.2 | 160.2 | 156.4 | 159.9 | 162.2 | 171.5 | 163.8 | 10.3 | 0.06 | -0.18 | 0.21 | 14.7 | 0 |
| 11 | 0.0 | 1.0 | 0.0 | 159.2 | 160.3 | 156.5 | 160.0 | 162.3 | 171.6 | 164.0 | 10.3 | 0.06 | -0.18 | 0.22 | 14.3 | 0 |
| 12 | 0.0 | 1.0 | 0.0 | 159.7 | 160.7 | 156.9 | 160.5 | 162.9 | 172.6 | 164.5 | 10.3 | 0.06 | -0.18 | 0.2 | 14.2 | 0 |
| 13 | 0.0 | 1.0 | 0.0 | 186.2 | 187.2 | 183.9 | 186.9 | 188.4 | 196.3 | 190.5 | 7.8 | 0.04 | -0.11 | 0.12 | 5.5 | 0 |
| 14 | 0.0 | 1.0 | 0.0 | 186.3 | 187.2 | 183.9 | 187.0 | 188.5 | 196.5 | 190.6 | 7.8 | 0.04 | -0.11 | 0.11 | 5.0 | 0 |
| 15 | 0.0 | 1.0 | 0.0 | 187.2 | 187.9 | 184.9 | 187.9 | 189.3 | 197.1 | 191.3 | 7.7 | 0.04 | -0.11 | 0.12 | 5.3 | 0 |
| 16 | 0.0 | 1.0 | 0.0 | 187.2 | 187.9 | 185.0 | 187.9 | 189.3 | 197.3 | 191.4 | 7.7 | 0.04 | -0.11 | 0.11 | 4.8 | 0 |
| 17 | 1.0 | 0.0 | 0.0 | 368.5 | 367.6 | 367.1 | 369.4 | 367.5 | 365.4 | 372.3 | 6.3 | 0.02 | -0.06 | 0.09 | 1.8 | 0 |
| 18 | 1.0 | 0.0 | 0.0 | 368.5 | 367.6 | 367.1 | 369.5 | 367.5 | 365.7 | 372.4 | 6.3 | 0.02 | -0.06 | 0.09 | 1.9 | 0 |
| 19 | 1.0 | 0.0 | 0.0 | 382.4 | 380.9 | 382.1 | 383.0 | 379.7 | 376.8 | 385.2 | 7.9 | 0.02 | -0.09 | 0.08 | 3.0 | 0 |
| 20 | 1.0 | 0.0 | 0.0 | 382.4 | 381.0 | 382.2 | 383.0 | 379.9 | 376.9 | 385.3 | 7.9 | 0.02 | -0.09 | 0.08 | 3.0 | 0 |
| 21 | 1.0 | 0.0 | 0.0 | 382.6 | 381.0 | 382.2 | 383.2 | 380.0 | 377.1 | 385.5 | 7.9 | 0.02 | -0.09 | 0.08 | 3.0 | 0 |
| 22 | 1.0 | 0.0 | 0.0 | 382.6 | 381.2 | 382.3 | 383.2 | 380.1 | 377.3 | 385.6 | 7.9 | 0.02 | -0.09 | 0.08 | 3.1 | 0 |
| 23 | 1.0 | 0.0 | 0.0 | 382.8 | 381.8 | 382.7 | 383.6 | 380.6 | 377.8 | 386.2 | 8.0 | 0.02 | -0.08 | 0.07 | 2.8 | 0 |
| 24 | 1.0 | 0.0 | 0.0 | 382.9 | 381.9 | 382.8 | 383.7 | 380.7 | 378.0 | 386.3 | 8.0 | 0.02 | -0.08 | 0.07 | 2.8 | 0 |

Table S3: NH$_3$ dimer harmonic vibrational frequencies.

|    | S   | B   | T   | PBE   | RPBE  | PBEsol | PW91  | optPBE-vdW | BEEF-vdW | $\mu$ | $\sigma$ | COV  | Skew  | Kurt. | JB    | Imag. |
|----|-----|-----|-----|-------|-------|--------|-------|------------|----------|-------|----------|------|-------|-------|-------|-------|
| 1  |     |     |     | 2.1   | 0.5   | 0.1    | 2.6   | 0.9        | 27.5     | 30.1  | 23.5     | 0.78 | 0.17  | -1.09 | 108.4 | 729   |
| 2  |     |     |     | 11.5  | 9.1   | 10.8   | 12.3  | 9.8        | 31.7     | 23.3  | 24.3     | 1.04 | 0.58  | -0.98 | 191.8 | 272   |
| 3  |     |     |     | 14.3  | 11.6  | 15.2   | 15.1  | 11.9       | 32.3     | 21.7  | 24.4     | 1.12 | 0.71  | -0.85 | 228.9 | 21    |
| 4  |     |     |     | 16.4  | 13.4  | 18.8   | 16.2  | 14.8       | 33.2     | 23.2  | 24.1     | 1.04 | 0.69  | -0.83 | 217.9 | 0     |
| 5  |     |     |     | 26.1  | 22.2  | 29.0   | 25.9  | 22.6       | 39.8     | 30.5  | 20.6     | 0.67 | 0.8   | -0.54 | 237.7 | 0     |
| 6  |     |     |     | 52.3  | 46.0  | 57.3   | 52.5  | 48.8       | 55.9     | 45.6  | 20.8     | 0.46 | 0.21  | -0.97 | 93.2  | 0     |
| 7  | 0.0 | 1.0 | 0.0 | 126.4 | 129.2 | 123.2  | 126.3 | 128.1      | 135.4    | 130.0 | 13.0     | 0.1  | -0.37 | 0.67  | 83.2  | 0     |
| 8  | 0.0 | 1.0 | 0.0 | 129.5 | 131.3 | 127.6  | 129.6 | 130.7      | 137.1    | 132.3 | 12.6     | 0.1  | -0.35 | 0.58  | 68.2  | 0     |
| 9  | 0.0 | 1.0 | 0.0 | 198.6 | 200.4 | 195.4  | 199.2 | 201.2      | 208.4    | 204.1 | 8.9      | 0.04 | -0.14 | 0.14  | 8.2   | 0     |
| 10 | 0.0 | 1.0 | 0.0 | 200.8 | 202.0 | 198.3  | 201.3 | 202.9      | 209.2    | 205.4 | 8.7      | 0.04 | -0.13 | 0.18  | 8.6   | 0     |
| 11 | 0.0 | 1.0 | 0.0 | 201.1 | 202.3 | 198.4  | 201.6 | 203.3      | 210.2    | 206.0 | 8.7      | 0.04 | -0.13 | 0.14  | 7.2   | 0     |
| 12 | 0.0 | 1.0 | 0.0 | 202.1 | 202.9 | 199.8  | 202.5 | 203.9      | 210.2    | 206.6 | 8.6      | 0.04 | -0.13 | 0.15  | 7.4   | 0     |
| 13 | 1.0 | 0.0 | 0.0 | 418.3 | 418.1 | 416.4  | 419.3 | 417.4      | 420.7    | 425.0 | 7.5      | 0.02 | -0.06 | 0.05  | 1.5   | 0     |
| 14 | 1.0 | 0.0 | 0.0 | 418.6 | 418.2 | 417.1  | 419.6 | 417.6      | 421.0    | 425.1 | 7.5      | 0.02 | -0.06 | 0.04  | 1.4   | 0     |
| 15 | 1.0 | 0.0 | 0.0 | 433.1 | 432.7 | 432.3  | 434.0 | 431.7      | 433.9    | 439.7 | 8.5      | 0.02 | -0.08 | 0.06  | 2.3   | 0     |
| 16 | 1.0 | 0.0 | 0.0 | 433.2 | 432.7 | 432.4  | 434.0 | 431.8      | 434.0    | 439.9 | 8.5      | 0.02 | -0.08 | 0.06  | 2.3   | 0     |
| 17 | 1.0 | 0.0 | 0.0 | 435.8 | 433.8 | 436.3  | 436.5 | 433.0      | 434.9    | 440.5 | 8.4      | 0.02 | -0.08 | 0.06  | 2.3   | 0     |
| 18 | 1.0 | 0.0 | 0.0 | 435.8 | 433.8 | 436.4  | 436.6 | 433.0      | 434.9    | 440.5 | 8.4      | 0.02 | -0.08 | 0.06  | 2.3   | 0     |

Table S4: H$_2$O dimer harmonic vibrational frequencies.

|    | S   | B   | T   | PBE   | RPBE  | PBEsol | PW91  | optPBE-vdW | BEEF-vdW | $\mu$ | $\sigma$ | COV  | Skew  | Kurt. | JB    | Imag. |
|----|-----|-----|-----|-------|-------|--------|-------|------------|----------|-------|----------|------|-------|-------|-------|-------|
| 1  |     |     |     | 17.0  | 13.2  | 17.2   | 14.2  | 13.4       | 13.5     | 26.3  | 25.2     | 0.96 | 0.49  | -1.0  | 164.7 | 458   |
| 2  |     |     |     | 19.7  | 15.3  | 21.3   | 18.8  | 17.4       | 25.3     | 23.3  | 25.3     | 1.09 | 0.73  | -0.77 | 228.9 | 23    |
| 3  |     |     |     | 21.1  | 18.5  | 22.3   | 20.0  | 17.8       | 35.7     | 26.1  | 25.6     | 0.98 | 0.63  | -0.9  | 199.7 | 0     |
| 4  |     |     |     | 23.4  | 19.2  | 27.0   | 23.4  | 20.9       | 39.8     | 31.1  | 23.6     | 0.76 | 0.64  | -0.79 | 188.6 | 0     |
| 5  |     |     |     | 45.9  | 38.4  | 49.0   | 44.0  | 42.1       | 47.7     | 40.3  | 22.6     | 0.56 | 0.51  | -0.9  | 154.6 | 0     |
| 6  |     |     |     | 79.5  | 69.1  | 88.2   | 80.2  | 72.2       | 74.6     | 64.8  | 21.1     | 0.33 | -0.32 | -0.47 | 52.5  | 0     |
| 7  | 0.0 | 1.0 | 0.0 | 196.9 | 198.7 | 193.9  | 196.8 | 199.0      | 204.6    | 201.6 | 9.2      | 0.05 | -0.15 | 0.19  | 10.5  | 0     |
| 8  | 0.0 | 1.0 | 0.0 | 200.0 | 201.1 | 198.2  | 200.4 | 201.0      | 206.9    | 203.7 | 9.1      | 0.04 | -0.16 | 0.17  | 10.9  | 0     |
| 9  | 1.0 | 0.0 | 0.0 | 440.4 | 447.8 | 431.6  | 440.3 | 444.6      | 456.5    | 457.0 | 9.2      | 0.02 | -0.06 | -0.03 | 1.4   | 0     |
| 10 | 1.0 | 0.0 | 0.0 | 460.3 | 459.9 | 460.7  | 461.7 | 457.9      | 465.6    | 466.6 | 8.2      | 0.02 | -0.06 | 0.03  | 1.2   | 0     |
| 11 | 1.0 | 0.0 | 0.0 | 470.0 | 469.6 | 470.6  | 471.3 | 467.8      | 475.3    | 476.9 | 8.8      | 0.02 | -0.06 | 0.03  | 1.3   | 0     |
| 12 | 1.0 | 0.0 | 0.0 | 472.5 | 471.9 | 473.2  | 473.8 | 469.8      | 476.6    | 478.7 | 8.6      | 0.02 | -0.07 | 0.04  | 1.6   | 0     |

Table S5: Ethene-ethyne dimer harmonic vibrational frequencies.

| | S | B | T | PBE | RPBE | PBEsol | PW91 | optPBE-vdW | BEEF-vdW | $\mu$ | $\sigma$ | COV | Skew | Kurt. | JB | Imag. |
|---|---|---|---|---|---|---|---|---|---|---|---|---|---|---|---|---|
| 1 | | | | 4.5 | 3.6 | 4.0 | 2.9 | 0.3 | 20.4 | 18.0 | 8.9 | 0.49 | -0.1 | -0.65 | 38.1 | 999 |
| 2 | | | | 4.7 | 3.7 | 6.0 | 4.1 | 3.2 | 23.5 | 15.2 | 13.3 | 0.87 | 0.29 | -1.16 | 140.1 | 551 |
| 3 | | | | 8.5 | 5.9 | 7.5 | 8.3 | 9.1 | 26.9 | 14.3 | 13.8 | 0.97 | 0.67 | -0.88 | 213.1 | 114 |
| 4 | | | | 10.4 | 8.7 | 11.2 | 10.2 | 9.8 | 27.2 | 14.6 | 13.2 | 0.91 | 0.73 | -0.75 | 225.8 | 14 |
| 5 | | | | 12.5 | 10.4 | 13.3 | 12.0 | 10.5 | 36.9 | 23.3 | 19.0 | 0.82 | 0.9 | -0.33 | 280.6 | 4 |
| 6 | | | | 76.8 | 74.3 | 77.6 | 77.2 | 75.5 | 85.9 | 73.1 | 17.2 | 0.24 | -0.72 | 1.44 | 346.5 | 1 |
| 7 | 0.35 | 0.64 | 0.0 | 76.8 | 74.3 | 77.7 | 77.2 | 75.6 | 85.9 | 73.2 | 17.0 | 0.23 | -0.64 | 1.02 | 223.3 | 1 |
| 8 | 0.01 | 0.0 | 0.99 | 93.7 | 92.8 | 93.4 | 93.8 | 93.2 | 101.7 | 92.6 | 14.6 | 0.16 | -0.66 | 1.81 | 419.1 | 0 |
| 9 | 0.05 | 0.95 | 0.0 | 93.9 | 92.8 | 93.6 | 93.9 | 93.2 | 101.8 | 92.7 | 14.5 | 0.16 | -0.63 | 1.58 | 340.5 | 0 |
| 10 | 0.0 | 1.0 | 0.0 | 99.7 | 99.8 | 98.6 | 100.1 | 101.1 | 109.9 | 100.7 | 12.4 | 0.12 | -0.44 | 0.91 | 134.3 | 0 |
| 11 | 0.0 | 0.0 | 1.0 | 116.8 | 115.7 | 116.2 | 117.3 | 116.8 | 124.8 | 116.6 | 11.3 | 0.1 | -0.28 | 0.47 | 44.8 | 0 |
| 12 | 0.0 | 0.0 | 1.0 | 117.4 | 117.1 | 116.3 | 117.7 | 117.7 | 125.3 | 118.0 | 11.5 | 0.1 | -0.28 | 0.37 | 36.9 | 0 |
| 13 | 0.0 | 0.0 | 1.0 | 128.6 | 128.2 | 128.0 | 128.9 | 128.6 | 135.3 | 129.5 | 9.0 | 0.07 | -0.2 | 0.26 | 19.0 | 0 |
| 14 | 0.0 | 1.0 | 0.0 | 148.9 | 149.2 | 147.3 | 149.5 | 150.6 | 157.6 | 151.8 | 7.7 | 0.05 | -0.11 | -0.01 | 4.0 | 0 |
| 15 | 0.35 | 0.65 | 0.0 | 166.0 | 166.0 | 164.6 | 166.4 | 167.0 | 169.8 | 167.6 | 3.7 | 0.02 | -1.41 | 3.21 | 1517.7 | 0 |
| 16 | 0.0 | 1.0 | 0.0 | 176.4 | 177.0 | 173.9 | 177.2 | 178.9 | 187.0 | 180.5 | 8.7 | 0.05 | -0.13 | 0.13 | 7.5 | 0 |
| 17 | 0.65 | 0.35 | 0.0 | 203.0 | 201.9 | 203.2 | 203.4 | 203.2 | 205.4 | 206.0 | 2.1 | 0.01 | 0.89 | 1.75 | 519.5 | 0 |
| 18 | 1.0 | 0.0 | 0.0 | 248.5 | 247.3 | 248.3 | 249.0 | 248.2 | 245.3 | 250.1 | 4.7 | 0.02 | -0.07 | 0.04 | 1.7 | 0 |
| 19 | 1.0 | 0.0 | 0.0 | 379.4 | 378.8 | 377.9 | 380.4 | 378.7 | 377.0 | 384.2 | 6.8 | 0.02 | -0.07 | 0.08 | 2.4 | 0 |
| 20 | 1.0 | 0.0 | 0.0 | 380.9 | 380.4 | 379.4 | 381.9 | 380.0 | 378.1 | 385.7 | 7.1 | 0.02 | -0.08 | 0.08 | 2.6 | 0 |
| 21 | 1.0 | 0.0 | 0.0 | 388.3 | 387.6 | 387.3 | 389.2 | 387.0 | 385.3 | 393.0 | 7.5 | 0.02 | -0.08 | 0.06 | 2.7 | 0 |
| 22 | 1.0 | 0.0 | 0.0 | 391.7 | 391.1 | 390.6 | 392.7 | 390.5 | 388.7 | 396.5 | 7.5 | 0.02 | -0.09 | 0.07 | 2.8 | 0 |
| 23 | 1.0 | 0.0 | 0.0 | 410.9 | 413.2 | 406.9 | 412.1 | 412.4 | 412.3 | 419.2 | 7.2 | 0.02 | -0.08 | 0.05 | 2.6 | 0 |
| 24 | 1.0 | 0.0 | 0.0 | 424.5 | 425.1 | 421.9 | 425.7 | 425.1 | 424.3 | 431.4 | 7.4 | 0.02 | -0.08 | 0.05 | 2.4 | 0 |

Table S6: Ethene dimer harmonic vibrational frequencies.

| | S | B | T | PBE | RPBE | PBEsol | PW91 | optPBE-vdW | BEEF-vdW | $\mu$ | $\sigma$ | COV | Skew | Kurt. | JB | Imag. |
|---|---|---|---|---|---|---|---|---|---|---|---|---|---|---|---|---|
| 1 | | | | 2.8 | 3.8 | 5.7 | 0.0 | 3.0 | 19.1 | 17.4 | 9.0 | 0.52 | -0.14 | -0.62 | 39.0 | 995 |
| 2 | | | | 2.8 | 3.8 | 5.7 | 0.5 | 6.4 | 20.7 | 14.2 | 10.7 | 0.75 | 0.07 | -1.14 | 109.5 | 673 |
| 3 | | | | 5.6 | 3.9 | 8.0 | 5.6 | 8.7 | 26.6 | 13.8 | 13.8 | 1.01 | 0.45 | -1.18 | 182.7 | 249 |
| 4 | | | | 5.8 | 4.6 | 8.2 | 5.7 | 8.8 | 26.7 | 14.3 | 12.2 | 0.85 | 0.71 | -0.79 | 218.9 | 61 |
| 5 | | | | 8.1 | 7.0 | 10.5 | 7.0 | 11.3 | 36.0 | 21.1 | 20.7 | 0.98 | 0.83 | -0.57 | 256.2 | 14 |
| 6 | | | | 8.1 | 7.0 | 10.5 | 7.0 | 11.6 | 36.1 | 23.1 | 19.1 | 0.83 | 0.91 | -0.32 | 284.3 | 0 |
| 7 | 0.0 | 1.0 | 0.0 | 99.5 | 99.8 | 98.3 | 99.8 | 100.9 | 109.5 | 100.7 | 12.3 | 0.12 | -0.44 | 0.9 | 133.2 | 0 |
| 8 | 0.0 | 1.0 | 0.0 | 100.0 | 100.2 | 98.9 | 100.3 | 101.8 | 110.4 | 101.6 | 12.1 | 0.12 | -0.42 | 0.82 | 115.4 | 0 |
| 9 | 0.0 | 0.0 | 1.0 | 115.8 | 115.1 | 115.3 | 116.3 | 116.1 | 123.8 | 115.8 | 11.5 | 0.1 | -0.3 | 0.5 | 51.0 | 0 |
| 10 | 0.0 | 0.0 | 1.0 | 116.3 | 115.4 | 115.5 | 116.8 | 116.7 | 124.3 | 116.2 | 11.5 | 0.1 | -0.31 | 0.52 | 53.5 | 0 |
| 11 | 0.0 | 0.0 | 1.0 | 116.5 | 116.6 | 115.5 | 116.8 | 117.0 | 124.6 | 117.4 | 11.7 | 0.1 | -0.29 | 0.4 | 41.8 | 0 |
| 12 | 0.0 | 0.0 | 1.0 | 116.5 | 116.6 | 116.1 | 116.8 | 117.2 | 124.6 | 117.5 | 11.6 | 0.1 | -0.29 | 0.39 | 39.7 | 0 |
| 13 | 0.0 | 0.0 | 1.0 | 127.7 | 127.7 | 127.0 | 128.1 | 127.9 | 134.5 | 128.7 | 9.3 | 0.07 | -0.22 | 0.29 | 22.7 | 0 |
| 14 | 0.0 | 0.0 | 1.0 | 128.8 | 128.2 | 128.8 | 129.1 | 129.2 | 135.5 | 129.6 | 9.2 | 0.07 | -0.21 | 0.28 | 21.6 | 0 |
| 15 | 0.0 | 1.0 | 0.0 | 148.9 | 149.2 | 147.3 | 149.5 | 150.6 | 157.4 | 151.8 | 7.6 | 0.05 | -0.12 | 0.0 | 4.8 | 0 |
| 16 | 0.0 | 1.0 | 0.0 | 148.9 | 149.2 | 147.3 | 149.5 | 150.6 | 157.4 | 151.8 | 7.6 | 0.05 | -0.12 | 0.01 | 4.6 | 0 |
| 17 | 0.3 | 0.7 | 0.0 | 165.4 | 165.7 | 163.8 | 166.0 | 166.5 | 169.2 | 167.2 | 3.7 | 0.02 | -1.43 | 3.25 | 1563.7 | 0 |
| 18 | 0.32 | 0.68 | 0.0 | 166.4 | 166.3 | 165.3 | 167.0 | 167.6 | 169.9 | 168.1 | 3.4 | 0.02 | -1.5 | 3.61 | 1834.4 | 0 |
| 19 | 0.0 | 1.0 | 0.0 | 176.5 | 177.2 | 173.9 | 177.3 | 179.1 | 186.9 | 180.7 | 8.6 | 0.05 | -0.13 | 0.15 | 7.7 | 0 |
| 20 | 0.0 | 1.0 | 0.0 | 176.5 | 177.2 | 173.9 | 177.3 | 179.1 | 186.9 | 180.7 | 8.6 | 0.05 | -0.13 | 0.14 | 7.6 | 0 |
| 21 | 0.69 | 0.31 | 0.0 | 202.8 | 202.0 | 203.0 | 203.4 | 202.9 | 204.8 | 205.6 | 2.1 | 0.01 | 0.98 | 2.11 | 693.8 | 0 |
| 22 | 0.68 | 0.32 | 0.0 | 203.2 | 202.3 | 203.3 | 203.7 | 203.4 | 205.5 | 206.1 | 2.2 | 0.01 | 1.03 | 2.09 | 714.4 | 0 |
| 23 | 1.0 | 0.0 | 0.0 | 379.3 | 378.5 | 377.4 | 380.4 | 378.8 | 377.2 | 384.2 | 6.9 | 0.02 | -0.07 | 0.07 | 2.2 | 0 |
| 24 | 1.0 | 0.0 | 0.0 | 379.3 | 378.5 | 377.4 | 380.4 | 378.8 | 377.3 | 384.3 | 6.9 | 0.02 | -0.07 | 0.07 | 2.2 | 0 |
| 25 | 1.0 | 0.0 | 0.0 | 380.8 | 380.1 | 378.7 | 381.8 | 380.1 | 378.2 | 385.6 | 7.2 | 0.02 | -0.08 | 0.06 | 2.4 | 0 |
| 26 | 1.0 | 0.0 | 0.0 | 380.9 | 380.2 | 379.0 | 381.9 | 380.1 | 378.4 | 385.9 | 7.2 | 0.02 | -0.07 | 0.06 | 2.2 | 0 |
| 27 | 1.0 | 0.0 | 0.0 | 388.2 | 387.2 | 386.7 | 389.2 | 387.1 | 385.4 | 393.0 | 7.6 | 0.02 | -0.08 | 0.06 | 2.6 | 0 |
| 28 | 1.0 | 0.0 | 0.0 | 388.2 | 387.2 | 386.7 | 389.2 | 387.1 | 385.5 | 393.0 | 7.6 | 0.02 | -0.08 | 0.06 | 2.6 | 0 |
| 29 | 1.0 | 0.0 | 0.0 | 391.6 | 390.6 | 389.9 | 392.6 | 390.4 | 388.7 | 396.3 | 7.6 | 0.02 | -0.08 | 0.07 | 2.7 | 0 |
| 30 | 1.0 | 0.0 | 0.0 | 391.6 | 390.7 | 390.1 | 392.6 | 390.5 | 389.0 | 396.6 | 7.6 | 0.02 | -0.08 | 0.06 | 2.7 | 0 |

Table S7: Formamide dimer harmonic vibrational frequencies.

| | S | B | T | PBE | RPBE | PBEsol | PW91 | optPBE-vdW | BEEF-vdW | $\mu$ | $\sigma$ | COV | Skew | Kurt. | JB | Imag. |
|---|---|---|---|---|---|---|---|---|---|---|---|---|---|---|---|---|
| 1 | | | | 7.8 | 7.4 | 8.2 | 8.0 | 6.0 | 11.3 | 10.3 | 6.8 | 0.66 | -0.26 | -1.15 | 132.6 | 694 |
| 2 | | | | 17.5 | 15.9 | 19.2 | 17.7 | 15.2 | 14.1 | 14.0 | 6.7 | 0.47 | -0.81 | -0.34 | 230.9 | 545 |
| 3 | | | | 18.7 | 16.9 | 19.6 | 18.9 | 17.8 | 19.5 | 15.8 | 9.1 | 0.58 | -0.43 | -1.0 | 145.5 | 329 |
| 4 | | | | 21.8 | 17.8 | 24.1 | 22.4 | 20.1 | 21.3 | 18.9 | 8.8 | 0.46 | -0.2 | -0.46 | 30.5 | 94 |
| 5 | | | | 22.4 | 20.5 | 25.4 | 22.6 | 21.0 | 29.8 | 23.1 | 11.4 | 0.49 | 0.28 | -0.09 | 27.4 | 5 |
| 6 | | | | 28.2 | 22.2 | 33.3 | 28.9 | 26.5 | 33.5 | 26.1 | 10.9 | 0.42 | 0.51 | -0.27 | 94.0 | 1 |
| 7 | 0.0 | 0.0 | 1.0 | 61.4 | 56.2 | 65.1 | 61.8 | 58.9 | 66.2 | 52.9 | 21.2 | 0.4 | -0.45 | -0.83 | 124.6 | 0 |
| 8 | 0.0 | 0.0 | 1.0 | 63.2 | 58.0 | 66.9 | 63.6 | 60.6 | 67.4 | 55.8 | 19.1 | 0.34 | -0.33 | -0.85 | 95.5 | 0 |
| 9 | 0.06 | 0.94 | 0.0 | 74.1 | 72.8 | 74.9 | 74.3 | 73.9 | 75.6 | 73.7 | 7.6 | 0.1 | 0.33 | 3.31 | 949.3 | 0 |
| 10 | 0.06 | 0.94 | 0.0 | 77.8 | 75.6 | 79.8 | 78.1 | 76.8 | 78.7 | 75.9 | 7.1 | 0.09 | 0.12 | 3.39 | 964.6 | 0 |
| 11 | 0.0 | 0.0 | 1.0 | 103.6 | 95.9 | 110.8 | 104.4 | 98.5 | 101.0 | 94.1 | 12.6 | 0.13 | -0.36 | 0.43 | 58.9 | 0 |
| 12 | 0.0 | 0.0 | 1.0 | 106.8 | 99.7 | 112.8 | 107.5 | 102.3 | 104.9 | 98.4 | 11.5 | 0.12 | -0.3 | 0.45 | 46.1 | 0 |
| 13 | 0.0 | 0.0 | 1.0 | 124.8 | 124.3 | 124.8 | 125.2 | 124.1 | 132.2 | 125.3 | 8.9 | 0.07 | -0.38 | 0.01 | 48.8 | 0 |
| 14 | 0.0 | 0.0 | 1.0 | 126.6 | 125.5 | 127.7 | 127.0 | 125.6 | 133.1 | 126.2 | 8.9 | 0.07 | -0.44 | 0.04 | 65.3 | 0 |
| 15 | 0.3 | 0.7 | 0.0 | 132.2 | 131.0 | 132.5 | 132.7 | 132.3 | 134.2 | 132.7 | 5.6 | 0.04 | -0.22 | 0.73 | 60.6 | 0 |
| 16 | 0.35 | 0.65 | 0.0 | 133.2 | 131.7 | 133.9 | 133.7 | 133.2 | 134.8 | 133.5 | 5.5 | 0.04 | -0.13 | 0.95 | 80.6 | 0 |
| 17 | 0.58 | 0.42 | 0.0 | 161.5 | 158.7 | 163.4 | 161.8 | 160.0 | 159.4 | 160.8 | 1.7 | 0.01 | -0.63 | 9.95 | 8390.4 | 0 |
| 18 | 0.54 | 0.46 | 0.0 | 163.2 | 160.2 | 164.7 | 163.5 | 161.9 | 160.9 | 162.2 | 1.9 | 0.01 | -0.82 | 7.42 | 4815.5 | 0 |
| 19 | 0.18 | 0.82 | 0.0 | 169.8 | 170.1 | 168.7 | 170.4 | 170.7 | 176.8 | 172.9 | 5.0 | 0.03 | -0.19 | -0.67 | 50.3 | 0 |
| 20 | 0.17 | 0.83 | 0.0 | 169.9 | 170.2 | 169.4 | 170.4 | 170.8 | 176.9 | 173.1 | 4.8 | 0.03 | -0.11 | -0.76 | 51.6 | 0 |
| 21 | 0.32 | 0.68 | 0.0 | 195.9 | 196.9 | 194.0 | 196.2 | 197.4 | 200.8 | 198.9 | 4.6 | 0.02 | -1.36 | 1.88 | 912.0 | 0 |
| 22 | 0.1 | 0.9 | 0.0 | 197.5 | 197.8 | 195.9 | 198.0 | 199.0 | 203.9 | 201.0 | 5.9 | 0.03 | -0.62 | 0.03 | 130.3 | 0 |
| 23 | 0.61 | 0.39 | 0.0 | 209.2 | 208.3 | 209.9 | 209.4 | 208.2 | 210.7 | 213.1 | 2.8 | 0.01 | 0.79 | 1.64 | 433.1 | 0 |
| 24 | 0.8 | 0.2 | 0.0 | 213.3 | 211.9 | 215.0 | 213.3 | 211.5 | 212.2 | 216.1 | 3.1 | 0.01 | 0.74 | 0.87 | 247.2 | 0 |
| 25 | 1.0 | 0.0 | 0.0 | 357.3 | 356.1 | 355.9 | 358.6 | 357.8 | 354.6 | 362.4 | 7.4 | 0.02 | -0.09 | 0.08 | 3.1 | 0 |
| 26 | 1.0 | 0.0 | 0.0 | 358.0 | 356.7 | 356.6 | 359.1 | 358.6 | 356.4 | 364.3 | 7.4 | 0.02 | -0.09 | 0.08 | 3.1 | 0 |
| 27 | 0.99 | 0.01 | 0.0 | 380.8 | 399.2 | 358.3 | 380.0 | 390.5 | 401.8 | 410.6 | 8.6 | 0.02 | -0.07 | 0.02 | 1.8 | 0 |
| 28 | 1.0 | 0.0 | 0.0 | 388.4 | 403.9 | 369.4 | 387.9 | 396.5 | 406.8 | 414.7 | 8.4 | 0.02 | -0.07 | 0.02 | 1.8 | 0 |
| 29 | 1.0 | 0.0 | 0.0 | 443.9 | 444.6 | 442.1 | 444.6 | 442.2 | 445.8 | 450.6 | 7.8 | 0.02 | -0.07 | 0.06 | 1.8 | 0 |
| 30 | 1.0 | 0.0 | 0.0 | 444.1 | 444.8 | 442.2 | 444.7 | 442.2 | 445.9 | 450.8 | 7.8 | 0.02 | -0.07 | 0.06 | 1.8 | 0 |

Table S8: Formic acid dimer harmonic vibrational frequencies.

| | S | B | T | PBE | RPBE | PBEsol | PW91 | optPBE-vdW | BEEF-vdW | $\mu$ | $\sigma$ | COV | Skew | Kurt. | JB | Imag. |
|---|---|---|---|---|---|---|---|---|---|---|---|---|---|---|---|---|
| 1 | | | | 9.6 | 9.0 | 9.7 | 9.5 | 8.3 | 15.6 | 10.4 | 7.1 | 0.68 | -0.24 | -1.19 | 137.0 | 583 |
| 2 | | | | 21.8 | 19.7 | 23.1 | 22.1 | 21.0 | 22.7 | 13.8 | 8.5 | 0.61 | -0.52 | -1.11 | 192.5 | 203 |
| 3 | | | | 23.5 | 22.2 | 25.3 | 23.6 | 21.3 | 25.5 | 19.7 | 7.2 | 0.37 | -0.55 | -0.43 | 116.2 | 9 |
| 4 | | | | 26.6 | 23.6 | 32.3 | 27.2 | 25.4 | 32.6 | 24.4 | 7.7 | 0.31 | 0.08 | -0.66 | 39.1 | 0 |
| 5 | | | | 32.8 | 31.3 | 34.4 | 32.9 | 30.8 | 34.7 | 29.6 | 7.6 | 0.26 | 0.53 | 0.42 | 108.1 | 0 |
| 6 | | | | 37.0 | 31.4 | 47.8 | 38.2 | 34.4 | 39.7 | 33.1 | 8.1 | 0.24 | 0.59 | -0.09 | 118.0 | 0 |
| 7 | 0.12 | 0.88 | 0.0 | 83.2 | 81.6 | 85.1 | 83.6 | 82.5 | 83.4 | 81.9 | 2.5 | 0.03 | -0.96 | 3.9 | 1575.2 | 0 |
| 8 | 0.15 | 0.85 | 0.0 | 88.7 | 86.0 | 92.7 | 89.3 | 86.9 | 87.2 | 85.4 | 2.3 | 0.03 | -1.05 | 3.94 | 1663.9 | 0 |
| 9 | 0.0 | 0.0 | 1.0 | 122.0 | 116.6 | 124.4 | 122.6 | 117.9 | 117.8 | 112.3 | 11.2 | 0.1 | -0.27 | 0.13 | 25.6 | 0 |
| 10 | 0.0 | 0.0 | 1.0 | 122.7 | 117.9 | 125.5 | 123.5 | 118.9 | 120.5 | 115.1 | 10.3 | 0.09 | -0.24 | 0.22 | 22.8 | 0 |
| 11 | 0.0 | 0.0 | 1.0 | 129.6 | 127.2 | 139.3 | 130.4 | 127.3 | 134.4 | 127.8 | 9.1 | 0.07 | -0.29 | 0.0 | 27.7 | 0 |
| 12 | 0.0 | 0.0 | 1.0 | 135.0 | 130.6 | 145.1 | 136.0 | 131.2 | 136.4 | 130.0 | 9.1 | 0.07 | -0.36 | 0.02 | 43.7 | 0 |
| 13 | 0.79 | 0.21 | 0.0 | 152.7 | 147.8 | 159.4 | 153.5 | 149.4 | 146.7 | 147.5 | 1.9 | 0.01 | -1.49 | 9.14 | 7705.0 | 0 |
| 14 | 0.69 | 0.31 | 0.0 | 153.0 | 148.4 | 159.4 | 153.8 | 149.8 | 147.5 | 148.2 | 1.7 | 0.01 | -0.9 | 9.09 | 7149.3 | 0 |
| 15 | 0.09 | 0.91 | 0.0 | 167.4 | 167.2 | 165.9 | 168.0 | 168.2 | 171.9 | 169.5 | 5.7 | 0.03 | 0.06 | -0.18 | 3.8 | 0 |
| 16 | 0.15 | 0.85 | 0.0 | 167.9 | 167.5 | 167.7 | 168.4 | 168.4 | 173.8 | 170.3 | 5.9 | 0.03 | 0.16 | -0.37 | 20.4 | 0 |
| 17 | 0.14 | 0.86 | 0.0 | 176.2 | 174.3 | 179.2 | 176.7 | 175.0 | 178.1 | 175.6 | 5.0 | 0.03 | -0.14 | -0.58 | 34.0 | 0 |
| 18 | 0.3 | 0.7 | 0.0 | 179.9 | 176.8 | 183.3 | 180.7 | 178.1 | 178.9 | 177.3 | 4.4 | 0.02 | -0.17 | -0.33 | 19.1 | 0 |
| 19 | 0.81 | 0.19 | 0.0 | 200.7 | 202.3 | 192.6 | 200.1 | 201.2 | 203.5 | 207.2 | 3.8 | 0.02 | 0.69 | 0.5 | 181.8 | 0 |
| 20 | 0.88 | 0.12 | 0.0 | 211.9 | 210.9 | 212.9 | 212.1 | 210.3 | 210.8 | 214.3 | 4.2 | 0.02 | 0.43 | 0.04 | 61.1 | 0 |
| 21 | 0.99 | 0.01 | 0.0 | 336.9 | 365.4 | 279.5 | 332.5 | 356.2 | 367.6 | 374.5 | 7.1 | 0.02 | -0.09 | 0.08 | 3.5 | 0 |
| 22 | 1.0 | 0.0 | 0.0 | 354.1 | 367.7 | 306.4 | 350.8 | 366.4 | 368.2 | 375.2 | 7.1 | 0.02 | -0.09 | 0.08 | 3.4 | 0 |
| 23 | 1.0 | 0.0 | 0.0 | 370.0 | 369.4 | 368.2 | 371.2 | 370.3 | 390.3 | 389.6 | 10.2 | 0.03 | -0.02 | -0.08 | 0.6 | 0 |
| 24 | 1.0 | 0.0 | 0.0 | 370.7 | 379.9 | 368.4 | 371.7 | 374.3 | 401.5 | 400.6 | 9.7 | 0.02 | -0.04 | -0.03 | 0.7 | 0 |

Table S9: Benzene methane harmonic vibrational frequencies.

| | S | B | T | PBE | RPBE | PBEsol | PW91 | optPBE-vdW | BEEF-vdW | $\mu$ | $\sigma$ | COV | Skew | Kurt. | JB | Imag. |
|---|---|---|---|---|---|---|---|---|---|---|---|---|---|---|---|---|
| 1 | | | | 0.1 | 0.0 | 0.2 | 0.5 | 0.1 | 13.2 | 11.7 | 6.4 | 0.54 | -0.13 | -0.69 | 45.5 | 964 |
| 2 | | | | 0.8 | 0.2 | 3.7 | 0.9 | 0.2 | 13.5 | 10.1 | 7.3 | 0.72 | 0.01 | -1.13 | 107.0 | 754 |
| 3 | | | | 2.6 | 1.0 | 4.0 | 1.2 | 2.1 | 13.5 | 7.8 | 7.6 | 0.97 | 0.41 | -1.15 | 166.7 | 289 |
| 4 | | | | 4.0 | 2.8 | 7.4 | 3.6 | 7.0 | 35.7 | 19.3 | 19.7 | 1.02 | 0.82 | -0.8 | 276.0 | 25 |
| 5 | | | | 4.6 | 3.1 | 7.6 | 3.8 | 7.2 | 36.9 | 19.5 | 19.8 | 1.01 | 0.8 | -0.83 | 272.3 | 6 |
| 6 | | | | 6.0 | 5.3 | 8.2 | 5.6 | 10.2 | 37.7 | 22.7 | 18.6 | 0.82 | 0.93 | -0.37 | 302.9 | 3 |
| 7 | 0.0 | 0.0 | 1.0 | 49.0 | 48.8 | 48.6 | 49.2 | 48.9 | 53.7 | 48.7 | 7.9 | 0.16 | -0.54 | 4.23 | 1588.5 | 0 |
| 8 | 0.0 | 0.0 | 1.0 | 49.0 | 48.9 | 48.6 | 49.2 | 48.9 | 53.7 | 48.8 | 7.8 | 0.16 | -0.25 | 3.18 | 861.3 | 0 |
| 9 | 0.06 | 0.94 | 0.0 | 74.5 | 74.2 | 74.0 | 74.8 | 75.0 | 77.4 | 72.2 | 9.5 | 0.13 | -2.68 | 9.26 | 9539.8 | 0 |
| 10 | 0.06 | 0.94 | 0.0 | 74.5 | 74.2 | 74.0 | 74.8 | 75.0 | 77.4 | 75.3 | 3.2 | 0.04 | -0.84 | 3.67 | 1360.0 | 0 |
| 11 | 0.0 | 0.0 | 1.0 | 82.1 | 82.1 | 81.6 | 82.3 | 81.9 | 90.0 | 81.7 | 8.2 | 0.1 | -0.04 | -1.2 | 121.1 | 0 |
| 12 | 0.0 | 0.0 | 1.0 | 87.3 | 86.5 | 87.5 | 87.5 | 86.7 | 92.0 | 88.9 | 7.5 | 0.08 | 0.95 | 1.78 | 561.1 | 0 |
| 13 | 0.0 | 0.0 | 1.0 | 103.6 | 103.3 | 103.2 | 103.8 | 103.4 | 111.1 | 103.0 | 10.9 | 0.11 | -0.55 | 0.13 | 102.6 | 0 |
| 14 | 0.0 | 0.0 | 1.0 | 103.6 | 103.4 | 103.2 | 103.9 | 103.5 | 111.4 | 103.7 | 11.3 | 0.11 | -0.14 | -0.25 | 12.1 | 0 |
| 15 | 0.0 | 0.0 | 1.0 | 118.5 | 118.0 | 118.1 | 118.9 | 118.4 | 120.2 | 116.1 | 7.2 | 0.06 | -1.26 | 2.08 | 892.7 | 0 |
| 16 | 0.0 | 0.0 | 1.0 | 118.5 | 118.0 | 118.1 | 118.9 | 118.5 | 125.8 | 118.0 | 8.8 | 0.07 | -0.76 | 0.46 | 207.5 | 0 |
| 17 | 0.0 | 0.0 | 1.0 | 122.0 | 121.4 | 121.7 | 122.4 | 121.9 | 126.0 | 120.3 | 8.1 | 0.07 | -0.9 | 1.22 | 392.2 | 0 |
| 18 | 1.0 | 0.0 | 0.0 | 123.3 | 122.0 | 122.8 | 123.5 | 122.4 | 126.8 | 124.9 | 3.4 | 0.03 | 0.52 | 0.27 | 95.5 | 0 |
| 19 | 0.0 | 1.0 | 0.0 | 123.4 | 123.0 | 124.4 | 123.9 | 124.2 | 129.2 | 127.1 | 3.8 | 0.03 | 1.2 | 2.06 | 833.3 | 0 |
| 20 | 0.54 | 0.46 | 0.0 | 128.3 | 127.7 | 128.4 | 128.6 | 128.2 | 129.3 | 129.1 | 3.5 | 0.03 | 0.81 | 4.32 | 1773.8 | 0 |
| 21 | 0.53 | 0.47 | 0.0 | 128.4 | 127.8 | 128.4 | 128.7 | 128.2 | 129.4 | 130.0 | 3.8 | 0.03 | 1.32 | 3.08 | 1375.5 | 0 |
| 22 | 0.09 | 0.91 | 0.0 | 141.2 | 141.5 | 139.7 | 141.7 | 142.7 | 146.4 | 142.0 | 5.4 | 0.04 | -0.96 | 0.25 | 314.5 | 0 |
| 23 | 0.23 | 0.77 | 0.0 | 144.1 | 144.2 | 142.9 | 144.5 | 145.0 | 149.0 | 145.6 | 5.1 | 0.03 | -0.66 | 0.66 | 182.3 | 0 |
| 24 | 0.23 | 0.77 | 0.0 | 144.1 | 144.2 | 143.0 | 144.6 | 145.0 | 149.3 | 145.9 | 5.0 | 0.03 | -0.65 | 0.57 | 168.9 | 0 |
| 25 | 0.0 | 1.0 | 0.0 | 158.6 | 159.8 | 155.6 | 159.4 | 161.5 | 160.7 | 159.4 | 6.5 | 0.04 | -1.55 | 3.04 | 1563.7 | 0 |
| 26 | 0.0 | 1.0 | 0.0 | 159.2 | 160.3 | 156.4 | 160.0 | 162.1 | 171.1 | 163.5 | 10.0 | 0.06 | -0.19 | 0.28 | 19.1 | 0 |
| 27 | 0.0 | 1.0 | 0.0 | 159.3 | 160.5 | 156.4 | 160.0 | 162.2 | 171.1 | 163.9 | 10.1 | 0.06 | -0.15 | 0.05 | 7.2 | 0 |
| 28 | 0.0 | 1.0 | 0.0 | 165.4 | 164.0 | 163.6 | 166.1 | 164.1 | 171.6 | 167.1 | 7.4 | 0.04 | 0.22 | 1.0 | 99.4 | 0 |
| 29 | 0.91 | 0.09 | 0.0 | 166.4 | 165.6 | 169.4 | 166.3 | 167.3 | 174.6 | 172.9 | 4.7 | 0.03 | 0.95 | 1.13 | 404.9 | 0 |
| 30 | 0.29 | 0.71 | 0.0 | 181.8 | 181.4 | 181.2 | 182.4 | 182.8 | 187.1 | 184.5 | 4.6 | 0.02 | -0.06 | 1.33 | 149.1 | 0 |
| 31 | 0.29 | 0.71 | 0.0 | 181.9 | 181.5 | 181.3 | 182.5 | 182.8 | 187.2 | 184.9 | 4.3 | 0.02 | 0.39 | 0.02 | 51.8 | 0 |
| 32 | 0.0 | 1.0 | 0.0 | 186.8 | 187.6 | 184.3 | 187.5 | 188.8 | 195.9 | 190.4 | 6.2 | 0.03 | -0.45 | -0.78 | 119.2 | 0 |
| 33 | 0.0 | 1.0 | 0.0 | 186.8 | 187.6 | 184.3 | 187.5 | 188.8 | 196.0 | 190.6 | 6.0 | 0.03 | -0.38 | -0.92 | 118.7 | 0 |
| 34 | 0.71 | 0.29 | 0.0 | 197.2 | 195.7 | 198.6 | 197.5 | 196.3 | 196.7 | 199.4 | 2.4 | 0.01 | 1.66 | 4.73 | 2783.8 | 0 |
| 35 | 0.71 | 0.29 | 0.0 | 197.3 | 195.8 | 198.7 | 197.5 | 196.4 | 196.8 | 199.5 | 2.4 | 0.01 | 1.67 | 4.75 | 2805.5 | 0 |
| 36 | 1.0 | 0.0 | 0.0 | 368.3 | 367.5 | 366.5 | 369.2 | 367.5 | 365.5 | 372.5 | 6.3 | 0.02 | -0.06 | 0.09 | 1.9 | 0 |
| 37 | 1.0 | 0.0 | 0.0 | 382.1 | 380.6 | 381.6 | 382.5 | 378.9 | 376.1 | 384.7 | 7.9 | 0.02 | -0.09 | 0.08 | 3.0 | 0 |
| 38 | 1.0 | 0.0 | 0.0 | 382.2 | 380.7 | 381.8 | 382.8 | 379.2 | 376.5 | 385.1 | 7.9 | 0.02 | -0.09 | 0.08 | 3.1 | 0 |
| 39 | 1.0 | 0.0 | 0.0 | 383.2 | 382.4 | 382.0 | 383.8 | 382.0 | 379.8 | 387.4 | 7.5 | 0.02 | -0.2 | 0.19 | 16.1 | 0 |
| 40 | 1.0 | 0.0 | 0.0 | 383.3 | 382.6 | 382.1 | 384.3 | 383.1 | 380.2 | 388.4 | 7.4 | 0.02 | -0.09 | -0.0 | 2.5 | 0 |
| 41 | 1.0 | 0.0 | 0.0 | 384.4 | 383.6 | 383.1 | 385.4 | 383.2 | 381.2 | 388.9 | 7.3 | 0.02 | -0.07 | 0.06 | 1.9 | 0 |
| 42 | 1.0 | 0.0 | 0.0 | 384.6 | 383.7 | 383.2 | 385.5 | 383.5 | 381.4 | 389.4 | 7.5 | 0.02 | -0.01 | 0.04 | 0.2 | 0 |
| 43 | 1.0 | 0.0 | 0.0 | 386.3 | 385.5 | 384.9 | 387.3 | 384.9 | 383.1 | 390.7 | 7.3 | 0.02 | -0.07 | 0.09 | 2.2 | 0 |
| 44 | 1.0 | 0.0 | 0.0 | 386.4 | 385.6 | 385.0 | 387.4 | 385.0 | 383.2 | 391.0 | 7.3 | 0.02 | -0.06 | 0.1 | 2.2 | 0 |
| 45 | 1.0 | 0.0 | 0.0 | 387.5 | 386.8 | 386.1 | 388.5 | 386.1 | 384.4 | 392.0 | 7.3 | 0.02 | -0.08 | 0.08 | 2.7 | 0 |

Table S10: Benzene ammonia harmonic vibrational frequencies.

| | S | B | T | PBE | RPBE | PBEsol | PW91 | optPBE-vdW | BEEF-vdW | $\mu$ | $\sigma$ | COV | Skew | Kurt. | JB | Imag. |
|---|---|---|---|---|---|---|---|---|---|---|---|---|---|---|---|---|
| 1 | | | | 0.9 | 0.6 | 0.9 | 1.6 | 0.3 | 13.2 | 9.8 | 7.2 | 0.74 | 0.04 | -1.11 | 103.4 | 746 |
| 2 | | | | 1.5 | 1.8 | 1.5 | 3.3 | 3.9 | 13.5 | 7.7 | 7.6 | 0.98 | 0.44 | -1.17 | 179.7 | 301 |
| 3 | | | | 4.0 | 2.3 | 4.0 | 3.9 | 4.1 | 13.8 | 7.5 | 7.5 | 1.0 | 0.49 | -1.16 | 191.9 | 57 |
| 4 | | | | 5.1 | 3.4 | 5.1 | 4.9 | 8.8 | 32.4 | 20.2 | 20.8 | 1.03 | 0.76 | -1.0 | 276.3 | 7 |
| 5 | | | | 9.0 | 5.7 | 9.0 | 8.6 | 9.8 | 38.4 | 22.1 | 21.3 | 0.96 | 0.61 | -1.25 | 252.3 | 3 |
| 6 | | | | 13.6 | 10.1 | 13.6 | 14.7 | 11.6 | 39.5 | 27.2 | 20.1 | 0.74 | 0.79 | -0.69 | 244.8 | 0 |
| 7 | 0.0 | 0.0 | 1.0 | 49.0 | 48.9 | 49.0 | 49.2 | 49.0 | 53.7 | 49.4 | 9.1 | 0.18 | 0.35 | 3.32 | 961.1 | 0 |
| 8 | 0.0 | 0.0 | 1.0 | 49.0 | 48.9 | 49.0 | 49.2 | 49.0 | 53.9 | 49.7 | 9.3 | 0.19 | 0.6 | 2.92 | 828.9 | 0 |
| 9 | 0.06 | 0.94 | 0.0 | 74.4 | 74.2 | 74.4 | 74.8 | 75.0 | 77.3 | 72.4 | 9.3 | 0.13 | -2.69 | 9.38 | 9745.0 | 0 |
| 10 | 0.06 | 0.94 | 0.0 | 74.5 | 74.2 | 74.5 | 74.8 | 75.0 | 77.4 | 75.3 | 3.3 | 0.04 | -0.51 | 4.14 | 1512.9 | 0 |
| 11 | 0.0 | 0.0 | 1.0 | 82.5 | 82.3 | 82.5 | 82.9 | 82.4 | 90.1 | 81.9 | 8.3 | 0.1 | -0.09 | -1.21 | 125.1 | 0 |
| 12 | 0.0 | 0.0 | 1.0 | 87.3 | 86.5 | 87.3 | 87.5 | 86.7 | 92.2 | 89.0 | 7.6 | 0.09 | 0.95 | 1.63 | 519.1 | 0 |
| 13 | 0.0 | 0.0 | 1.0 | 103.8 | 103.5 | 103.8 | 104.2 | 103.7 | 111.2 | 103.3 | 10.8 | 0.1 | -0.57 | 0.15 | 109.8 | 0 |
| 14 | 0.0 | 0.0 | 1.0 | 103.9 | 103.6 | 103.9 | 104.3 | 103.8 | 111.6 | 104.1 | 11.3 | 0.11 | -0.18 | -0.17 | 12.9 | 0 |
| 15 | 0.0 | 0.0 | 1.0 | 118.7 | 118.1 | 118.7 | 119.2 | 118.6 | 120.1 | 116.3 | 7.3 | 0.06 | -1.46 | 3.48 | 1718.0 | 0 |
| 16 | 0.0 | 0.0 | 1.0 | 118.7 | 118.2 | 118.7 | 119.2 | 118.7 | 125.8 | 118.3 | 8.7 | 0.07 | -0.79 | 0.53 | 229.6 | 0 |
| 17 | 0.0 | 0.0 | 1.0 | 122.2 | 121.5 | 122.2 | 122.7 | 122.1 | 126.1 | 120.5 | 8.2 | 0.07 | -1.02 | 1.73 | 599.8 | 0 |
| 18 | 1.0 | 0.0 | 0.0 | 123.2 | 121.9 | 123.2 | 123.4 | 122.3 | 126.7 | 122.7 | 7.4 | 0.06 | -1.6 | 3.7 | 1990.4 | 0 |
| 19 | 0.0 | 1.0 | 0.0 | 123.4 | 123.0 | 123.4 | 123.9 | 124.2 | 129.1 | 126.3 | 4.5 | 0.04 | 0.82 | 0.79 | 274.7 | 0 |
| 20 | 0.06 | 0.94 | 0.0 | 127.4 | 127.7 | 127.4 | 127.1 | 128.0 | 129.4 | 128.4 | 4.0 | 0.03 | 0.87 | 2.47 | 761.4 | 0 |
| 21 | 0.54 | 0.46 | 0.0 | 128.4 | 127.7 | 128.4 | 128.6 | 128.2 | 129.4 | 129.7 | 4.2 | 0.03 | 0.96 | 2.22 | 717.0 | 0 |
| 22 | 0.48 | 0.52 | 0.0 | 128.4 | 130.6 | 128.4 | 128.7 | 128.8 | 137.4 | 133.2 | 6.8 | 0.05 | 0.62 | -0.79 | 181.6 | 0 |
| 23 | 0.09 | 0.91 | 0.0 | 141.3 | 141.6 | 141.3 | 141.9 | 142.8 | 146.4 | 142.6 | 5.9 | 0.04 | -0.64 | 0.03 | 135.2 | 0 |
| 24 | 0.23 | 0.77 | 0.0 | 144.1 | 144.2 | 144.1 | 144.6 | 145.0 | 149.1 | 145.7 | 5.0 | 0.03 | -0.66 | 0.62 | 176.5 | 0 |
| 25 | 0.23 | 0.77 | 0.0 | 144.2 | 144.2 | 144.2 | 144.6 | 145.1 | 149.2 | 146.0 | 5.1 | 0.04 | -0.29 | 0.48 | 47.0 | 0 |
| 26 | 0.0 | 1.0 | 0.0 | 165.4 | 163.9 | 165.4 | 166.1 | 163.9 | 160.7 | 162.9 | 3.8 | 0.02 | -1.8 | 6.3 | 4382.5 | 0 |
| 27 | 0.91 | 0.09 | 0.0 | 166.2 | 165.6 | 166.2 | 166.2 | 167.4 | 174.6 | 172.9 | 4.7 | 0.03 | 0.89 | 0.8 | 317.1 | 0 |
| 28 | 0.29 | 0.71 | 0.0 | 181.7 | 181.4 | 181.7 | 182.3 | 182.7 | 187.1 | 184.7 | 4.3 | 0.02 | 0.43 | 0.08 | 62.2 | 0 |
| 29 | 0.29 | 0.71 | 0.0 | 181.9 | 181.5 | 181.9 | 182.4 | 182.8 | 187.2 | 184.9 | 4.2 | 0.02 | 0.53 | 0.03 | 95.0 | 0 |
| 30 | 0.71 | 0.29 | 0.0 | 197.1 | 195.6 | 197.1 | 197.3 | 196.2 | 195.9 | 196.9 | 3.2 | 0.02 | -2.66 | 9.68 | 10171.2 | 0 |
| 31 | 0.71 | 0.29 | 0.0 | 197.1 | 195.7 | 197.1 | 197.4 | 196.3 | 196.0 | 197.2 | 2.9 | 0.01 | -2.24 | 6.4 | 5088.2 | 0 |
| 32 | 0.0 | 1.0 | 0.0 | 200.1 | 201.6 | 200.1 | 200.7 | 202.1 | 208.9 | 206.3 | 6.6 | 0.03 | 0.83 | 0.05 | 228.0 | 0 |
| 33 | 0.0 | 1.0 | 0.0 | 200.2 | 201.9 | 200.2 | 200.7 | 202.3 | 209.8 | 206.8 | 6.7 | 0.03 | 0.78 | -0.03 | 200.7 | 0 |
| 34 | 1.0 | 0.0 | 0.0 | 383.5 | 382.4 | 383.5 | 384.3 | 382.0 | 380.5 | 387.9 | 7.1 | 0.02 | -0.08 | 0.08 | 2.5 | 0 |
| 35 | 1.0 | 0.0 | 0.0 | 384.6 | 383.6 | 384.6 | 385.4 | 383.0 | 381.5 | 388.8 | 7.2 | 0.02 | -0.08 | 0.08 | 2.6 | 0 |
| 36 | 1.0 | 0.0 | 0.0 | 384.7 | 383.8 | 384.7 | 385.6 | 383.2 | 381.7 | 389.3 | 7.2 | 0.02 | -0.08 | 0.09 | 2.7 | 0 |
| 37 | 1.0 | 0.0 | 0.0 | 386.5 | 385.5 | 386.5 | 387.2 | 384.8 | 383.3 | 390.7 | 7.2 | 0.02 | -0.08 | 0.09 | 2.8 | 0 |
| 38 | 1.0 | 0.0 | 0.0 | 386.5 | 385.6 | 386.5 | 387.3 | 385.0 | 383.5 | 391.2 | 7.3 | 0.02 | -0.08 | 0.09 | 2.8 | 0 |
| 39 | 1.0 | 0.0 | 0.0 | 387.6 | 386.8 | 387.6 | 388.4 | 386.1 | 384.6 | 392.1 | 7.3 | 0.02 | -0.08 | 0.09 | 2.9 | 0 |
| 40 | 1.0 | 0.0 | 0.0 | 420.0 | 418.9 | 420.0 | 420.7 | 418.2 | 421.7 | 426.1 | 7.7 | 0.02 | -0.07 | 0.05 | 1.6 | 0 |
| 41 | 1.0 | 0.0 | 0.0 | 434.5 | 433.4 | 434.5 | 435.0 | 432.3 | 434.7 | 440.2 | 8.5 | 0.02 | -0.08 | 0.06 | 2.5 | 0 |
| 42 | 1.0 | 0.0 | 0.0 | 435.4 | 433.5 | 435.4 | 436.2 | 432.7 | 435.4 | 442.2 | 8.9 | 0.02 | -0.07 | 0.04 | 1.7 | 0 |

Table S11: Benzene water harmonic vibrational frequencies.

| | S | B | T | PBE | RPBE | PBEsol | PW91 | optPBE-vdW | BEEF-vdW | $\mu$ | $\sigma$ | COV | Skew | Kurt. | JB | Imag. |
|---|---|---|---|---|---|---|---|---|---|---|---|---|---|---|---|---|
| 1 | | | | 1.4 | 0.2 | 1.5 | 1.3 | 0.1 | 13.0 | 9.1 | 7.1 | 0.78 | 0.22 | -1.12 | 120.9 | 660 |
| 2 | | | | 3.7 | 0.4 | 3.7 | 3.7 | 3.2 | 13.8 | 7.7 | 7.3 | 0.95 | 0.5 | -1.07 | 179.0 | 248 |
| 3 | | | | 7.3 | 4.3 | 9.1 | 5.1 | 4.3 | 14.2 | 7.9 | 7.2 | 0.91 | 0.53 | -1.04 | 182.3 | 36 |
| 4 | | | | 10.4 | 6.7 | 10.1 | 10.2 | 11.3 | 30.9 | 21.2 | 21.6 | 1.02 | 0.67 | -1.2 | 267.1 | 4 |
| 5 | | | | 16.4 | 7.3 | 18.6 | 16.6 | 14.7 | 32.4 | 24.1 | 21.7 | 0.9 | 0.49 | -1.43 | 250.8 | 1 |
| 6 | | | | 26.8 | 24.2 | 30.1 | 26.8 | 26.5 | 41.1 | 30.2 | 21.8 | 0.72 | 0.51 | -1.13 | 192.7 | 0 |
| 7 | 0.0 | 0.0 | 1.0 | 48.9 | 48.8 | 48.5 | 49.1 | 48.9 | 53.9 | 50.0 | 10.1 | 0.2 | 0.59 | 2.46 | 620.8 | 0 |
| 8 | 0.0 | 0.0 | 1.0 | 49.1 | 49.0 | 48.7 | 49.3 | 49.2 | 54.1 | 51.1 | 11.0 | 0.22 | 0.85 | 1.87 | 534.6 | 0 |
| 9 | 0.06 | 0.94 | 0.0 | 74.4 | 74.2 | 74.0 | 74.7 | 74.9 | 77.2 | 72.6 | 9.2 | 0.13 | -2.62 | 9.59 | 9943.5 | 0 |
| 10 | 0.06 | 0.94 | 0.0 | 74.5 | 74.2 | 74.0 | 74.8 | 75.0 | 77.3 | 75.4 | 3.7 | 0.05 | 0.88 | 7.66 | 5142.0 | 0 |
| 11 | 0.0 | 0.0 | 1.0 | 82.9 | 82.6 | 82.4 | 83.2 | 82.7 | 90.2 | 82.0 | 8.3 | 0.1 | -0.11 | -1.19 | 122.3 | 0 |
| 12 | 0.0 | 0.0 | 1.0 | 87.3 | 86.5 | 87.5 | 87.5 | 86.7 | 92.9 | 89.2 | 7.7 | 0.09 | 0.95 | 1.52 | 493.5 | 0 |
| 13 | 0.0 | 0.0 | 1.0 | 104.1 | 103.8 | 103.7 | 104.5 | 103.9 | 111.8 | 103.6 | 10.8 | 0.1 | -0.59 | 0.18 | 117.0 | 0 |
| 14 | 0.0 | 0.0 | 1.0 | 104.3 | 103.8 | 103.8 | 104.6 | 104.1 | 112.1 | 104.3 | 11.3 | 0.11 | -0.16 | -0.24 | 13.7 | 0 |
| 15 | 0.0 | 0.0 | 1.0 | 119.0 | 118.3 | 118.5 | 119.4 | 118.9 | 120.0 | 116.4 | 7.1 | 0.06 | -1.29 | 2.24 | 974.7 | 0 |
| 16 | 0.0 | 0.0 | 1.0 | 119.0 | 118.4 | 118.5 | 119.5 | 118.9 | 126.2 | 118.4 | 8.7 | 0.07 | -0.8 | 0.52 | 233.5 | 0 |
| 17 | 0.0 | 0.0 | 1.0 | 122.4 | 121.6 | 122.0 | 122.9 | 122.2 | 126.6 | 120.7 | 8.0 | 0.07 | -0.92 | 1.27 | 418.8 | 0 |
| 18 | 1.0 | 0.0 | 0.0 | 123.2 | 121.9 | 122.9 | 123.4 | 122.3 | 126.7 | 125.0 | 3.5 | 0.03 | 0.5 | 0.09 | 84.6 | 0 |
| 19 | 0.0 | 1.0 | 0.0 | 123.5 | 123.0 | 124.2 | 124.0 | 124.2 | 129.0 | 127.1 | 3.9 | 0.03 | 1.19 | 1.83 | 753.7 | 0 |
| 20 | 0.54 | 0.46 | 0.0 | 128.2 | 127.7 | 128.3 | 128.6 | 128.0 | 129.3 | 129.1 | 3.6 | 0.03 | 0.92 | 4.32 | 1837.4 | 0 |
| 21 | 0.54 | 0.46 | 0.0 | 128.4 | 127.7 | 128.5 | 128.7 | 128.2 | 129.9 | 130.0 | 4.0 | 0.03 | 1.33 | 2.83 | 1257.0 | 0 |
| 22 | 0.09 | 0.91 | 0.0 | 141.4 | 141.7 | 139.9 | 142.0 | 142.8 | 146.4 | 142.2 | 5.3 | 0.04 | -0.98 | 0.26 | 323.8 | 0 |
| 23 | 0.23 | 0.77 | 0.0 | 144.2 | 144.2 | 143.1 | 144.6 | 145.0 | 149.1 | 145.7 | 5.0 | 0.03 | -0.67 | 0.62 | 179.7 | 0 |
| 24 | 0.23 | 0.77 | 0.0 | 144.2 | 144.3 | 143.1 | 144.7 | 145.1 | 149.1 | 145.9 | 4.9 | 0.03 | -0.56 | 0.11 | 106.5 | 0 |
| 25 | 0.0 | 1.0 | 0.0 | 165.5 | 163.8 | 163.7 | 166.1 | 163.9 | 160.7 | 162.9 | 3.8 | 0.02 | -1.8 | 6.34 | 4424.2 | 0 |
| 26 | 0.91 | 0.09 | 0.0 | 166.2 | 165.6 | 169.1 | 166.2 | 167.4 | 174.6 | 172.9 | 4.7 | 0.03 | 0.9 | 0.81 | 323.2 | 0 |
| 27 | 0.28 | 0.72 | 0.0 | 181.7 | 181.3 | 181.0 | 182.3 | 182.7 | 187.0 | 184.7 | 4.3 | 0.02 | 0.45 | 0.02 | 68.6 | 0 |
| 28 | 0.29 | 0.71 | 0.0 | 181.9 | 181.4 | 181.3 | 182.5 | 182.8 | 187.1 | 184.9 | 4.2 | 0.02 | 0.51 | 0.04 | 86.6 | 0 |
| 29 | 0.71 | 0.29 | 0.0 | 197.0 | 195.5 | 195.3 | 197.3 | 196.1 | 195.7 | 196.1 | 3.8 | 0.02 | -1.95 | 3.77 | 2454.6 | 0 |
| 30 | 0.69 | 0.31 | 0.0 | 197.1 | 195.6 | 198.3 | 197.4 | 196.3 | 195.9 | 198.6 | 1.6 | 0.01 | 0.68 | 1.15 | 266.0 | 0 |
| 31 | 0.03 | 0.97 | 0.0 | 198.0 | 199.7 | 198.5 | 198.3 | 200.1 | 205.5 | 205.1 | 6.4 | 0.03 | 1.01 | 0.52 | 362.3 | 0 |
| 32 | 1.0 | 0.0 | 0.0 | 383.8 | 382.7 | 382.4 | 384.7 | 382.3 | 380.7 | 388.0 | 7.1 | 0.02 | -0.08 | 0.08 | 2.6 | 0 |
| 33 | 1.0 | 0.0 | 0.0 | 384.9 | 383.8 | 383.5 | 385.8 | 383.4 | 381.8 | 389.1 | 7.2 | 0.02 | -0.08 | 0.09 | 2.8 | 0 |
| 34 | 1.0 | 0.0 | 0.0 | 385.0 | 383.9 | 383.6 | 385.9 | 383.6 | 381.8 | 389.3 | 7.2 | 0.02 | -0.08 | 0.08 | 2.6 | 0 |
| 35 | 1.0 | 0.0 | 0.0 | 386.7 | 385.7 | 385.2 | 387.6 | 385.1 | 383.5 | 390.9 | 7.3 | 0.02 | -0.08 | 0.09 | 2.9 | 0 |
| 36 | 1.0 | 0.0 | 0.0 | 386.8 | 385.8 | 385.4 | 387.7 | 385.3 | 383.6 | 391.1 | 7.3 | 0.02 | -0.08 | 0.09 | 2.7 | 0 |
| 37 | 1.0 | 0.0 | 0.0 | 387.8 | 387.0 | 386.4 | 388.8 | 386.3 | 384.7 | 392.2 | 7.3 | 0.02 | -0.08 | 0.09 | 2.9 | 0 |
| 38 | 1.0 | 0.0 | 0.0 | 457.3 | 460.1 | 454.6 | 457.8 | 455.8 | 464.6 | 466.3 | 8.7 | 0.02 | -0.07 | 0.03 | 1.5 | 0 |
| 39 | 1.0 | 0.0 | 0.0 | 471.4 | 471.8 | 471.4 | 472.2 | 468.1 | 476.0 | 478.4 | 8.9 | 0.02 | -0.07 | 0.03 | 1.5 | 0 |

Table S12: Benzene HCN harmonic vibrational frequencies.

| | S | B | T | PBE | RPBE | PBEsol | PW91 | optPBE-vdW | BEEF-vdW | $\mu$ | $\sigma$ | COV | Skew | Kurt. | JB | Imag. |
|---|---|---|---|---|---|---|---|---|---|---|---|---|---|---|---|---|
| 1 | | | | 3.4 | 3.5 | 5.6 | 2.6 | 0.2 | 13.4 | 8.1 | 7.4 | 0.91 | 0.36 | -1.16 | 154.4 | 453 |
| 2 | | | | 3.5 | 3.7 | 5.6 | 3.1 | 0.2 | 13.7 | 7.2 | 7.5 | 1.04 | 0.57 | -1.07 | 202.9 | 144 |
| 3 | | | | 7.0 | 5.7 | 7.5 | 6.8 | 7.8 | 13.9 | 8.8 | 6.1 | 0.7 | 0.75 | -0.53 | 210.8 | 17 |
| 4 | | | | 7.1 | 6.8 | 7.7 | 6.9 | 8.0 | 27.7 | 14.9 | 13.1 | 0.88 | 0.83 | -0.58 | 255.9 | 5 |
| 5 | | | | 10.1 | 7.3 | 10.4 | 10.1 | 11.1 | 28.3 | 17.5 | 11.3 | 0.64 | 0.94 | -0.1 | 297.2 | 1 |
| 6 | | | | 49.0 | 49.0 | 48.6 | 49.2 | 49.0 | 54.0 | 48.6 | 7.3 | 0.15 | -1.15 | 4.6 | 2206.4 | 0 |
| 7 | 0.0 | 0.0 | 1.0 | 49.0 | 49.0 | 48.7 | 49.2 | 49.1 | 54.1 | 48.7 | 7.1 | 0.15 | -0.94 | 3.3 | 1199.7 | 0 |
| 8 | 0.06 | 0.94 | 0.0 | 74.4 | 74.2 | 73.9 | 74.7 | 74.9 | 77.3 | 72.6 | 8.9 | 0.12 | -2.94 | 11.67 | 14234.6 | 0 |
| 9 | 0.06 | 0.94 | 0.0 | 74.4 | 74.2 | 74.0 | 74.8 | 75.0 | 77.4 | 74.4 | 6.1 | 0.08 | -4.19 | 29.28 | 77271.0 | 0 |
| 10 | 0.0 | 0.0 | 1.0 | 83.3 | 83.0 | 82.9 | 83.7 | 83.0 | 90.3 | 81.3 | 10.3 | 0.13 | -1.29 | 3.82 | 1771.7 | 0 |
| 11 | 0.0 | 0.0 | 1.0 | 87.3 | 86.6 | 87.5 | 87.6 | 86.8 | 93.1 | 86.9 | 10.4 | 0.12 | 0.21 | -0.05 | 15.2 | 0 |
| 12 | 0.23 | 0.06 | 0.71 | 89.7 | 90.7 | 88.9 | 89.9 | 88.5 | 99.7 | 90.0 | 12.6 | 0.14 | 0.1 | -0.7 | 43.7 | 0 |
| 13 | 0.0 | 1.0 | 0.0 | 90.1 | 90.8 | 89.2 | 90.2 | 88.9 | 99.9 | 92.5 | 10.1 | 0.11 | 0.63 | 0.06 | 134.2 | 0 |
| 14 | 0.0 | 0.0 | 1.0 | 104.6 | 104.1 | 104.2 | 104.9 | 104.3 | 112.2 | 104.1 | 10.7 | 0.1 | -0.57 | 0.2 | 111.9 | 0 |
| 15 | 0.0 | 0.0 | 1.0 | 104.7 | 104.1 | 104.3 | 105.1 | 104.4 | 112.3 | 104.7 | 11.2 | 0.11 | -0.17 | -0.21 | 13.5 | 0 |
| 16 | 0.0 | 0.0 | 1.0 | 119.3 | 118.5 | 118.9 | 119.7 | 119.1 | 120.0 | 116.7 | 7.0 | 0.06 | -1.31 | 2.35 | 1032.1 | 0 |
| 17 | 0.0 | 0.0 | 1.0 | 119.3 | 118.5 | 118.9 | 119.8 | 119.1 | 126.5 | 118.7 | 8.6 | 0.07 | -0.81 | 0.56 | 245.8 | 0 |
| 18 | 0.0 | 0.01 | 0.99 | 122.6 | 121.8 | 122.3 | 123.2 | 122.2 | 126.6 | 120.9 | 7.9 | 0.07 | -0.93 | 1.3 | 428.7 | 0 |
| 19 | 1.0 | 0.0 | 0.0 | 123.1 | 121.8 | 123.0 | 123.3 | 122.4 | 126.8 | 125.1 | 3.5 | 0.03 | 0.48 | 0.06 | 76.6 | 0 |
| 20 | 0.0 | 0.99 | 0.01 | 123.6 | 123.1 | 124.2 | 124.1 | 124.3 | 129.3 | 127.2 | 3.9 | 0.03 | 1.18 | 1.74 | 713.7 | 0 |
| 21 | 0.54 | 0.46 | 0.0 | 128.3 | 127.7 | 128.4 | 128.6 | 128.1 | 129.3 | 129.2 | 3.6 | 0.03 | 0.96 | 4.29 | 1837.4 | 0 |
| 22 | 0.54 | 0.46 | 0.0 | 128.4 | 127.7 | 128.4 | 128.6 | 128.2 | 130.0 | 130.1 | 4.0 | 0.03 | 1.35 | 2.76 | 1245.5 | 0 |
| 23 | 0.09 | 0.91 | 0.0 | 141.5 | 141.7 | 140.1 | 142.1 | 142.9 | 146.6 | 142.3 | 5.4 | 0.04 | -0.98 | 0.28 | 329.1 | 0 |
| 24 | 0.23 | 0.77 | 0.0 | 144.3 | 144.3 | 143.2 | 144.7 | 145.1 | 149.2 | 145.8 | 5.0 | 0.03 | -0.67 | 0.63 | 183.0 | 0 |
| 25 | 0.23 | 0.77 | 0.0 | 144.3 | 144.3 | 143.2 | 144.7 | 145.2 | 149.4 | 146.0 | 4.9 | 0.03 | -0.57 | 0.1 | 107.4 | 0 |
| 26 | 0.0 | 1.0 | 0.0 | 165.5 | 163.7 | 163.7 | 166.1 | 163.8 | 160.8 | 162.9 | 3.8 | 0.02 | -1.8 | 6.38 | 4474.0 | 0 |
| 27 | 0.91 | 0.09 | 0.0 | 166.2 | 165.7 | 169.1 | 166.3 | 167.4 | 174.8 | 172.9 | 4.7 | 0.03 | 0.89 | 0.79 | 316.4 | 0 |
| 28 | 0.28 | 0.72 | 0.0 | 181.8 | 181.3 | 181.2 | 182.3 | 182.7 | 187.2 | 184.7 | 4.3 | 0.02 | 0.45 | -0.0 | 68.6 | 0 |
| 29 | 0.28 | 0.72 | 0.0 | 181.9 | 181.4 | 181.2 | 182.4 | 182.8 | 187.3 | 185.0 | 4.2 | 0.02 | 0.55 | 0.0 | 101.3 | 0 |
| 30 | 0.71 | 0.29 | 0.0 | 196.9 | 195.4 | 198.2 | 197.1 | 196.0 | 195.8 | 198.4 | 1.6 | 0.01 | 0.7 | 1.11 | 264.5 | 0 |
| 31 | 0.71 | 0.29 | 0.0 | 197.0 | 195.4 | 198.3 | 197.2 | 196.0 | 195.9 | 198.4 | 1.6 | 0.01 | 0.7 | 1.16 | 277.9 | 0 |
| 32 | 1.0 | 0.0 | 0.0 | 262.9 | 260.9 | 263.2 | 263.5 | 262.2 | 259.8 | 264.3 | 5.4 | 0.02 | -0.07 | -0.01 | 1.6 | 0 |
| 33 | 1.0 | 0.0 | 0.0 | 384.0 | 382.9 | 382.6 | 384.9 | 382.5 | 380.8 | 388.1 | 7.1 | 0.02 | -0.08 | 0.08 | 2.6 | 0 |
| 34 | 1.0 | 0.0 | 0.0 | 385.0 | 384.0 | 383.7 | 385.9 | 383.6 | 381.9 | 389.2 | 7.1 | 0.02 | -0.08 | 0.08 | 2.7 | 0 |
| 35 | 1.0 | 0.0 | 0.0 | 385.1 | 384.1 | 383.8 | 386.1 | 383.7 | 381.9 | 389.4 | 7.2 | 0.02 | -0.08 | 0.08 | 2.7 | 0 |
| 36 | 1.0 | 0.0 | 0.0 | 386.8 | 385.9 | 385.4 | 387.6 | 385.3 | 383.6 | 391.0 | 7.2 | 0.02 | -0.08 | 0.09 | 2.9 | 0 |
| 37 | 1.0 | 0.0 | 0.0 | 386.9 | 385.9 | 385.4 | 387.8 | 385.4 | 383.6 | 391.2 | 7.3 | 0.02 | -0.08 | 0.08 | 2.7 | 0 |
| 38 | 1.0 | 0.0 | 0.0 | 387.9 | 387.1 | 386.5 | 388.8 | 386.5 | 384.7 | 392.2 | 7.3 | 0.02 | -0.08 | 0.08 | 2.9 | 0 |
| 39 | 1.0 | 0.0 | 0.0 | 410.6 | 413.8 | 405.1 | 411.5 | 412.9 | 412.4 | 420.9 | 7.8 | 0.02 | -0.08 | 0.02 | 2.0 | 0 |

Table S13: Benzene dimer (parallel) harmonic vibrational frequencies.

| | S | B | T | PBE | RPBE | PBEsol | PW91 | optPBE-vdW | BEEF-vdW | $\mu$ | $\sigma$ | COV | Skew | Kurt. | JB | Imag. |
|---|---|---|---|---|---|---|---|---|---|---|---|---|---|---|---|---|
| 1 | | | | 0.0 | 1.1 | 0.7 | 0.2 | 0.3 | 13.4 | 11.5 | 6.4 | 0.55 | -0.08 | -0.74 | 48.4 | 951 |
| 2 | | | | 0.1 | 1.3 | 1.4 | 0.7 | 0.6 | 13.6 | 10.1 | 7.1 | 0.7 | 0.05 | -1.05 | 92.6 | 767 |
| 3 | | | | 0.4 | 1.4 | 1.5 | 2.8 | 4.7 | 13.7 | 8.4 | 7.5 | 0.89 | 0.35 | -1.15 | 151.8 | 451 |
| 4 | | | | 1.2 | 1.5 | 3.3 | 3.2 | 4.9 | 14.1 | 7.5 | 7.3 | 0.98 | 0.6 | -0.97 | 197.3 | 138 |
| 5 | | | | 1.5 | 2.8 | 4.4 | 3.8 | 6.7 | 14.6 | 8.0 | 6.9 | 0.86 | 0.7 | -0.81 | 217.4 | 11 |
| 6 | | | | 2.3 | 3.5 | 4.6 | 4.1 | 8.2 | 14.8 | 10.6 | 5.2 | 0.5 | 0.81 | -0.01 | 220.3 | 4 |
| 7 | 0.0 | 0.0 | 1.0 | 48.8 | 48.7 | 48.4 | 49.0 | 48.8 | 53.9 | 48.3 | 7.5 | 0.16 | -1.36 | 6.06 | 3669.0 | 3 |
| 8 | 0.0 | 0.0 | 1.0 | 48.9 | 48.7 | 48.5 | 49.1 | 48.8 | 54.0 | 48.4 | 7.4 | 0.15 | -1.1 | 4.07 | 1785.8 | 2 |
| 9 | 0.0 | 0.0 | 1.0 | 48.9 | 48.8 | 48.5 | 49.2 | 48.9 | 54.1 | 48.5 | 7.2 | 0.15 | -0.88 | 2.79 | 907.6 | 1 |
| 10 | 0.0 | 0.0 | 1.0 | 48.9 | 48.8 | 48.6 | 49.2 | 49.1 | 54.1 | 48.6 | 7.3 | 0.15 | -0.97 | 3.39 | 1273.7 | 0 |
| 11 | 0.06 | 0.94 | 0.0 | 74.4 | 74.2 | 73.9 | 74.7 | 74.9 | 77.2 | 71.7 | 10.1 | 0.14 | -2.61 | 8.85 | 8806.8 | 0 |
| 12 | 0.06 | 0.94 | 0.0 | 74.5 | 74.2 | 74.0 | 74.7 | 74.9 | 77.3 | 72.1 | 9.6 | 0.13 | -2.69 | 9.26 | 9542.1 | 0 |
| 13 | 0.06 | 0.94 | 0.0 | 74.5 | 74.2 | 74.0 | 74.8 | 75.0 | 77.3 | 75.2 | 3.2 | 0.04 | -0.96 | 4.86 | 2278.7 | 0 |
| 14 | 0.06 | 0.94 | 0.0 | 74.5 | 74.2 | 74.0 | 74.8 | 75.0 | 77.4 | 75.3 | 3.2 | 0.04 | -0.94 | 4.63 | 2079.2 | 0 |
| 15 | 0.0 | 0.0 | 1.0 | 81.3 | 81.4 | 80.6 | 81.5 | 81.3 | 90.0 | 81.4 | 8.3 | 0.1 | 0.0 | -1.18 | 115.2 | 0 |
| 16 | 0.0 | 0.0 | 1.0 | 81.9 | 81.9 | 81.3 | 82.3 | 82.0 | 90.1 | 81.7 | 8.2 | 0.1 | -0.06 | -1.19 | 119.8 | 0 |
| 17 | 0.0 | 0.0 | 1.0 | 87.1 | 86.3 | 87.3 | 87.4 | 86.5 | 92.0 | 88.5 | 7.5 | 0.08 | 0.93 | 1.89 | 584.1 | 0 |
| 18 | 0.0 | 0.0 | 1.0 | 87.1 | 86.4 | 87.3 | 87.4 | 86.5 | 92.6 | 88.8 | 7.7 | 0.09 | 0.92 | 1.68 | 517.2 | 0 |
| 19 | 0.0 | 0.0 | 1.0 | 102.9 | 102.8 | 102.3 | 103.2 | 102.8 | 110.9 | 102.4 | 11.1 | 0.11 | -0.52 | 0.04 | 91.0 | 0 |
| 20 | 0.0 | 0.0 | 1.0 | 103.0 | 102.8 | 102.4 | 103.3 | 103.0 | 111.4 | 102.6 | 11.0 | 0.11 | -0.52 | 0.04 | 91.2 | 0 |
| 21 | 0.0 | 0.0 | 1.0 | 103.2 | 103.1 | 102.6 | 103.6 | 103.3 | 111.4 | 103.3 | 11.4 | 0.11 | -0.15 | -0.28 | 14.1 | 0 |
| 22 | 0.0 | 0.0 | 1.0 | 103.4 | 103.2 | 103.0 | 103.8 | 103.4 | 111.7 | 103.6 | 11.3 | 0.11 | -0.15 | -0.26 | 13.6 | 0 |
| 23 | 0.0 | 0.0 | 1.0 | 118.0 | 117.6 | 117.5 | 118.5 | 118.0 | 120.0 | 115.8 | 7.5 | 0.06 | -1.24 | 1.95 | 829.4 | 0 |
| 24 | 0.0 | 0.0 | 1.0 | 118.1 | 117.6 | 117.6 | 118.5 | 118.1 | 120.1 | 115.9 | 7.4 | 0.06 | -1.24 | 1.95 | 828.9 | 0 |
| 25 | 0.0 | 0.0 | 1.0 | 118.2 | 117.7 | 117.7 | 118.7 | 118.2 | 125.7 | 117.6 | 8.9 | 0.08 | -0.75 | 0.45 | 203.7 | 0 |
| 26 | 0.0 | 0.0 | 1.0 | 118.2 | 117.7 | 117.7 | 118.7 | 118.3 | 125.8 | 117.8 | 8.9 | 0.08 | -0.75 | 0.46 | 207.7 | 0 |
| 27 | 0.0 | 0.0 | 1.0 | 121.6 | 121.0 | 121.1 | 122.1 | 121.5 | 125.9 | 120.0 | 8.2 | 0.07 | -0.89 | 1.17 | 378.1 | 0 |
| 28 | 0.0 | 0.0 | 1.0 | 121.7 | 121.1 | 121.2 | 122.2 | 121.6 | 126.1 | 120.1 | 8.2 | 0.07 | -0.89 | 1.18 | 379.9 | 0 |
| 29 | 1.0 | 0.0 | 0.0 | 123.3 | 122.0 | 122.8 | 123.5 | 122.4 | 126.7 | 124.8 | 3.4 | 0.03 | 0.54 | 0.31 | 104.6 | 0 |
| 30 | 1.0 | 0.0 | 0.0 | 123.4 | 122.0 | 122.8 | 123.6 | 122.5 | 126.7 | 124.9 | 3.4 | 0.03 | 0.53 | 0.3 | 102.3 | 0 |
| 31 | 0.0 | 1.0 | 0.0 | 123.4 | 122.9 | 124.4 | 123.9 | 124.1 | 129.1 | 126.9 | 3.7 | 0.03 | 1.22 | 2.12 | 867.8 | 0 |
| 32 | 0.0 | 1.0 | 0.0 | 123.4 | 123.0 | 124.5 | 123.9 | 124.1 | 129.1 | 127.0 | 3.7 | 0.03 | 1.22 | 2.11 | 864.0 | 0 |
| 33 | 0.55 | 0.45 | 0.0 | 128.3 | 127.7 | 128.3 | 128.5 | 128.1 | 129.2 | 129.1 | 3.5 | 0.03 | 0.79 | 4.4 | 1821.7 | 0 |
| 34 | 0.55 | 0.45 | 0.0 | 128.4 | 127.7 | 128.4 | 128.6 | 128.2 | 129.3 | 129.1 | 3.5 | 0.03 | 0.79 | 4.39 | 1813.1 | 0 |
| 35 | 0.55 | 0.45 | 0.0 | 128.4 | 127.7 | 128.4 | 128.6 | 128.2 | 129.5 | 129.9 | 3.8 | 0.03 | 1.33 | 3.19 | 1434.2 | 0 |
| 36 | 0.55 | 0.45 | 0.0 | 128.4 | 127.8 | 128.5 | 128.7 | 128.2 | 129.5 | 130.0 | 3.8 | 0.03 | 1.33 | 3.2 | 1445.0 | 0 |
| 37 | 0.11 | 0.89 | 0.0 | 141.1 | 141.5 | 139.6 | 141.6 | 142.5 | 146.1 | 141.9 | 5.4 | 0.04 | -0.96 | 0.2 | 308.9 | 0 |
| 38 | 0.11 | 0.89 | 0.0 | 141.2 | 141.5 | 139.7 | 141.6 | 142.6 | 146.2 | 142.0 | 5.4 | 0.04 | -0.95 | 0.19 | 306.9 | 0 |
| 39 | 0.24 | 0.76 | 0.0 | 144.0 | 144.1 | 142.8 | 144.4 | 144.8 | 148.8 | 145.5 | 5.1 | 0.03 | -0.66 | 0.62 | 176.3 | 0 |
| 40 | 0.24 | 0.76 | 0.0 | 144.0 | 144.1 | 142.8 | 144.4 | 144.9 | 148.9 | 145.6 | 5.0 | 0.03 | -0.66 | 0.62 | 176.5 | 0 |
| 41 | 0.24 | 0.76 | 0.0 | 144.1 | 144.2 | 142.9 | 144.4 | 144.9 | 148.9 | 145.7 | 5.0 | 0.03 | -0.55 | 0.08 | 100.4 | 0 |
| 42 | 0.24 | 0.76 | 0.0 | 144.1 | 144.2 | 143.0 | 144.5 | 145.0 | 148.9 | 145.8 | 5.0 | 0.03 | -0.55 | 0.07 | 99.9 | 0 |
| 43 | 0.89 | 0.11 | 0.0 | 165.3 | 163.9 | 163.5 | 166.0 | 164.1 | 160.5 | 162.8 | 3.8 | 0.02 | -1.81 | 6.34 | 4443.8 | 0 |
| 44 | 0.89 | 0.11 | 0.0 | 165.3 | 164.0 | 163.5 | 166.0 | 164.2 | 160.6 | 162.9 | 3.8 | 0.02 | -1.81 | 6.27 | 4371.8 | 0 |
| 45 | 0.0 | 1.0 | 0.0 | 166.4 | 165.5 | 169.4 | 166.3 | 167.2 | 174.3 | 172.8 | 4.7 | 0.03 | 0.9 | 0.83 | 328.0 | 0 |
| 46 | 0.0 | 1.0 | 0.0 | 166.5 | 165.5 | 169.5 | 166.3 | 167.2 | 174.3 | 172.9 | 4.7 | 0.03 | 0.9 | 0.83 | 325.1 | 0 |
| 47 | 0.27 | 0.73 | 0.0 | 181.8 | 181.4 | 181.2 | 182.4 | 182.7 | 186.8 | 184.7 | 4.3 | 0.02 | 0.47 | 0.05 | 73.4 | 0 |
| 48 | 0.27 | 0.73 | 0.0 | 181.9 | 181.4 | 181.2 | 182.4 | 182.7 | 186.9 | 184.7 | 4.2 | 0.02 | 0.47 | 0.05 | 73.7 | 0 |
| 49 | 0.27 | 0.73 | 0.0 | 181.9 | 181.4 | 181.2 | 182.4 | 182.8 | 187.0 | 184.9 | 4.1 | 0.02 | 0.57 | 0.06 | 108.6 | 0 |
| 50 | 0.27 | 0.73 | 0.0 | 181.9 | 181.5 | 181.3 | 182.4 | 182.8 | 187.0 | 185.0 | 4.1 | 0.02 | 0.57 | 0.06 | 109.8 | 0 |
| 51 | 0.7 | 0.3 | 0.0 | 197.3 | 195.7 | 198.6 | 197.5 | 196.3 | 195.8 | 198.5 | 1.7 | 0.01 | 0.72 | 1.11 | 273.1 | 0 |
| 52 | 0.7 | 0.3 | 0.0 | 197.3 | 195.7 | 198.6 | 197.5 | 196.4 | 195.8 | 198.6 | 1.7 | 0.01 | 0.72 | 1.1 | 272.1 | 0 |
| 53 | 0.7 | 0.3 | 0.0 | 197.3 | 195.7 | 198.7 | 197.6 | 196.4 | 195.8 | 198.7 | 1.7 | 0.01 | 0.72 | 1.11 | 275.1 | 0 |
| 54 | 0.7 | 0.3 | 0.0 | 197.4 | 195.7 | 198.7 | 197.6 | 196.5 | 195.9 | 198.7 | 1.7 | 0.01 | 0.72 | 1.1 | 273.0 | 0 |
| 55 | 1.0 | 0.0 | 0.0 | 383.1 | 382.1 | 381.7 | 384.1 | 381.4 | 379.8 | 387.2 | 7.1 | 0.02 | -0.08 | 0.08 | 2.5 | 0 |
| 56 | 1.0 | 0.0 | 0.0 | 383.2 | 382.2 | 381.8 | 384.1 | 381.5 | 379.9 | 387.2 | 7.1 | 0.02 | -0.08 | 0.08 | 2.5 | 0 |
| 57 | 1.0 | 0.0 | 0.0 | 384.2 | 383.4 | 382.8 | 385.2 | 382.4 | 380.9 | 388.2 | 7.2 | 0.02 | -0.08 | 0.08 | 2.6 | 0 |
| 58 | 1.0 | 0.0 | 0.0 | 384.3 | 383.4 | 382.9 | 385.2 | 382.6 | 381.0 | 388.3 | 7.2 | 0.02 | -0.08 | 0.08 | 2.6 | 0 |
| 59 | 1.0 | 0.0 | 0.0 | 384.4 | 383.5 | 382.9 | 385.4 | 382.7 | 381.0 | 388.5 | 7.2 | 0.02 | -0.08 | 0.08 | 2.6 | 0 |
| 60 | 1.0 | 0.0 | 0.0 | 384.5 | 383.5 | 383.0 | 385.5 | 382.7 | 381.1 | 388.6 | 7.2 | 0.02 | -0.08 | 0.08 | 2.7 | 0 |
| 61 | 1.0 | 0.0 | 0.0 | 386.1 | 385.3 | 384.7 | 387.0 | 384.3 | 382.7 | 390.1 | 7.3 | 0.02 | -0.08 | 0.08 | 2.8 | 0 |

| 62 | 1.0 | 0.0 | 0.0 | 386.2 | 385.4 | 384.7 | 387.1 | 384.4 | 382.9 | 390.3 | 7.3 | 0.02 | -0.08 | 0.08 | 2.8 | 0 |
| 63 | 1.0 | 0.0 | 0.0 | 386.3 | 385.4 | 384.7 | 387.2 | 384.5 | 382.9 | 390.4 | 7.3 | 0.02 | -0.08 | 0.08 | 2.7 | 0 |
| 64 | 1.0 | 0.0 | 0.0 | 386.3 | 385.4 | 384.8 | 387.3 | 384.5 | 382.9 | 390.5 | 7.3 | 0.02 | -0.08 | 0.08 | 2.8 | 0 |
| 65 | 1.0 | 0.0 | 0.0 | 387.4 | 386.7 | 385.9 | 388.4 | 385.6 | 384.1 | 391.5 | 7.3 | 0.02 | -0.08 | 0.08 | 2.8 | 0 |
| 66 | 1.0 | 0.0 | 0.0 | 387.5 | 386.7 | 385.9 | 388.4 | 385.7 | 384.1 | 391.6 | 7.3 | 0.02 | -0.08 | 0.08 | 2.9 | 0 |

Table S14: Benzene dimer (t-shape) harmonic vibrational frequencies.

| | S | B | T | PBE | RPBE | PBEsol | PW91 | optPBE-vdW | BEEF-vdW | $\mu$ | $\sigma$ | COV | Skew | Kurt. | JB | Imag. |
|---|---|---|---|---|---|---|---|---|---|---|---|---|---|---|---|---|
| 1 | | | | 0.4 | 0.2 | 0.7 | 0.5 | 0.1 | 12.7 | 12.1 | 5.9 | 0.49 | -0.02 | -0.66 | 36.8 | 1028 |
| 2 | | | | 1.1 | 1.6 | 1.7 | 0.9 | 0.1 | 12.8 | 12.2 | 5.9 | 0.48 | -0.04 | -0.65 | 35.6 | 1008 |
| 3 | | | | 1.8 | 1.7 | 2.2 | 1.8 | 0.5 | 13.6 | 8.4 | 7.5 | 0.89 | 0.3 | -1.18 | 145.5 | 515 |
| 4 | | | | 4.0 | 2.7 | 4.8 | 3.9 | 6.4 | 13.6 | 7.9 | 6.4 | 0.81 | 0.85 | -0.52 | 264.6 | 23 |
| 5 | | | | 4.3 | 2.9 | 5.9 | 4.0 | 6.7 | 14.2 | 8.2 | 6.5 | 0.79 | 0.77 | -0.68 | 233.9 | 9 |
| 6 | | | | 4.5 | 3.3 | 5.9 | 4.1 | 6.8 | 14.2 | 9.9 | 5.4 | 0.55 | 0.95 | -0.05 | 302.6 | 3 |
| 7 | 0.0 | 0.0 | 1.0 | 49.0 | 48.9 | 48.6 | 49.2 | 48.9 | 53.7 | 48.5 | 7.3 | 0.15 | -1.21 | 5.19 | 2736.8 | 3 |
| 8 | 0.0 | 0.0 | 1.0 | 49.0 | 48.9 | 48.7 | 49.2 | 49.0 | 53.7 | 48.6 | 7.2 | 0.15 | -1.01 | 3.64 | 1444.9 | 3 |
| 9 | 0.0 | 0.0 | 1.0 | 49.1 | 48.9 | 48.9 | 49.3 | 49.1 | 53.9 | 48.7 | 7.1 | 0.15 | -0.89 | 3.01 | 1018.2 | 1 |
| 10 | 0.0 | 0.0 | 1.0 | 49.1 | 49.0 | 48.9 | 49.3 | 49.2 | 54.0 | 48.8 | 7.1 | 0.15 | -0.94 | 3.28 | 1193.0 | 0 |
| 11 | 0.06 | 0.94 | 0.0 | 74.4 | 74.2 | 73.9 | 74.7 | 75.0 | 77.3 | 72.1 | 9.5 | 0.13 | -2.76 | 10.18 | 11187.2 | 0 |
| 12 | 0.06 | 0.94 | 0.0 | 74.5 | 74.2 | 73.9 | 74.8 | 75.0 | 77.4 | 72.3 | 9.3 | 0.13 | -2.77 | 10.04 | 10955.1 | 0 |
| 13 | 0.06 | 0.94 | 0.0 | 74.5 | 74.2 | 74.0 | 74.8 | 75.0 | 77.4 | 75.3 | 3.2 | 0.04 | -0.86 | 3.72 | 1400.2 | 0 |
| 14 | 0.06 | 0.94 | 0.0 | 74.5 | 74.3 | 74.1 | 74.8 | 75.1 | 77.4 | 75.3 | 3.2 | 0.04 | -0.84 | 3.72 | 1386.4 | 0 |
| 15 | 0.0 | 0.0 | 1.0 | 82.1 | 82.2 | 81.6 | 82.5 | 82.0 | 90.0 | 81.7 | 8.2 | 0.1 | -0.05 | -1.21 | 123.4 | 0 |
| 16 | 0.0 | 0.0 | 1.0 | 82.4 | 82.2 | 81.8 | 82.6 | 82.0 | 90.2 | 81.9 | 8.3 | 0.1 | -0.08 | -1.21 | 123.6 | 0 |
| 17 | 0.0 | 0.0 | 1.0 | 87.3 | 86.5 | 87.5 | 87.5 | 86.7 | 92.2 | 88.8 | 7.6 | 0.09 | 0.94 | 1.76 | 554.0 | 0 |
| 18 | 0.0 | 0.0 | 1.0 | 87.3 | 86.5 | 87.6 | 87.6 | 86.7 | 92.4 | 89.1 | 7.6 | 0.09 | 0.94 | 1.69 | 535.0 | 0 |
| 19 | 0.0 | 0.0 | 1.0 | 103.5 | 103.3 | 103.0 | 103.8 | 103.2 | 111.3 | 102.9 | 10.9 | 0.11 | -0.55 | 0.1 | 100.0 | 0 |
| 20 | 0.0 | 0.0 | 1.0 | 103.6 | 103.4 | 103.1 | 103.9 | 103.5 | 111.4 | 103.2 | 10.8 | 0.1 | -0.55 | 0.12 | 103.0 | 0 |
| 21 | 0.0 | 0.0 | 1.0 | 103.9 | 103.4 | 103.4 | 104.1 | 103.6 | 111.4 | 103.9 | 11.2 | 0.11 | -0.16 | -0.25 | 14.3 | 0 |
| 22 | 0.0 | 0.0 | 1.0 | 104.0 | 103.6 | 103.5 | 104.3 | 103.7 | 111.8 | 104.0 | 11.2 | 0.11 | -0.16 | -0.25 | 13.8 | 0 |
| 23 | 0.0 | 0.0 | 1.0 | 118.4 | 118.0 | 118.0 | 118.9 | 118.3 | 120.0 | 116.1 | 7.3 | 0.06 | -1.26 | 2.09 | 894.7 | 0 |
| 24 | 0.0 | 0.0 | 1.0 | 118.7 | 118.0 | 118.2 | 119.1 | 118.5 | 120.2 | 116.2 | 7.2 | 0.06 | -1.26 | 2.09 | 896.2 | 0 |
| 25 | 0.0 | 0.0 | 1.0 | 118.7 | 118.1 | 118.3 | 119.1 | 118.5 | 125.9 | 118.1 | 8.7 | 0.07 | -0.77 | 0.5 | 221.3 | 0 |
| 26 | 0.0 | 0.0 | 1.0 | 118.8 | 118.2 | 118.3 | 119.2 | 118.6 | 126.0 | 118.2 | 8.7 | 0.07 | -0.78 | 0.52 | 226.8 | 0 |
| 27 | 0.0 | 0.0 | 1.0 | 122.1 | 121.4 | 121.7 | 122.6 | 121.9 | 126.0 | 120.3 | 8.1 | 0.07 | -0.91 | 1.22 | 398.0 | 0 |
| 28 | 0.0 | 0.0 | 1.0 | 122.2 | 121.5 | 121.8 | 122.6 | 122.0 | 126.3 | 120.6 | 8.0 | 0.07 | -0.92 | 1.29 | 422.6 | 0 |
| 29 | 0.85 | 0.15 | 0.0 | 123.2 | 121.8 | 122.7 | 123.4 | 122.2 | 126.7 | 124.9 | 3.4 | 0.03 | 0.51 | 0.2 | 89.5 | 0 |
| 30 | 1.0 | 0.0 | 0.0 | 123.4 | 122.0 | 122.8 | 123.5 | 122.4 | 126.7 | 125.0 | 3.4 | 0.03 | 0.5 | 0.18 | 85.3 | 0 |
| 31 | 0.15 | 0.85 | 0.0 | 123.4 | 122.9 | 124.2 | 123.9 | 124.2 | 129.2 | 127.0 | 3.8 | 0.03 | 1.19 | 1.99 | 799.3 | 0 |
| 32 | 0.0 | 1.0 | 0.0 | 123.4 | 123.0 | 124.4 | 123.9 | 124.2 | 129.3 | 127.1 | 3.8 | 0.03 | 1.18 | 1.93 | 774.5 | 0 |
| 33 | 0.53 | 0.47 | 0.0 | 128.3 | 127.6 | 128.2 | 128.5 | 128.1 | 129.3 | 129.0 | 3.5 | 0.03 | 0.85 | 4.25 | 1748.0 | 0 |
| 34 | 0.53 | 0.47 | 0.0 | 128.3 | 127.7 | 128.3 | 128.5 | 128.1 | 129.4 | 129.1 | 3.5 | 0.03 | 0.89 | 4.32 | 1817.9 | 0 |
| 35 | 0.54 | 0.46 | 0.0 | 128.4 | 127.7 | 128.4 | 128.7 | 128.2 | 129.4 | 129.9 | 3.9 | 0.03 | 1.34 | 3.05 | 1368.0 | 0 |
| 36 | 0.53 | 0.47 | 0.0 | 128.4 | 127.7 | 128.4 | 128.7 | 128.2 | 129.8 | 130.0 | 3.9 | 0.03 | 1.36 | 2.96 | 1344.9 | 0 |
| 37 | 0.09 | 0.91 | 0.0 | 141.1 | 141.5 | 139.4 | 141.6 | 142.5 | 146.4 | 141.8 | 5.4 | 0.04 | -0.95 | 0.2 | 306.5 | 0 |
| 38 | 0.09 | 0.91 | 0.0 | 141.3 | 141.6 | 139.8 | 141.8 | 142.7 | 146.4 | 142.1 | 5.3 | 0.04 | -0.98 | 0.26 | 323.3 | 0 |
| 39 | 0.22 | 0.78 | 0.0 | 143.9 | 144.0 | 142.6 | 144.3 | 144.8 | 149.1 | 145.4 | 5.1 | 0.03 | -0.65 | 0.57 | 168.0 | 0 |
| 40 | 0.22 | 0.77 | 0.0 | 144.1 | 144.2 | 142.9 | 144.6 | 145.0 | 149.2 | 145.6 | 5.0 | 0.03 | -0.65 | 0.56 | 169.3 | 0 |
| 41 | 0.23 | 0.77 | 0.0 | 144.1 | 144.2 | 142.9 | 144.6 | 145.0 | 149.2 | 145.8 | 5.0 | 0.03 | -0.56 | 0.08 | 103.4 | 0 |
| 42 | 0.23 | 0.77 | 0.0 | 144.1 | 144.3 | 143.1 | 144.6 | 145.0 | 149.2 | 146.0 | 4.9 | 0.03 | -0.56 | 0.09 | 104.4 | 0 |
| 43 | 0.0 | 1.0 | 0.0 | 165.3 | 163.7 | 163.4 | 166.0 | 163.9 | 160.7 | 162.7 | 3.8 | 0.02 | -1.8 | 6.25 | 4331.4 | 0 |
| 44 | 0.0 | 1.0 | 0.0 | 165.4 | 164.0 | 163.6 | 166.2 | 164.1 | 160.8 | 163.0 | 3.8 | 0.02 | -1.79 | 6.28 | 4354.5 | 0 |
| 45 | 0.91 | 0.09 | 0.0 | 166.3 | 165.6 | 169.2 | 166.3 | 167.2 | 174.7 | 172.7 | 4.7 | 0.03 | 0.89 | 0.81 | 317.9 | 0 |
| 46 | 0.91 | 0.09 | 0.0 | 166.4 | 165.6 | 169.3 | 166.3 | 167.3 | 174.8 | 173.0 | 4.7 | 0.03 | 0.89 | 0.79 | 313.9 | 0 |
| 47 | 0.29 | 0.71 | 0.0 | 181.8 | 181.3 | 181.1 | 182.3 | 182.6 | 187.1 | 184.5 | 4.3 | 0.02 | 0.47 | 0.02 | 73.3 | 0 |
| 48 | 0.29 | 0.71 | 0.0 | 181.8 | 181.3 | 181.1 | 182.3 | 182.7 | 187.2 | 184.7 | 4.2 | 0.02 | 0.47 | 0.03 | 73.3 | 0 |
| 49 | 0.29 | 0.71 | 0.0 | 181.8 | 181.4 | 181.1 | 182.4 | 182.7 | 187.3 | 184.9 | 4.1 | 0.02 | 0.56 | 0.03 | 105.1 | 0 |
| 50 | 0.29 | 0.71 | 0.0 | 182.0 | 181.5 | 181.3 | 182.5 | 182.9 | 187.3 | 185.1 | 4.2 | 0.02 | 0.56 | 0.02 | 103.8 | 0 |
| 51 | 0.71 | 0.29 | 0.0 | 197.1 | 195.5 | 198.3 | 197.4 | 196.1 | 195.9 | 198.4 | 1.7 | 0.01 | 0.69 | 1.11 | 260.7 | 0 |
| 52 | 0.71 | 0.29 | 0.0 | 197.2 | 195.5 | 198.5 | 197.4 | 196.2 | 195.9 | 198.5 | 1.7 | 0.01 | 0.7 | 1.07 | 257.3 | 0 |
| 53 | 0.72 | 0.28 | 0.0 | 197.3 | 195.7 | 198.5 | 197.5 | 196.3 | 195.9 | 198.6 | 1.7 | 0.01 | 0.7 | 1.09 | 263.8 | 0 |
| 54 | 0.71 | 0.29 | 0.0 | 197.3 | 195.7 | 198.6 | 197.5 | 196.4 | 195.9 | 198.7 | 1.7 | 0.01 | 0.7 | 1.08 | 261.6 | 0 |
| 55 | 1.0 | 0.0 | 0.0 | 383.2 | 382.3 | 381.6 | 384.0 | 381.6 | 380.0 | 387.4 | 7.1 | 0.02 | -0.08 | 0.08 | 2.4 | 0 |
| 56 | 1.0 | 0.0 | 0.0 | 383.2 | 382.3 | 382.0 | 384.3 | 381.9 | 380.2 | 387.6 | 7.1 | 0.02 | -0.08 | 0.08 | 2.4 | 0 |
| 57 | 1.0 | 0.0 | 0.0 | 384.2 | 383.3 | 382.7 | 385.0 | 382.3 | 380.7 | 388.1 | 7.2 | 0.02 | -0.08 | 0.08 | 2.5 | 0 |
| 58 | 1.0 | 0.0 | 0.0 | 384.2 | 383.5 | 382.9 | 385.4 | 382.9 | 381.3 | 388.6 | 7.2 | 0.02 | -0.08 | 0.08 | 2.6 | 0 |
| 59 | 1.0 | 0.0 | 0.0 | 384.5 | 383.6 | 383.0 | 385.5 | 383.2 | 381.4 | 388.9 | 7.2 | 0.02 | -0.08 | 0.08 | 2.5 | 0 |
| 60 | 1.0 | 0.0 | 0.0 | 384.7 | 383.9 | 383.3 | 385.5 | 383.2 | 381.6 | 389.0 | 7.2 | 0.02 | -0.08 | 0.07 | 2.4 | 0 |
| 61 | 1.0 | 0.0 | 0.0 | 386.0 | 385.2 | 384.6 | 386.9 | 384.3 | 382.7 | 390.1 | 7.3 | 0.02 | -0.08 | 0.08 | 2.7 | 0 |

| 62 | 1.0 | 0.0 | 0.0 | 386.1 | 385.4 | 384.6 | 387.2 | 384.7 | 383.1 | 390.5 | 7.2 | 0.02 | -0.08 | 0.08 | 2.7 | 0 |
| 63 | 1.0 | 0.0 | 0.0 | 386.3 | 385.5 | 384.8 | 387.3 | 385.0 | 383.2 | 390.8 | 7.3 | 0.02 | -0.08 | 0.08 | 2.6 | 0 |
| 64 | 1.0 | 0.0 | 0.0 | 386.7 | 385.9 | 385.1 | 387.3 | 385.4 | 383.7 | 391.2 | 7.3 | 0.02 | -0.08 | 0.08 | 2.7 | 0 |
| 65 | 1.0 | 0.0 | 0.0 | 387.4 | 386.7 | 385.7 | 388.3 | 386.0 | 384.3 | 391.9 | 7.3 | 0.02 | -0.08 | 0.08 | 2.8 | 0 |
| 66 | 1.0 | 0.0 | 0.0 | 387.6 | 387.1 | 386.1 | 388.4 | 389.0 | 385.8 | 394.9 | 7.9 | 0.02 | -0.08 | 0.02 | 2.2 | 0 |

Table S15: Pyrazine dimer harmonic vibrational frequencies.

| | S | B | T | PBE | RPBE | PBEsol | PW91 | optPBE-vdW | BEEF-vdW | $\mu$ | $\sigma$ | COV | Skew | Kurt. | JB | Imag. |
|---|---|---|---|---|---|---|---|---|---|---|---|---|---|---|---|---|
| 1 | | | | 2.2 | 1.0 | 1.5 | 1.1 | 2.0 | 12.7 | 10.7 | 6.1 | 0.57 | -0.08 | -0.8 | 55.2 | 923 |
| 2 | | | | 2.5 | 1.3 | 3.0 | 1.5 | 3.2 | 12.9 | 9.1 | 6.8 | 0.75 | 0.11 | -1.14 | 111.6 | 674 |
| 3 | | | | 2.5 | 2.1 | 3.3 | 2.6 | 4.6 | 13.1 | 7.6 | 7.3 | 0.95 | 0.51 | -1.02 | 173.7 | 294 |
| 4 | | | | 3.7 | 2.2 | 5.4 | 3.6 | 7.5 | 13.4 | 7.6 | 7.0 | 0.92 | 0.69 | -0.79 | 210.1 | 78 |
| 5 | | | | 4.7 | 3.1 | 5.5 | 4.5 | 8.2 | 14.1 | 8.4 | 7.1 | 0.85 | 0.71 | -0.76 | 216.6 | 19 |
| 6 | | | | 5.1 | 4.0 | 6.0 | 4.8 | 9.4 | 14.7 | 10.9 | 5.6 | 0.52 | 0.73 | -0.01 | 177.2 | 1 |
| 7 | 0.0 | 0.0 | 1.0 | 38.3 | 38.6 | 37.2 | 38.4 | 39.0 | 45.6 | 37.1 | 12.0 | 0.32 | -0.95 | 0.91 | 370.5 | 0 |
| 8 | 0.0 | 0.0 | 1.0 | 38.7 | 38.9 | 37.8 | 39.0 | 39.6 | 46.2 | 37.9 | 11.4 | 0.3 | -0.8 | 0.48 | 231.0 | 0 |
| 9 | 0.0 | 0.0 | 1.0 | 50.5 | 50.2 | 50.1 | 50.7 | 50.7 | 55.2 | 50.5 | 6.5 | 0.13 | -0.68 | 2.28 | 586.3 | 0 |
| 10 | 0.0 | 0.0 | 1.0 | 50.6 | 50.4 | 50.2 | 50.9 | 50.8 | 55.3 | 50.7 | 6.4 | 0.13 | -0.58 | 1.51 | 299.6 | 0 |
| 11 | 0.06 | 0.94 | 0.0 | 72.9 | 72.7 | 72.3 | 73.2 | 73.4 | 75.4 | 73.5 | 4.5 | 0.06 | -3.4 | 28.35 | 70806.3 | 0 |
| 12 | 0.06 | 0.94 | 0.0 | 73.0 | 72.7 | 72.4 | 73.3 | 73.5 | 75.7 | 73.8 | 4.4 | 0.06 | -3.32 | 27.7 | 67616.9 | 0 |
| 13 | 0.08 | 0.92 | 0.0 | 86.3 | 85.8 | 86.1 | 86.5 | 86.4 | 87.7 | 85.1 | 5.6 | 0.07 | -2.25 | 4.85 | 3643.8 | 0 |
| 14 | 0.08 | 0.92 | 0.0 | 86.4 | 85.9 | 86.3 | 86.7 | 86.6 | 87.9 | 85.4 | 5.5 | 0.06 | -2.32 | 5.28 | 4114.8 | 0 |
| 15 | 0.0 | 0.0 | 1.0 | 93.9 | 93.1 | 94.1 | 94.1 | 93.2 | 96.5 | 92.2 | 5.4 | 0.06 | -0.24 | -0.51 | 40.5 | 0 |
| 16 | 0.0 | 0.0 | 1.0 | 93.9 | 93.1 | 94.1 | 94.1 | 93.2 | 96.6 | 92.4 | 5.3 | 0.06 | -0.26 | -0.43 | 38.0 | 0 |
| 17 | 0.0 | 0.0 | 1.0 | 95.3 | 95.3 | 94.6 | 95.7 | 95.4 | 104.4 | 98.0 | 9.1 | 0.09 | 0.62 | -0.28 | 134.3 | 0 |
| 18 | 0.0 | 0.0 | 1.0 | 95.4 | 95.4 | 94.8 | 95.9 | 95.6 | 105.5 | 98.3 | 9.2 | 0.09 | 0.59 | -0.35 | 126.1 | 0 |
| 19 | 0.0 | 0.0 | 1.0 | 112.8 | 112.7 | 112.3 | 113.3 | 112.8 | 120.0 | 112.1 | 9.2 | 0.08 | -0.72 | 0.03 | 172.3 | 0 |
| 20 | 0.0 | 0.0 | 1.0 | 113.0 | 112.8 | 112.4 | 113.3 | 112.9 | 120.9 | 112.5 | 9.1 | 0.08 | -0.74 | 0.08 | 184.1 | 0 |
| 21 | 0.0 | 0.0 | 1.0 | 118.4 | 117.8 | 118.1 | 118.9 | 118.2 | 123.2 | 117.1 | 8.3 | 0.07 | -0.85 | 0.39 | 251.9 | 0 |
| 22 | 0.0 | 0.0 | 1.0 | 118.5 | 117.9 | 118.1 | 119.0 | 118.4 | 123.4 | 117.5 | 8.3 | 0.07 | -0.87 | 0.47 | 273.3 | 0 |
| 23 | 0.0 | 0.0 | 1.0 | 119.5 | 118.9 | 119.1 | 119.9 | 119.3 | 125.4 | 118.6 | 8.1 | 0.07 | -0.75 | 0.35 | 195.7 | 0 |
| 24 | 0.0 | 0.0 | 1.0 | 119.6 | 119.0 | 119.3 | 120.1 | 119.5 | 125.8 | 119.0 | 8.1 | 0.07 | -0.76 | 0.39 | 204.3 | 0 |
| 25 | 0.18 | 0.82 | 0.0 | 124.1 | 123.4 | 124.0 | 124.4 | 124.3 | 126.2 | 125.4 | 3.3 | 0.03 | 0.98 | 1.11 | 419.7 | 0 |
| 26 | 0.18 | 0.82 | 0.0 | 124.2 | 123.5 | 124.0 | 124.6 | 124.4 | 126.4 | 125.7 | 3.4 | 0.03 | 0.99 | 1.08 | 423.6 | 0 |
| 27 | 0.93 | 0.07 | 0.0 | 125.8 | 124.4 | 127.0 | 126.1 | 124.9 | 126.4 | 127.6 | 3.0 | 0.02 | 1.57 | 3.93 | 2106.3 | 0 |
| 28 | 0.93 | 0.07 | 0.0 | 126.0 | 124.5 | 127.2 | 126.3 | 125.2 | 127.1 | 127.9 | 3.0 | 0.02 | 1.62 | 3.95 | 2168.3 | 0 |
| 29 | 0.5 | 0.5 | 0.0 | 131.2 | 130.6 | 131.4 | 131.6 | 131.4 | 132.2 | 132.0 | 2.6 | 0.02 | 0.76 | 2.8 | 843.8 | 0 |
| 30 | 0.5 | 0.5 | 0.0 | 131.4 | 130.7 | 131.4 | 131.8 | 131.5 | 132.4 | 132.3 | 2.5 | 0.02 | 0.73 | 2.77 | 819.1 | 0 |
| 31 | 0.41 | 0.59 | 0.0 | 140.6 | 139.7 | 140.8 | 140.9 | 140.2 | 141.8 | 140.3 | 2.2 | 0.02 | -1.23 | 3.4 | 1473.6 | 0 |
| 32 | 0.41 | 0.59 | 0.0 | 140.6 | 139.7 | 141.0 | 140.9 | 140.4 | 141.8 | 140.5 | 2.1 | 0.02 | -1.23 | 3.43 | 1486.3 | 0 |
| 33 | 0.34 | 0.66 | 0.0 | 150.6 | 149.5 | 149.9 | 151.0 | 149.7 | 141.9 | 147.5 | 3.4 | 0.02 | -0.14 | -1.08 | 104.6 | 0 |
| 34 | 0.34 | 0.66 | 0.0 | 150.7 | 149.7 | 150.2 | 151.1 | 150.0 | 142.0 | 147.8 | 3.5 | 0.02 | -0.16 | -1.1 | 108.7 | 0 |
| 35 | 0.98 | 0.02 | 0.0 | 152.6 | 150.4 | 157.2 | 152.7 | 151.1 | 154.0 | 155.1 | 2.6 | 0.02 | -0.09 | 1.79 | 271.1 | 0 |
| 36 | 0.98 | 0.02 | 0.0 | 153.0 | 150.5 | 157.3 | 153.0 | 151.4 | 154.2 | 155.4 | 2.6 | 0.02 | -0.04 | 1.62 | 220.0 | 0 |
| 37 | 0.07 | 0.93 | 0.0 | 164.5 | 164.6 | 163.1 | 165.2 | 166.1 | 171.5 | 167.7 | 4.3 | 0.03 | -0.07 | -1.07 | 97.3 | 0 |
| 38 | 0.07 | 0.93 | 0.0 | 164.7 | 164.7 | 163.3 | 165.4 | 166.2 | 171.5 | 168.1 | 4.2 | 0.03 | -0.09 | -1.09 | 102.7 | 0 |
| 39 | 0.29 | 0.71 | 0.0 | 173.3 | 172.8 | 173.1 | 173.9 | 174.0 | 178.2 | 175.9 | 3.9 | 0.02 | 0.7 | 0.32 | 173.5 | 0 |
| 40 | 0.29 | 0.71 | 0.0 | 173.5 | 172.9 | 173.1 | 174.1 | 174.1 | 178.3 | 176.3 | 4.0 | 0.02 | 0.68 | 0.26 | 159.8 | 0 |
| 41 | 0.27 | 0.73 | 0.0 | 181.5 | 181.0 | 181.2 | 182.0 | 182.2 | 186.1 | 183.7 | 3.3 | 0.02 | -0.05 | -0.68 | 39.4 | 0 |
| 42 | 0.27 | 0.73 | 0.0 | 181.5 | 181.0 | 181.4 | 182.0 | 182.2 | 186.2 | 184.1 | 3.3 | 0.02 | -0.08 | -0.66 | 38.6 | 0 |
| 43 | 0.84 | 0.16 | 0.0 | 189.7 | 187.3 | 192.6 | 190.0 | 187.8 | 186.6 | 191.0 | 2.9 | 0.02 | 0.73 | 0.53 | 200.0 | 0 |
| 44 | 0.84 | 0.16 | 0.0 | 190.0 | 187.5 | 192.7 | 190.3 | 188.1 | 186.7 | 191.2 | 2.9 | 0.02 | 0.73 | 0.51 | 197.7 | 0 |
| 45 | 0.68 | 0.32 | 0.0 | 193.5 | 192.1 | 194.9 | 193.8 | 192.6 | 192.8 | 195.2 | 1.8 | 0.01 | 0.66 | 1.58 | 353.1 | 0 |
| 46 | 0.68 | 0.32 | 0.0 | 193.6 | 192.1 | 195.1 | 193.9 | 192.8 | 193.1 | 195.5 | 1.7 | 0.01 | 0.63 | 1.53 | 326.9 | 0 |
| 47 | 1.0 | 0.0 | 0.0 | 381.0 | 380.3 | 379.0 | 381.7 | 379.4 | 378.2 | 385.4 | 7.0 | 0.02 | -0.08 | 0.07 | 2.7 | 0 |
| 48 | 1.0 | 0.0 | 0.0 | 381.0 | 380.3 | 379.1 | 381.8 | 379.8 | 378.4 | 385.7 | 7.0 | 0.02 | -0.08 | 0.07 | 2.8 | 0 |
| 49 | 1.0 | 0.0 | 0.0 | 381.1 | 380.3 | 379.3 | 382.0 | 380.0 | 378.5 | 385.8 | 7.0 | 0.02 | -0.08 | 0.07 | 2.4 | 0 |
| 50 | 1.0 | 0.0 | 0.0 | 381.1 | 380.4 | 379.4 | 382.1 | 380.1 | 378.6 | 385.9 | 7.0 | 0.02 | -0.08 | 0.07 | 2.4 | 0 |
| 51 | 1.0 | 0.0 | 0.0 | 382.8 | 382.2 | 380.9 | 383.5 | 381.3 | 380.0 | 387.3 | 7.1 | 0.02 | -0.08 | 0.08 | 2.8 | 0 |
| 52 | 1.0 | 0.0 | 0.0 | 382.8 | 382.2 | 380.9 | 383.7 | 381.6 | 380.2 | 387.5 | 7.1 | 0.02 | -0.08 | 0.08 | 2.8 | 0 |
| 53 | 1.0 | 0.0 | 0.0 | 383.5 | 382.9 | 381.6 | 384.2 | 382.3 | 380.9 | 388.1 | 7.1 | 0.02 | -0.08 | 0.07 | 2.8 | 0 |
| 54 | 1.0 | 0.0 | 0.0 | 383.5 | 382.9 | 381.7 | 384.5 | 382.4 | 381.0 | 388.3 | 7.1 | 0.02 | -0.08 | 0.08 | 2.8 | 0 |

Table S16: Phenol dimer harmonic vibrational frequencies.

| | S | B | T | PBE | RPBE | PBEsol | PW91 | optPBE-vdW | BEEF-vdW | $\mu$ | $\sigma$ | COV | Skew | Kurt. | JB | Imag. |
|---|---|---|---|---|---|---|---|---|---|---|---|---|---|---|---|---|
| 1 | | | | 1.3 | 1.5 | 1.8 | 2.0 | 0.9 | 9.4 | 8.5 | 4.5 | 0.53 | -0.17 | -0.79 | 62.5 | 954 |
| 2 | | | | 3.7 | 3.1 | 4.0 | 3.9 | 2.6 | 10.8 | 8.2 | 5.2 | 0.64 | -0.22 | -1.1 | 117.2 | 776 |
| 3 | | | | 6.8 | 5.9 | 6.1 | 4.7 | 3.5 | 12.3 | 7.6 | 6.4 | 0.85 | 0.22 | -1.29 | 154.8 | 459 |
| 4 | | | | 9.4 | 6.6 | 9.1 | 8.1 | 7.7 | 13.7 | 8.3 | 6.1 | 0.74 | 0.33 | -0.98 | 116.9 | 176 |
| 5 | | | | 12.2 | 8.1 | 10.6 | 9.9 | 9.3 | 14.1 | 9.1 | 6.3 | 0.7 | 0.42 | -0.71 | 101.4 | 59 |
| 6 | | | | 15.2 | 12.1 | 15.9 | 13.4 | 13.3 | 14.6 | 11.2 | 5.6 | 0.5 | 0.17 | -0.69 | 48.7 | 12 |
| 7 | 0.0 | 0.0 | 0.99 | 28.2 | 27.8 | 28.4 | 28.1 | 27.9 | 31.6 | 23.8 | 9.8 | 0.41 | -0.58 | -1.08 | 207.9 | 9 |
| 8 | 0.01 | 0.01 | 0.98 | 29.6 | 28.0 | 28.6 | 28.4 | 28.4 | 31.7 | 27.1 | 6.6 | 0.24 | -0.98 | 1.44 | 490.7 | 3 |
| 9 | 0.02 | 0.01 | 0.97 | 43.7 | 44.2 | 43.5 | 43.7 | 43.6 | 48.9 | 39.0 | 13.9 | 0.36 | -0.3 | -1.39 | 191.2 | 2 |
| 10 | 0.07 | 0.86 | 0.07 | 49.2 | 48.8 | 49.5 | 49.5 | 49.6 | 52.5 | 46.9 | 9.5 | 0.2 | -1.95 | 4.05 | 2636.0 | 0 |
| 11 | 0.01 | 0.06 | 0.94 | 50.2 | 50.1 | 50.0 | 50.4 | 50.3 | 52.6 | 49.7 | 7.2 | 0.14 | -1.55 | 6.75 | 4603.8 | 0 |
| 12 | 0.08 | 0.9 | 0.03 | 52.4 | 51.2 | 51.6 | 50.6 | 50.4 | 54.8 | 50.3 | 6.9 | 0.14 | -1.56 | 6.44 | 4263.7 | 0 |
| 13 | 0.0 | 0.02 | 0.97 | 53.3 | 51.7 | 52.3 | 51.6 | 51.4 | 55.0 | 54.0 | 8.5 | 0.16 | -0.19 | 0.8 | 66.2 | 0 |
| 14 | 0.0 | 0.01 | 0.99 | 61.6 | 61.5 | 61.5 | 61.8 | 61.3 | 65.9 | 58.8 | 8.3 | 0.14 | -1.39 | 2.37 | 1111.2 | 0 |
| 15 | 0.07 | 0.29 | 0.64 | 63.6 | 61.5 | 62.9 | 63.3 | 61.6 | 66.0 | 60.4 | 7.8 | 0.13 | -0.64 | 0.29 | 144.6 | 0 |
| 16 | 0.14 | 0.62 | 0.23 | 64.4 | 63.9 | 64.2 | 64.6 | 64.5 | 66.6 | 62.3 | 7.5 | 0.12 | -1.02 | 0.78 | 399.3 | 0 |
| 17 | 0.16 | 0.71 | 0.12 | 64.6 | 64.1 | 64.5 | 64.8 | 64.7 | 66.9 | 66.6 | 5.0 | 0.08 | 1.44 | 2.84 | 1357.7 | 0 |
| 18 | 0.1 | 0.9 | 0.0 | 75.7 | 72.7 | 75.3 | 76.0 | 76.1 | 78.1 | 71.1 | 6.9 | 0.1 | -0.12 | -1.18 | 120.5 | 0 |
| 19 | 0.11 | 0.89 | 0.0 | 75.8 | 75.4 | 75.5 | 76.3 | 76.2 | 78.2 | 76.4 | 4.7 | 0.06 | -0.31 | 7.84 | 5154.7 | 0 |
| 20 | 0.01 | 0.04 | 0.95 | 82.7 | 75.9 | 84.5 | 84.3 | 77.9 | 80.8 | 79.1 | 7.4 | 0.09 | 0.33 | 0.73 | 80.9 | 0 |
| 21 | 0.0 | 0.0 | 1.0 | 84.6 | 84.1 | 86.3 | 84.8 | 84.3 | 89.3 | 83.6 | 6.9 | 0.08 | -0.45 | 0.23 | 72.8 | 0 |
| 22 | 0.0 | 0.0 | 1.0 | 87.9 | 84.3 | 88.4 | 85.6 | 84.6 | 89.7 | 84.2 | 7.5 | 0.09 | -0.18 | 0.05 | 11.5 | 0 |
| 23 | 0.0 | 0.0 | 0.99 | 91.8 | 91.3 | 91.7 | 92.1 | 91.3 | 98.9 | 91.1 | 7.6 | 0.08 | -0.73 | 0.21 | 183.2 | 0 |
| 24 | 0.01 | 0.0 | 0.99 | 92.2 | 91.5 | 92.2 | 93.0 | 91.5 | 99.2 | 91.6 | 8.0 | 0.09 | -0.44 | 0.2 | 66.8 | 0 |
| 25 | 0.01 | 0.0 | 0.98 | 98.7 | 98.3 | 98.1 | 99.0 | 98.8 | 99.3 | 96.0 | 8.9 | 0.09 | -0.42 | 0.51 | 81.0 | 0 |
| 26 | 0.63 | 0.37 | 0.01 | 99.8 | 98.5 | 100.5 | 99.6 | 99.1 | 100.0 | 96.5 | 8.9 | 0.09 | -0.46 | 0.48 | 88.0 | 0 |
| 27 | 0.6 | 0.38 | 0.02 | 100.8 | 98.7 | 101.3 | 99.9 | 99.2 | 107.5 | 102.7 | 8.0 | 0.08 | 0.34 | 0.65 | 73.7 | 0 |
| 28 | 0.01 | 0.0 | 0.98 | 103.9 | 99.5 | 101.9 | 101.0 | 100.1 | 108.1 | 103.2 | 7.9 | 0.08 | 0.3 | 0.72 | 73.0 | 0 |
| 29 | 0.0 | 0.0 | 0.99 | 107.8 | 106.9 | 107.4 | 108.0 | 107.0 | 115.2 | 108.3 | 8.4 | 0.08 | 0.19 | -0.28 | 18.8 | 0 |
| 30 | 0.01 | 0.01 | 0.99 | 112.8 | 107.1 | 110.0 | 108.3 | 107.8 | 115.8 | 109.1 | 8.4 | 0.08 | 0.15 | -0.33 | 16.8 | 0 |
| 31 | 0.0 | 0.0 | 1.0 | 116.7 | 116.1 | 116.2 | 117.0 | 116.3 | 123.2 | 115.5 | 8.5 | 0.07 | -0.77 | 0.35 | 208.0 | 0 |
| 32 | 0.0 | 0.0 | 1.0 | 118.9 | 116.2 | 117.5 | 117.2 | 116.7 | 123.3 | 116.0 | 8.4 | 0.07 | -0.77 | 0.36 | 208.6 | 0 |
| 33 | 0.0 | 0.0 | 1.0 | 119.5 | 118.5 | 119.2 | 119.5 | 118.6 | 124.4 | 118.1 | 8.4 | 0.07 | -0.5 | -0.56 | 110.5 | 0 |
| 34 | 0.32 | 0.67 | 0.0 | 122.7 | 118.8 | 119.5 | 120.0 | 119.6 | 125.0 | 118.7 | 8.2 | 0.07 | -0.54 | -0.49 | 116.5 | 0 |
| 35 | 0.4 | 0.57 | 0.03 | 123.3 | 121.9 | 122.5 | 122.9 | 122.6 | 126.5 | 124.7 | 4.0 | 0.03 | 1.29 | 2.13 | 935.9 | 0 |
| 36 | 0.33 | 0.4 | 0.27 | 124.5 | 122.7 | 122.7 | 123.1 | 122.9 | 127.1 | 125.1 | 4.0 | 0.03 | 1.28 | 1.9 | 841.7 | 0 |
| 37 | 0.64 | 0.36 | 0.0 | 126.7 | 125.9 | 126.6 | 126.9 | 126.2 | 127.3 | 127.9 | 3.3 | 0.03 | 1.23 | 3.79 | 1698.8 | 0 |
| 38 | 0.38 | 0.33 | 0.29 | 128.6 | 126.2 | 126.9 | 127.0 | 126.4 | 127.6 | 128.3 | 3.4 | 0.03 | 1.27 | 3.29 | 1440.4 | 0 |
| 39 | 0.43 | 0.57 | 0.0 | 132.6 | 132.1 | 131.8 | 132.6 | 132.5 | 134.5 | 133.2 | 3.5 | 0.03 | -0.74 | 2.43 | 673.5 | 0 |
| 40 | 0.32 | 0.42 | 0.26 | 134.2 | 132.2 | 132.1 | 132.9 | 132.9 | 134.6 | 133.5 | 3.4 | 0.03 | -0.53 | 1.42 | 261.6 | 0 |
| 41 | 0.15 | 0.85 | 0.0 | 141.3 | 141.4 | 140.0 | 141.7 | 142.6 | 145.3 | 142.1 | 5.3 | 0.04 | -1.12 | 1.1 | 518.7 | 0 |
| 42 | 0.14 | 0.86 | 0.0 | 141.9 | 142.1 | 140.6 | 142.5 | 143.2 | 146.6 | 142.6 | 5.1 | 0.04 | -1.13 | 1.15 | 539.5 | 0 |
| 43 | 0.2 | 0.8 | 0.0 | 143.5 | 143.7 | 141.6 | 143.9 | 144.1 | 148.1 | 144.1 | 5.2 | 0.04 | -1.12 | 1.22 | 539.9 | 0 |
| 44 | 0.21 | 0.77 | 0.02 | 144.1 | 144.0 | 142.1 | 144.0 | 144.5 | 148.7 | 145.0 | 5.1 | 0.04 | -0.84 | 0.72 | 276.3 | 0 |
| 45 | 0.19 | 0.76 | 0.05 | 144.5 | 144.2 | 143.2 | 144.5 | 144.9 | 149.0 | 145.9 | 5.1 | 0.04 | -0.83 | 0.69 | 269.6 | 0 |
| 46 | 0.3 | 0.7 | 0.01 | 148.8 | 147.2 | 149.5 | 148.0 | 148.9 | 150.1 | 148.1 | 4.9 | 0.03 | -1.28 | 1.87 | 840.4 | 0 |
| 47 | 0.66 | 0.34 | 0.0 | 152.4 | 151.1 | 153.9 | 152.4 | 151.0 | 151.6 | 153.3 | 1.9 | 0.01 | -0.65 | 11.82 | 11791.3 | 0 |
| 48 | 0.66 | 0.33 | 0.01 | 156.4 | 154.2 | 158.0 | 156.1 | 154.8 | 154.7 | 155.9 | 1.8 | 0.01 | -3.05 | 23.08 | 47470.0 | 0 |
| 49 | 0.04 | 0.95 | 0.0 | 163.9 | 164.0 | 162.2 | 164.6 | 165.0 | 162.8 | 163.9 | 4.0 | 0.02 | -0.8 | 0.65 | 249.2 | 0 |
| 50 | 0.14 | 0.86 | 0.0 | 164.3 | 164.2 | 163.4 | 164.8 | 165.2 | 164.2 | 164.7 | 3.7 | 0.02 | -0.43 | 0.29 | 69.0 | 0 |
| 51 | 0.84 | 0.16 | 0.0 | 168.2 | 166.1 | 170.8 | 168.1 | 166.6 | 172.5 | 173.0 | 4.1 | 0.02 | 0.86 | 0.83 | 303.4 | 0 |
| 52 | 0.7 | 0.3 | 0.0 | 169.2 | 167.4 | 171.5 | 168.6 | 168.2 | 173.4 | 173.7 | 4.1 | 0.02 | 0.9 | 0.98 | 349.8 | 0 |
| 53 | 0.32 | 0.68 | 0.0 | 181.1 | 180.4 | 180.6 | 181.0 | 181.4 | 185.2 | 183.7 | 3.5 | 0.02 | 0.75 | 0.36 | 200.1 | 0 |
| 54 | 0.33 | 0.67 | 0.0 | 181.3 | 180.5 | 180.9 | 181.6 | 181.7 | 185.3 | 183.9 | 3.5 | 0.02 | 0.74 | 0.38 | 196.6 | 0 |
| 55 | 0.36 | 0.64 | 0.0 | 183.7 | 183.1 | 183.6 | 184.2 | 184.1 | 187.9 | 186.5 | 3.3 | 0.02 | 0.8 | 0.47 | 232.2 | 0 |
| 56 | 0.39 | 0.61 | 0.0 | 184.5 | 184.1 | 184.6 | 185.0 | 184.7 | 188.4 | 187.1 | 3.1 | 0.02 | 0.88 | 0.67 | 297.6 | 0 |
| 57 | 0.71 | 0.28 | 0.0 | 197.3 | 195.9 | 198.5 | 197.6 | 196.0 | 195.3 | 198.4 | 1.9 | 0.01 | 0.66 | 0.76 | 194.1 | 0 |
| 58 | 0.72 | 0.28 | 0.0 | 197.9 | 196.3 | 199.4 | 198.1 | 196.8 | 196.0 | 199.1 | 2.0 | 0.01 | 0.68 | 0.78 | 206.7 | 0 |
| 59 | 0.71 | 0.29 | 0.0 | 199.0 | 197.2 | 200.3 | 198.6 | 197.8 | 197.1 | 200.1 | 1.9 | 0.01 | 0.67 | 0.79 | 200.6 | 0 |
| 60 | 0.71 | 0.29 | 0.0 | 199.2 | 197.4 | 200.6 | 199.2 | 198.0 | 197.4 | 200.4 | 1.9 | 0.01 | 0.68 | 0.79 | 204.4 | 0 |
| 61 | 1.0 | 0.0 | 0.0 | 382.2 | 381.9 | 380.6 | 383.1 | 381.3 | 379.3 | 386.9 | 7.2 | 0.02 | -0.08 | 0.08 | 2.7 | 0 |

| | | | | | | | | | | | | | | | |
|---|---|---|---|---|---|---|---|---|---|---|---|---|---|---|---|
| 62 | 1.0 | 0.0 | 0.0 | 383.8 | 382.7 | 382.1 | 384.6 | 382.4 | 380.8 | 388.3 | 7.2 | 0.02 | -0.08 | 0.08 | 2.5 | 0 |
| 63 | 1.0 | 0.0 | 0.0 | 385.0 | 384.2 | 383.4 | 385.9 | 383.6 | 382.0 | 389.4 | 7.2 | 0.02 | -0.08 | 0.07 | 2.6 | 0 |
| 64 | 1.0 | 0.0 | 0.0 | 385.2 | 384.4 | 383.7 | 386.2 | 383.9 | 382.2 | 389.6 | 7.1 | 0.02 | -0.08 | 0.08 | 2.4 | 0 |
| 65 | 1.0 | 0.0 | 0.0 | 386.4 | 384.9 | 384.1 | 386.5 | 384.6 | 383.0 | 390.6 | 7.3 | 0.02 | -0.09 | 0.08 | 2.9 | 0 |
| 66 | 1.0 | 0.0 | 0.0 | 386.5 | 385.8 | 384.9 | 387.5 | 385.2 | 383.5 | 390.9 | 7.2 | 0.02 | -0.08 | 0.08 | 2.7 | 0 |
| 67 | 1.0 | 0.0 | 0.0 | 386.9 | 386.0 | 385.2 | 387.7 | 385.3 | 383.7 | 391.3 | 7.3 | 0.02 | -0.08 | 0.08 | 2.5 | 0 |
| 68 | 1.0 | 0.0 | 0.0 | 387.4 | 387.1 | 385.7 | 388.4 | 386.1 | 384.6 | 392.2 | 7.2 | 0.02 | -0.09 | 0.08 | 2.9 | 0 |
| 69 | 1.0 | 0.0 | 0.0 | 387.9 | 387.3 | 386.3 | 388.8 | 386.4 | 384.7 | 392.3 | 7.3 | 0.02 | -0.08 | 0.08 | 2.5 | 0 |
| 70 | 1.0 | 0.0 | 0.0 | 388.3 | 387.8 | 386.9 | 389.4 | 387.1 | 385.3 | 392.9 | 7.3 | 0.02 | -0.08 | 0.08 | 2.7 | 0 |
| 71 | 1.0 | 0.0 | 0.0 | 438.6 | 449.0 | 428.9 | 439.2 | 441.4 | 456.1 | 456.6 | 9.3 | 0.02 | -0.06 | -0.01 | 1.3 | 0 |
| 72 | 1.0 | 0.0 | 0.0 | 461.0 | 460.5 | 460.2 | 461.8 | 458.1 | 466.5 | 467.6 | 8.3 | 0.02 | -0.06 | 0.04 | 1.5 | 0 |

Table S17: 2-pyridoxine 2-aminopyridine harmonic vibrational frequencies.

|    | S    | B    | T    | PBE   | RPBE  | PBE sol | PW91  | optPBE -vdW | BEEF -vdW | $\mu$ | $\sigma$ | COV  | Skew  | Kurt. | JB     | Imag. |
|----|------|------|------|-------|-------|---------|-------|-------------|-----------|-------|----------|------|-------|-------|--------|-------|
| 1  |      |      |      | 3.1   | 3.3   | 3.5     | 3.1   | 2.6         | 10.0      | 9.2   | 4.6      | 0.49 | -0.44 | -0.49 | 85.3   | 920   |
| 2  |      |      |      | 5.0   | 4.7   | 5.2     | 4.9   | 4.3         | 11.0      | 8.4   | 5.9      | 0.7  | -0.18 | -1.23 | 135.7  | 639   |
| 3  |      |      |      | 9.8   | 8.8   | 10.4    | 9.8   | 9.1         | 13.1      | 9.2   | 6.4      | 0.7  | 0.07  | -0.98 | 81.6   | 330   |
| 4  |      |      |      | 11.7  | 10.7  | 12.6    | 11.8  | 11.3        | 14.4      | 11.1  | 5.8      | 0.52 | -0.06 | -0.6  | 31.5   | 190   |
| 5  |      |      |      | 14.3  | 13.7  | 14.7    | 14.5  | 13.5        | 16.0      | 13.0  | 5.1      | 0.39 | -0.37 | 0.23  | 51.2   | 108   |
| 6  |      |      |      | 18.3  | 14.5  | 20.6    | 18.6  | 17.1        | 17.0      | 14.8  | 5.1      | 0.34 | -1.02 | 1.4   | 510.5  | 48    |
| 7  | 0.0  | 0.0  | 1.0  | 22.4  | 22.0  | 22.6    | 22.4  | 22.1        | 27.1      | 20.8  | 7.9      | 0.38 | -0.39 | -0.31 | 58.0   | 20    |
| 8  | 0.0  | 0.0  | 0.99 | 24.4  | 24.2  | 24.7    | 24.5  | 24.2        | 28.9      | 23.4  | 7.2      | 0.31 | -0.41 | -0.15 | 59.1   | 8     |
| 9  | 0.0  | 0.0  | 1.0  | 48.4  | 47.8  | 48.7    | 48.6  | 47.9        | 52.1      | 41.4  | 14.2     | 0.34 | -0.8  | -0.72 | 255.7  | 5     |
| 10 | 0.0  | 0.01 | 0.99 | 50.0  | 49.8  | 49.9    | 50.2  | 49.8        | 54.3      | 48.2  | 8.1      | 0.17 | -1.23 | 3.69  | 1634.0 | 1     |
| 11 | 0.01 | 0.15 | 0.84 | 50.9  | 51.0  | 51.7    | 51.1  | 51.4        | 54.4      | 50.1  | 7.6      | 0.15 | -1.15 | 3.15  | 1266.7 | 0     |
| 12 | 0.07 | 0.81 | 0.12 | 53.9  | 52.7  | 54.4    | 54.0  | 53.5        | 59.3      | 53.7  | 7.3      | 0.14 | -1.29 | 4.1   | 1959.1 | 0     |
| 13 | 0.09 | 0.9  | 0.01 | 58.2  | 57.0  | 59.0    | 58.4  | 57.9        | 63.7      | 58.4  | 8.5      | 0.14 | -0.85 | 2.05  | 591.4  | 0     |
| 14 | 0.0  | 0.01 | 0.99 | 62.5  | 61.9  | 62.6    | 62.7  | 62.1        | 67.7      | 60.8  | 8.2      | 0.13 | -1.08 | 1.88  | 683.2  | 0     |
| 15 | 0.0  | 0.0  | 1.0  | 63.5  | 63.1  | 63.5    | 63.7  | 63.1        | 68.4      | 62.7  | 7.3      | 0.12 | -0.61 | 1.57  | 329.3  | 0     |
| 16 | 0.14 | 0.86 | 0.0  | 67.7  | 66.9  | 68.0    | 67.9  | 67.6        | 68.5      | 67.5  | 4.7      | 0.07 | -1.77 | 6.05  | 4092.2 | 0     |
| 17 | 0.15 | 0.84 | 0.01 | 69.2  | 68.4  | 69.3    | 69.4  | 69.2        | 70.1      | 70.0  | 4.0      | 0.06 | 0.06  | 4.19  | 1465.5 | 0     |
| 18 | 0.18 | 0.82 | 0.0  | 75.0  | 74.4  | 74.9    | 75.3  | 75.2        | 76.3      | 74.9  | 4.3      | 0.06 | -1.17 | 4.65  | 2257.5 | 0     |
| 19 | 0.09 | 0.91 | 0.0  | 78.0  | 77.4  | 77.9    | 78.3  | 78.4        | 79.8      | 78.1  | 4.7      | 0.06 | 0.13  | 4.36  | 1590.5 | 0     |
| 20 | 0.0  | 0.0  | 1.0  | 88.8  | 86.6  | 88.8    | 89.0  | 88.0        | 92.7      | 84.9  | 8.5      | 0.1  | -0.17 | -0.79 | 61.5   | 0     |
| 21 | 0.0  | 0.0  | 1.0  | 89.2  | 88.2  | 89.1    | 89.5  | 88.3        | 93.5      | 87.6  | 7.8      | 0.09 | -0.47 | -0.3  | 81.5   | 0     |
| 22 | 0.0  | 0.0  | 1.0  | 93.0  | 89.7  | 93.2    | 93.3  | 90.6        | 95.4      | 89.0  | 8.3      | 0.09 | -0.4  | -0.61 | 85.1   | 0     |
| 23 | 0.0  | 0.0  | 1.0  | 94.1  | 92.4  | 94.1    | 94.4  | 92.7        | 99.9      | 92.5  | 7.8      | 0.08 | -0.57 | -0.28 | 114.4  | 0     |
| 24 | 0.01 | 0.05 | 0.94 | 95.8  | 93.5  | 101.1   | 96.5  | 93.6        | 100.1     | 94.6  | 8.2      | 0.09 | -0.15 | 0.28  | 14.0   | 0     |
| 25 | 0.79 | 0.2  | 0.01 | 102.0 | 100.2 | 103.2   | 102.2 | 101.1       | 100.6     | 99.8  | 6.5      | 0.07 | -0.19 | 1.93  | 323.0  | 0     |
| 26 | 0.0  | 0.0  | 0.99 | 103.2 | 102.6 | 103.4   | 103.5 | 102.8       | 103.7     | 101.2 | 6.7      | 0.07 | -0.3  | 0.89  | 96.5   | 0     |
| 27 | 0.0  | 0.0  | 1.0  | 103.3 | 102.6 | 105.1   | 103.6 | 102.9       | 110.4     | 105.8 | 7.5      | 0.07 | 0.26  | 0.55  | 47.6   | 0     |
| 28 | 0.69 | 0.3  | 0.01 | 104.8 | 103.6 | 106.0   | 105.0 | 104.3       | 110.8     | 107.5 | 6.6      | 0.06 | 0.43  | 1.33  | 211.3  | 0     |
| 29 | 0.0  | 0.0  | 1.0  | 112.8 | 111.4 | 112.8   | 113.3 | 112.3       | 119.0     | 112.3 | 7.8      | 0.07 | -0.18 | -0.28 | 17.2   | 0     |
| 30 | 0.0  | 0.0  | 1.0  | 115.6 | 115.2 | 114.9   | 115.9 | 115.6       | 122.1     | 115.2 | 7.7      | 0.07 | -0.68 | 0.24  | 161.1  | 0     |
| 31 | 0.0  | 0.0  | 1.0  | 119.2 | 118.7 | 118.6   | 119.7 | 119.0       | 122.7     | 117.5 | 7.3      | 0.06 | -1.05 | 1.1   | 469.0  | 0     |
| 32 | 0.14 | 0.84 | 0.02 | 120.1 | 119.6 | 119.7   | 120.5 | 120.0       | 123.7     | 118.7 | 7.2      | 0.06 | -0.6  | -0.44 | 135.3  | 0     |
| 33 | 0.0  | 0.02 | 0.98 | 120.3 | 119.8 | 119.8   | 120.7 | 120.5       | 124.3     | 120.1 | 7.3      | 0.06 | -0.44 | -0.46 | 81.1   | 0     |
| 34 | 0.3  | 0.69 | 0.01 | 121.9 | 120.9 | 122.4   | 122.2 | 122.0       | 126.8     | 123.6 | 4.9      | 0.04 | 0.65  | 0.21  | 144.0  | 0     |
| 35 | 0.69 | 0.3  | 0.0  | 125.7 | 121.7 | 126.4   | 126.0 | 124.0       | 127.1     | 125.0 | 4.8      | 0.04 | 0.96  | 0.71  | 347.7  | 0     |
| 36 | 0.15 | 0.08 | 0.77 | 128.2 | 124.6 | 129.3   | 128.8 | 125.0       | 128.2     | 127.2 | 4.2      | 0.03 | 1.07  | 1.62  | 601.4  | 0     |
| 37 | 0.56 | 0.3  | 0.13 | 129.4 | 128.6 | 130.2   | 129.7 | 128.9       | 129.2     | 129.7 | 3.7      | 0.03 | 0.68  | 1.57  | 356.8  | 0     |
| 38 | 0.34 | 0.61 | 0.05 | 130.2 | 129.4 | 135.4   | 130.6 | 130.3       | 132.1     | 131.7 | 3.4      | 0.03 | 0.51  | 1.4   | 248.6  | 0     |
| 39 | 0.41 | 0.59 | 0.0  | 136.1 | 135.5 | 136.1   | 136.6 | 136.2       | 137.9     | 136.7 | 3.3      | 0.02 | -1.31 | 4.24  | 2074.7 | 0     |
| 40 | 0.27 | 0.73 | 0.0  | 138.7 | 138.6 | 137.9   | 139.1 | 139.4       | 142.8     | 139.8 | 4.6      | 0.03 | -0.92 | 0.93  | 353.0  | 0     |
| 41 | 0.15 | 0.85 | 0.0  | 140.4 | 140.3 | 139.0   | 140.9 | 141.3       | 144.8     | 141.2 | 4.8      | 0.03 | -0.94 | 0.27  | 302.8  | 0     |
| 42 | 0.14 | 0.86 | 0.0  | 141.5 | 141.8 | 140.0   | 142.0 | 142.7       | 147.2     | 143.3 | 5.6      | 0.04 | -0.64 | -0.06 | 135.2  | 0     |
| 43 | 0.38 | 0.62 | 0.0  | 149.9 | 148.3 | 150.4   | 150.4 | 149.7       | 149.0     | 149.6 | 2.7      | 0.02 | -1.94 | 7.12  | 5476.1 | 0     |
| 44 | 0.64 | 0.36 | 0.0  | 155.4 | 153.2 | 157.9   | 155.6 | 154.2       | 156.2     | 155.9 | 2.0      | 0.01 | -2.54 | 16.03 | 23558.3| 0     |
| 45 | 0.32 | 0.68 | 0.0  | 161.6 | 160.3 | 160.4   | 162.0 | 160.9       | 157.0     | 159.9 | 2.3      | 0.01 | -0.08 | -0.4  | 15.4   | 0     |
| 46 | 0.6  | 0.4  | 0.0  | 164.0 | 162.9 | 165.6   | 164.2 | 163.9       | 165.0     | 165.2 | 1.2      | 0.01 | -0.38 | 0.26  | 53.9   | 0     |
| 47 | 0.58 | 0.42 | 0.0  | 166.5 | 164.5 | 168.0   | 166.8 | 165.6       | 169.3     | 168.4 | 2.9      | 0.02 | 1.07  | 1.09  | 478.8  | 0     |
| 48 | 0.26 | 0.74 | 0.0  | 168.7 | 168.2 | 168.4   | 169.3 | 169.6       | 174.3     | 172.7 | 3.3      | 0.02 | 0.48  | -0.4  | 88.6   | 0     |
| 49 | 0.38 | 0.62 | 0.0  | 178.6 | 177.6 | 178.7   | 179.1 | 178.8       | 181.4     | 180.2 | 2.4      | 0.01 | -0.22 | -0.54 | 39.9   | 0     |
| 50 | 0.37 | 0.63 | 0.0  | 178.9 | 178.1 | 179.1   | 179.4 | 179.3       | 182.1     | 181.2 | 2.6      | 0.01 | 0.47  | -0.21 | 77.5   | 0     |
| 51 | 0.51 | 0.49 | 0.0  | 183.6 | 182.3 | 184.6   | 184.0 | 183.1       | 184.4     | 185.1 | 1.9      | 0.01 | 1.29  | 3.23  | 1420.4 | 0     |
| 52 | 0.58 | 0.42 | 0.0  | 185.2 | 183.0 | 187.1   | 185.4 | 183.9       | 186.5     | 186.4 | 2.0      | 0.01 | 0.87  | 1.56  | 456.2  | 0     |
| 53 | 0.75 | 0.25 | 0.0  | 192.3 | 190.7 | 193.6   | 192.6 | 191.0       | 190.8     | 193.2 | 1.6      | 0.01 | 0.31  | -0.15 | 35.0   | 0     |
| 54 | 0.7  | 0.3  | 0.0  | 192.5 | 191.3 | 193.9   | 192.7 | 191.8       | 191.6     | 194.4 | 1.7      | 0.01 | 0.74  | 1.45  | 357.3  | 0     |
| 55 | 0.59 | 0.41 | 0.0  | 198.7 | 197.1 | 199.4   | 199.0 | 197.9       | 197.5     | 199.3 | 1.5      | 0.01 | -0.1  | 0.2   | 6.3    | 0     |
| 56 | 0.59 | 0.41 | 0.0  | 200.6 | 199.0 | 201.4   | 200.9 | 200.0       | 200.7     | 202.2 | 1.3      | 0.01 | 0.27  | 0.95  | 98.8   | 0     |
| 57 | 0.19 | 0.8  | 0.01 | 203.3 | 202.8 | 203.2   | 203.7 | 203.8       | 205.8     | 206.2 | 3.1      | 0.01 | 0.35  | -0.03 | 41.4   | 0     |
| 58 | 0.54 | 0.46 | 0.0  | 207.5 | 205.9 | 209.2   | 207.9 | 206.3       | 209.8     | 211.6 | 3.7      | 0.02 | 1.26  | 2.25  | 951.0  | 0     |
| 59 | 0.99 | 0.01 | 0.0  | 345.1 | 368.1 | 314.8   | 344.8 | 356.7       | 373.4     | 380.0 | 8.6      | 0.02 | -0.18 | -0.02 | 10.5   | 0     |

| | | | | | | | | | | | | | | | |
|---|---|---|---|---|---|---|---|---|---|---|---|---|---|---|---|
| 60 | 1.0 | 0.0 | 0.0 | 380.5 | 379.8 | 363.4 | 381.6 | 378.9 | 377.6 | 385.1 | 7.4 | 0.02 | -0.04 | 0.07 | 0.9 | 0 |
| 61 | 1.0 | 0.0 | 0.0 | 383.8 | 383.1 | 379.2 | 384.2 | 383.2 | 381.5 | 388.9 | 7.2 | 0.02 | -0.08 | 0.08 | 2.5 | 0 |
| 62 | 1.0 | 0.0 | 0.0 | 384.1 | 383.6 | 382.6 | 384.9 | 383.3 | 381.9 | 389.3 | 7.2 | 0.02 | -0.08 | 0.06 | 2.6 | 0 |
| 63 | 1.0 | 0.0 | 0.0 | 384.6 | 385.6 | 382.8 | 385.4 | 385.2 | 383.5 | 391.0 | 7.2 | 0.02 | -0.11 | 0.08 | 4.3 | 0 |
| 64 | 1.0 | 0.0 | 0.0 | 386.1 | 386.2 | 384.6 | 387.5 | 385.3 | 384.1 | 391.2 | 7.1 | 0.02 | -0.06 | 0.07 | 1.5 | 0 |
| 65 | 1.0 | 0.0 | 0.0 | 386.5 | 388.3 | 384.7 | 387.5 | 387.6 | 385.9 | 393.4 | 7.2 | 0.02 | -0.08 | 0.08 | 2.7 | 0 |
| 66 | 1.0 | 0.0 | 0.0 | 389.0 | 388.4 | 387.7 | 390.0 | 388.0 | 386.3 | 393.7 | 7.2 | 0.02 | -0.08 | 0.09 | 2.6 | 0 |
| 67 | 1.0 | 0.0 | 0.0 | 389.2 | 390.5 | 387.9 | 390.2 | 389.8 | 388.0 | 395.6 | 7.2 | 0.02 | -0.08 | 0.08 | 2.6 | 0 |
| 68 | 1.0 | 0.0 | 0.0 | 391.1 | 402.1 | 389.4 | 391.8 | 395.7 | 405.5 | 413.8 | 8.5 | 0.02 | -0.07 | 0.0 | 1.5 | 0 |
| 69 | 1.0 | 0.0 | 0.0 | 445.5 | 445.1 | 444.3 | 446.1 | 442.7 | 445.9 | 451.0 | 7.9 | 0.02 | -0.07 | 0.05 | 2.0 | 0 |

Table S18: Indole benzene stack harmonic vibrational frequencies.

| | S | B | T | PBE | RPBE | PBEsol | PW91 | optPBE-vdW | BEEF-vdW | $\mu$ | $\sigma$ | COV | Skew | Kurt. | JB | Imag. |
|---|---|---|---|---|---|---|---|---|---|---|---|---|---|---|---|---|
| 1 | | | | 0.3 | 0.1 | 0.3 | 1.1 | 0.8 | 9.2 | 7.9 | 4.4 | 0.55 | -0.26 | -0.77 | 71.8 | 908 |
| 2 | | | | 0.7 | 0.3 | 1.4 | 1.6 | 3.0 | 9.8 | 7.7 | 5.4 | 0.71 | 0.05 | -1.05 | 92.3 | 730 |
| 3 | | | | 1.3 | 1.1 | 2.2 | 2.0 | 3.5 | 13.1 | 7.9 | 7.3 | 0.92 | 0.42 | -1.11 | 163.1 | 435 |
| 4 | | | | 2.1 | 1.8 | 3.2 | 3.4 | 6.5 | 13.6 | 7.8 | 7.1 | 0.91 | 0.55 | -0.96 | 179.5 | 191 |
| 5 | | | | 3.4 | 2.7 | 4.8 | 3.7 | 7.9 | 14.5 | 8.4 | 6.7 | 0.81 | 0.6 | -0.83 | 178.8 | 54 |
| 6 | | | | 3.7 | 3.0 | 5.1 | 4.0 | 8.6 | 14.8 | 9.9 | 6.1 | 0.62 | 0.42 | -0.58 | 86.3 | 20 |
| 7 | 0.0 | 0.0 | 1.0 | 25.7 | 25.3 | 26.0 | 25.8 | 25.9 | 30.7 | 22.6 | 10.4 | 0.46 | -0.58 | -0.76 | 157.9 | 10 |
| 8 | 0.0 | 0.0 | 1.0 | 29.5 | 29.2 | 29.7 | 29.6 | 30.0 | 33.7 | 27.6 | 8.6 | 0.31 | -1.06 | 0.62 | 406.6 | 7 |
| 9 | 0.17 | 0.82 | 0.01 | 48.6 | 46.8 | 48.3 | 48.7 | 48.1 | 51.7 | 40.0 | 13.8 | 0.34 | -0.62 | -0.9 | 195.2 | 3 |
| 10 | 0.0 | 0.0 | 1.0 | 48.9 | 48.6 | 48.5 | 49.2 | 48.7 | 54.1 | 48.2 | 8.0 | 0.17 | -1.58 | 6.57 | 4429.0 | 0 |
| 11 | 0.0 | 0.0 | 1.0 | 48.9 | 48.8 | 48.6 | 49.3 | 49.0 | 54.2 | 48.5 | 7.5 | 0.16 | -1.27 | 4.95 | 2578.6 | 0 |
| 12 | 0.0 | 0.0 | 1.0 | 49.6 | 48.8 | 50.6 | 50.4 | 49.2 | 56.1 | 49.7 | 7.8 | 0.16 | -0.81 | 2.54 | 756.2 | 0 |
| 13 | 0.0 | 0.0 | 1.0 | 51.6 | 51.2 | 52.4 | 51.9 | 51.3 | 59.2 | 53.2 | 9.4 | 0.18 | -0.05 | 0.36 | 11.4 | 0 |
| 14 | 0.28 | 0.72 | 0.0 | 66.5 | 66.2 | 66.3 | 66.7 | 66.6 | 67.3 | 65.7 | 6.8 | 0.1 | -2.62 | 10.14 | 10853.6 | 0 |
| 15 | 0.0 | 0.0 | 1.0 | 70.6 | 70.0 | 70.6 | 70.8 | 70.1 | 73.7 | 69.4 | 6.6 | 0.1 | -2.08 | 7.62 | 6271.2 | 0 |
| 16 | 0.04 | 0.17 | 0.8 | 74.2 | 73.6 | 74.0 | 74.4 | 73.6 | 75.6 | 72.2 | 5.2 | 0.07 | -1.38 | 4.24 | 2132.4 | 0 |
| 17 | 0.15 | 0.65 | 0.2 | 74.3 | 73.8 | 74.0 | 74.5 | 74.4 | 77.4 | 73.6 | 5.1 | 0.07 | -1.34 | 3.29 | 1504.2 | 0 |
| 18 | 0.06 | 0.94 | 0.0 | 74.5 | 74.2 | 74.2 | 74.8 | 74.9 | 77.4 | 74.8 | 4.1 | 0.05 | -1.59 | 5.74 | 3588.9 | 0 |
| 19 | 0.06 | 0.94 | 0.01 | 74.5 | 74.2 | 74.4 | 74.8 | 75.0 | 77.5 | 76.2 | 5.1 | 0.07 | 0.73 | 2.74 | 804.9 | 0 |
| 20 | 0.0 | 0.0 | 1.0 | 81.5 | 81.7 | 80.8 | 82.1 | 81.6 | 90.1 | 81.5 | 8.1 | 0.1 | -0.05 | -1.15 | 111.9 | 0 |
| 21 | 0.0 | 0.0 | 1.0 | 86.5 | 86.2 | 86.4 | 86.7 | 86.1 | 92.6 | 84.4 | 8.2 | 0.1 | -0.41 | -0.77 | 104.2 | 0 |
| 22 | 0.0 | 0.0 | 1.0 | 87.1 | 86.4 | 87.3 | 87.5 | 86.5 | 93.7 | 86.8 | 8.4 | 0.1 | -0.3 | -0.55 | 55.1 | 0 |
| 23 | 0.0 | 0.0 | 1.0 | 89.5 | 89.2 | 89.2 | 89.8 | 89.4 | 93.9 | 89.9 | 8.0 | 0.09 | 0.17 | 0.57 | 37.0 | 0 |
| 24 | 0.0 | 0.0 | 1.0 | 93.3 | 92.6 | 93.4 | 93.6 | 92.6 | 97.7 | 93.1 | 8.2 | 0.09 | -0.01 | 0.76 | 48.4 | 0 |
| 25 | 0.57 | 0.43 | 0.0 | 93.9 | 93.1 | 94.3 | 94.1 | 93.6 | 98.8 | 95.3 | 7.7 | 0.08 | -0.25 | 1.78 | 284.8 | 0 |
| 26 | 0.0 | 0.0 | 1.0 | 101.8 | 101.3 | 101.5 | 102.2 | 101.5 | 108.9 | 100.5 | 9.3 | 0.09 | -0.78 | 0.3 | 212.5 | 0 |
| 27 | 0.0 | 0.0 | 1.0 | 102.7 | 101.9 | 102.3 | 103.1 | 102.3 | 110.1 | 101.5 | 9.6 | 0.09 | -0.62 | -0.19 | 129.1 | 0 |
| 28 | 0.0 | 0.0 | 1.0 | 103.0 | 102.9 | 102.3 | 103.6 | 102.8 | 110.8 | 103.2 | 9.3 | 0.09 | -0.12 | -0.66 | 40.7 | 0 |
| 29 | 0.0 | 0.0 | 1.0 | 103.1 | 103.0 | 102.7 | 103.6 | 103.1 | 111.2 | 104.3 | 9.1 | 0.09 | 0.26 | -0.5 | 43.5 | 0 |
| 30 | 0.17 | 0.83 | 0.0 | 107.1 | 106.5 | 106.9 | 107.4 | 107.3 | 111.4 | 108.1 | 7.3 | 0.07 | -0.03 | 1.2 | 121.0 | 0 |
| 31 | 0.23 | 0.77 | 0.0 | 109.7 | 109.1 | 109.8 | 110.0 | 109.8 | 111.8 | 110.0 | 6.4 | 0.06 | -0.08 | 2.64 | 581.5 | 0 |
| 32 | 0.0 | 0.0 | 1.0 | 112.1 | 111.6 | 111.6 | 112.6 | 112.0 | 120.1 | 113.0 | 7.1 | 0.06 | -0.5 | 1.03 | 172.9 | 0 |
| 33 | 0.0 | 0.0 | 1.0 | 117.0 | 116.5 | 116.6 | 117.5 | 116.9 | 120.4 | 116.0 | 8.0 | 0.07 | -0.77 | 0.48 | 218.4 | 0 |
| 34 | 0.0 | 0.0 | 1.0 | 118.1 | 117.6 | 117.4 | 118.7 | 117.9 | 124.9 | 117.5 | 8.2 | 0.07 | -0.5 | -0.38 | 95.9 | 0 |
| 35 | 0.0 | 0.0 | 1.0 | 118.1 | 117.7 | 117.4 | 118.7 | 118.0 | 125.1 | 118.2 | 8.0 | 0.07 | -0.24 | -0.85 | 79.6 | 0 |
| 36 | 0.0 | 0.0 | 1.0 | 121.6 | 121.0 | 121.0 | 122.3 | 121.4 | 125.9 | 120.2 | 7.5 | 0.06 | -0.48 | -0.38 | 89.2 | 0 |
| 37 | 1.0 | 0.0 | 0.0 | 123.3 | 122.0 | 122.8 | 123.5 | 122.4 | 126.2 | 124.7 | 3.6 | 0.03 | 0.36 | 1.5 | 230.9 | 0 |
| 38 | 0.0 | 0.99 | 0.0 | 123.4 | 123.0 | 124.4 | 123.9 | 124.1 | 126.8 | 126.0 | 3.4 | 0.03 | 0.73 | 2.3 | 619.4 | 0 |
| 39 | 0.64 | 0.36 | 0.0 | 125.2 | 124.4 | 125.4 | 125.4 | 124.7 | 129.1 | 127.2 | 3.7 | 0.03 | 0.93 | 2.58 | 843.2 | 0 |
| 40 | 0.53 | 0.47 | 0.0 | 128.3 | 127.7 | 128.3 | 128.6 | 128.1 | 129.3 | 128.9 | 3.4 | 0.03 | 0.39 | 3.07 | 835.9 | 0 |
| 41 | 0.54 | 0.46 | 0.0 | 128.4 | 127.7 | 128.4 | 128.7 | 128.2 | 129.7 | 129.4 | 3.6 | 0.03 | 0.44 | 2.41 | 547.4 | 0 |
| 42 | 0.31 | 0.69 | 0.0 | 130.8 | 130.8 | 130.0 | 131.1 | 131.0 | 133.1 | 131.7 | 4.0 | 0.03 | 0.11 | 0.87 | 67.1 | 0 |
| 43 | 0.36 | 0.64 | 0.0 | 134.2 | 133.8 | 134.2 | 134.5 | 134.0 | 137.9 | 135.3 | 4.0 | 0.03 | -0.43 | 0.69 | 101.1 | 0 |
| 44 | 0.35 | 0.65 | 0.0 | 137.7 | 137.4 | 137.1 | 138.0 | 137.9 | 140.2 | 138.3 | 3.9 | 0.03 | -1.24 | 2.21 | 920.9 | 0 |
| 45 | 0.16 | 0.84 | 0.0 | 141.0 | 141.3 | 139.6 | 141.5 | 142.1 | 146.2 | 141.8 | 5.3 | 0.04 | -0.95 | 0.35 | 311.0 | 0 |
| 46 | 0.09 | 0.91 | 0.0 | 141.1 | 141.4 | 139.7 | 141.7 | 142.5 | 146.7 | 142.5 | 5.4 | 0.04 | -0.75 | -0.09 | 186.5 | 0 |
| 47 | 0.22 | 0.78 | 0.0 | 143.9 | 144.0 | 142.8 | 144.4 | 144.8 | 148.8 | 145.2 | 4.8 | 0.03 | -0.92 | 0.89 | 345.3 | 0 |
| 48 | 0.22 | 0.78 | 0.0 | 144.1 | 144.1 | 142.9 | 144.5 | 144.9 | 149.0 | 145.6 | 4.9 | 0.03 | -0.66 | 0.25 | 151.4 | 0 |
| 49 | 0.36 | 0.64 | 0.0 | 147.9 | 147.1 | 148.0 | 148.2 | 147.8 | 149.3 | 148.3 | 3.4 | 0.02 | -1.37 | 3.58 | 1695.8 | 0 |
| 50 | 0.28 | 0.72 | 0.0 | 152.6 | 152.3 | 151.8 | 153.1 | 153.1 | 154.7 | 153.2 | 3.3 | 0.02 | -1.32 | 2.97 | 1319.1 | 0 |
| 51 | 0.64 | 0.36 | 0.0 | 157.5 | 156.2 | 158.3 | 157.7 | 156.6 | 157.7 | 158.2 | 2.5 | 0.02 | -2.13 | 13.81 | 17418.1 | 0 |
| 52 | 0.0 | 1.0 | 0.0 | 165.3 | 163.9 | 163.4 | 166.1 | 164.1 | 160.7 | 163.0 | 2.8 | 0.02 | -0.59 | 0.04 | 117.9 | 0 |
| 53 | 0.28 | 0.72 | 0.0 | 165.9 | 165.3 | 165.4 | 166.3 | 166.0 | 163.6 | 165.4 | 2.1 | 0.01 | -0.37 | 0.37 | 56.8 | 0 |
| 54 | 0.91 | 0.09 | 0.0 | 166.4 | 165.5 | 169.5 | 166.4 | 167.2 | 171.5 | 171.0 | 2.9 | 0.02 | 0.72 | 0.82 | 228.6 | 0 |
| 55 | 0.68 | 0.32 | 0.0 | 167.9 | 166.3 | 169.9 | 167.9 | 167.2 | 174.3 | 172.9 | 4.0 | 0.02 | 0.82 | 0.26 | 229.0 | 0 |
| 56 | 0.51 | 0.48 | 0.0 | 175.2 | 173.9 | 176.0 | 175.3 | 174.1 | 175.8 | 177.0 | 2.5 | 0.01 | 1.83 | 4.94 | 3144.7 | 0 |
| 57 | 0.52 | 0.48 | 0.0 | 178.9 | 177.7 | 180.2 | 179.2 | 178.4 | 180.9 | 181.0 | 2.0 | 0.01 | 1.02 | 2.62 | 919.4 | 0 |
| 58 | 0.29 | 0.71 | 0.0 | 181.8 | 181.3 | 181.1 | 182.3 | 182.7 | 186.0 | 184.3 | 3.4 | 0.02 | 0.14 | -0.57 | 34.1 | 0 |
| 59 | 0.29 | 0.71 | 0.0 | 181.9 | 181.4 | 181.3 | 182.5 | 182.7 | 186.9 | 185.3 | 3.5 | 0.02 | 0.83 | 0.36 | 238.4 | 0 |
| 60 | 0.45 | 0.55 | 0.0 | 183.7 | 182.7 | 184.0 | 184.0 | 183.7 | 187.1 | 186.4 | 2.9 | 0.02 | 1.3 | 1.79 | 826.0 | 0 |
| 61 | 0.6 | 0.4 | 0.0 | 186.5 | 185.3 | 187.4 | 186.7 | 185.7 | 187.3 | 189.0 | 2.1 | 0.01 | 1.4 | 3.33 | 1579.8 | 0 |

| | | | | | | | | | | | | | | | |
|---|---|---|---|---|---|---|---|---|---|---|---|---|---|---|---|
| 62 | 0.69 | 0.31 | 0.0 | 195.0 | 193.3 | 196.4 | 195.1 | 193.9 | 193.5 | 196.5 | 1.8 | 0.01 | 0.83 | 1.42 | 400.8 | 0 |
| 63 | 0.71 | 0.29 | 0.0 | 197.3 | 195.6 | 198.6 | 197.6 | 196.3 | 195.9 | 198.6 | 1.7 | 0.01 | 0.71 | 1.1 | 268.8 | 0 |
| 64 | 0.71 | 0.29 | 0.0 | 197.3 | 195.7 | 198.7 | 197.6 | 196.3 | 195.9 | 198.7 | 1.7 | 0.01 | 0.72 | 1.12 | 275.2 | 0 |
| 65 | 0.71 | 0.29 | 0.0 | 200.3 | 198.4 | 202.2 | 200.6 | 199.1 | 197.9 | 201.5 | 2.3 | 0.01 | 0.75 | 0.76 | 236.4 | 0 |
| 66 | 1.0 | 0.0 | 0.0 | 382.9 | 382.1 | 381.6 | 383.8 | 381.1 | 379.4 | 386.9 | 7.1 | 0.02 | -0.08 | 0.07 | 2.5 | 0 |
| 67 | 1.0 | 0.0 | 0.0 | 383.7 | 382.8 | 382.3 | 384.7 | 382.0 | 380.2 | 387.8 | 7.2 | 0.02 | -0.08 | 0.08 | 2.6 | 0 |
| 68 | 1.0 | 0.0 | 0.0 | 384.3 | 383.3 | 382.9 | 385.1 | 382.4 | 380.6 | 388.2 | 7.2 | 0.02 | -0.08 | 0.07 | 2.6 | 0 |
| 69 | 1.0 | 0.0 | 0.0 | 384.5 | 383.4 | 383.2 | 385.4 | 382.7 | 380.9 | 388.4 | 7.2 | 0.02 | -0.08 | 0.08 | 2.8 | 0 |
| 70 | 1.0 | 0.0 | 0.0 | 384.5 | 383.6 | 383.2 | 385.6 | 382.8 | 381.1 | 388.6 | 7.2 | 0.02 | -0.08 | 0.08 | 2.5 | 0 |
| 71 | 1.0 | 0.0 | 0.0 | 385.7 | 384.9 | 384.3 | 386.7 | 383.9 | 382.2 | 389.8 | 7.3 | 0.02 | -0.08 | 0.08 | 2.7 | 0 |
| 72 | 1.0 | 0.0 | 0.0 | 386.1 | 385.3 | 384.7 | 386.9 | 384.2 | 382.5 | 390.1 | 7.3 | 0.02 | -0.08 | 0.07 | 2.7 | 0 |
| 73 | 1.0 | 0.0 | 0.0 | 386.3 | 385.3 | 385.0 | 387.2 | 384.5 | 382.7 | 390.3 | 7.3 | 0.02 | -0.08 | 0.08 | 2.8 | 0 |
| 74 | 1.0 | 0.0 | 0.0 | 387.1 | 386.4 | 385.8 | 388.1 | 385.4 | 383.6 | 391.3 | 7.3 | 0.02 | -0.08 | 0.08 | 2.8 | 0 |
| 75 | 1.0 | 0.0 | 0.0 | 387.5 | 386.6 | 386.2 | 388.5 | 385.7 | 383.9 | 391.5 | 7.3 | 0.02 | -0.08 | 0.08 | 2.8 | 0 |
| 76 | 1.0 | 0.0 | 0.0 | 393.2 | 392.3 | 391.4 | 394.3 | 391.7 | 390.1 | 397.5 | 7.2 | 0.02 | -0.08 | 0.07 | 2.3 | 0 |
| 77 | 1.0 | 0.0 | 0.0 | 395.5 | 394.7 | 393.7 | 396.6 | 394.0 | 392.7 | 400.0 | 7.3 | 0.02 | -0.08 | 0.06 | 2.3 | 0 |
| 78 | 1.0 | 0.0 | 0.0 | 445.1 | 445.1 | 443.9 | 446.0 | 442.2 | 445.8 | 450.4 | 7.7 | 0.02 | -0.07 | 0.05 | 1.8 | 0 |

Table S19: Indole benzene t-shaped harmonic vibrational frequencies.

| | S | B | T | PBE | RPBE | PBEsol | PW91 | optPBE-vdW | BEEF-vdW | $\mu$ | $\sigma$ | COV | Skew | Kurt. | JB | Imag. |
|---|---|---|---|---|---|---|---|---|---|---|---|---|---|---|---|---|
| 1 | | | | 0.1 | 1.2 | 0.8 | 0.5 | 0.2 | 8.1 | 7.4 | 4.2 | 0.56 | -0.22 | -0.67 | 54.2 | 918 |
| 2 | | | | 1.2 | 1.4 | 1.6 | 0.8 | 0.5 | 10.3 | 7.7 | 5.7 | 0.73 | 0.04 | -1.16 | 112.2 | 703 |
| 3 | | | | 1.6 | 1.4 | 1.7 | 1.6 | 2.6 | 12.3 | 7.7 | 7.1 | 0.92 | 0.45 | -1.07 | 163.5 | 411 |
| 4 | | | | 3.2 | 3.1 | 4.6 | 3.1 | 5.2 | 13.0 | 7.9 | 6.9 | 0.88 | 0.58 | -0.85 | 174.6 | 221 |
| 5 | | | | 4.9 | 4.4 | 5.6 | 4.9 | 6.5 | 13.9 | 8.0 | 6.9 | 0.86 | 0.64 | -0.84 | 195.8 | 84 |
| 6 | | | | 6.2 | 5.0 | 7.2 | 6.1 | 8.1 | 14.2 | 9.9 | 5.8 | 0.59 | 0.55 | -0.32 | 108.8 | 23 |
| 7 | 0.0 | 0.0 | 1.0 | 25.8 | 25.6 | 25.9 | 25.8 | 25.1 | 30.3 | 22.7 | 9.9 | 0.44 | -0.62 | -0.55 | 152.0 | 7 |
| 8 | 0.0 | 0.0 | 1.0 | 29.4 | 29.2 | 29.5 | 29.5 | 29.1 | 33.0 | 27.3 | 8.5 | 0.31 | -1.13 | 0.95 | 498.8 | 4 |
| 9 | 0.17 | 0.83 | 0.0 | 48.6 | 48.6 | 48.3 | 48.8 | 48.7 | 51.8 | 41.6 | 13.5 | 0.33 | -0.96 | -0.3 | 314.1 | 1 |
| 10 | 0.0 | 0.0 | 1.0 | 48.9 | 49.0 | 48.6 | 49.1 | 49.0 | 53.9 | 48.4 | 7.8 | 0.16 | -1.56 | 6.62 | 4458.6 | 0 |
| 11 | 0.0 | 0.0 | 1.0 | 49.1 | 49.1 | 48.8 | 49.4 | 49.3 | 54.1 | 48.8 | 7.4 | 0.15 | -1.21 | 4.77 | 2388.9 | 0 |
| 12 | 0.0 | 0.0 | 1.0 | 51.4 | 51.0 | 51.4 | 51.6 | 49.8 | 55.9 | 50.1 | 7.7 | 0.15 | -0.82 | 2.5 | 744.8 | 0 |
| 13 | 0.0 | 0.0 | 1.0 | 55.5 | 52.9 | 57.9 | 55.6 | 51.4 | 63.2 | 54.2 | 9.7 | 0.18 | -0.14 | -0.03 | 6.9 | 0 |
| 14 | 0.28 | 0.72 | 0.0 | 66.5 | 66.2 | 66.4 | 66.7 | 66.7 | 67.4 | 66.3 | 6.5 | 0.1 | -2.69 | 12.28 | 14977.6 | 0 |
| 15 | 0.0 | 0.0 | 1.0 | 70.8 | 70.3 | 71.0 | 71.0 | 70.3 | 73.8 | 69.7 | 6.6 | 0.09 | -2.11 | 7.86 | 6634.3 | 0 |
| 16 | 0.19 | 0.81 | 0.0 | 74.2 | 73.8 | 74.0 | 74.5 | 73.8 | 75.6 | 72.5 | 5.1 | 0.07 | -1.51 | 4.84 | 2719.1 | 0 |
| 17 | 0.06 | 0.94 | 0.0 | 74.4 | 74.1 | 74.0 | 74.7 | 74.5 | 77.2 | 73.8 | 5.0 | 0.07 | -1.48 | 4.02 | 2074.5 | 0 |
| 18 | 0.06 | 0.94 | 0.0 | 74.5 | 74.2 | 74.2 | 74.8 | 74.9 | 77.3 | 74.9 | 4.0 | 0.05 | -1.56 | 5.82 | 3634.0 | 0 |
| 19 | 0.0 | 0.0 | 1.0 | 74.6 | 74.2 | 74.8 | 74.8 | 75.0 | 77.9 | 76.5 | 5.3 | 0.07 | 0.79 | 2.14 | 591.8 | 0 |
| 20 | 0.0 | 0.0 | 1.0 | 83.2 | 82.9 | 82.7 | 83.5 | 83.1 | 90.1 | 82.2 | 8.1 | 0.1 | -0.16 | -1.18 | 124.4 | 0 |
| 21 | 0.0 | 0.0 | 1.0 | 86.8 | 86.5 | 86.5 | 87.1 | 86.4 | 93.3 | 84.8 | 8.3 | 0.1 | -0.42 | -0.8 | 113.5 | 0 |
| 22 | 0.0 | 0.0 | 1.0 | 87.3 | 86.6 | 87.6 | 87.6 | 86.8 | 93.8 | 87.0 | 8.4 | 0.1 | -0.3 | -0.58 | 59.3 | 0 |
| 23 | 0.0 | 0.0 | 1.0 | 89.8 | 89.4 | 89.3 | 90.0 | 89.4 | 93.9 | 90.3 | 8.0 | 0.09 | 0.26 | 0.53 | 45.7 | 0 |
| 24 | 0.0 | 0.0 | 1.0 | 93.6 | 92.8 | 93.7 | 93.9 | 92.9 | 97.6 | 93.4 | 8.1 | 0.09 | 0.04 | 0.79 | 52.1 | 0 |
| 25 | 0.57 | 0.43 | 0.0 | 93.9 | 93.1 | 94.3 | 94.1 | 93.7 | 98.7 | 95.6 | 7.6 | 0.08 | -0.16 | 1.67 | 239.5 | 0 |
| 26 | 0.0 | 0.0 | 1.0 | 102.1 | 101.5 | 101.8 | 102.5 | 101.8 | 109.1 | 101.0 | 8.9 | 0.09 | -0.76 | 0.38 | 202.3 | 0 |
| 27 | 0.0 | 0.0 | 1.0 | 102.9 | 102.1 | 102.8 | 103.4 | 102.5 | 109.7 | 102.1 | 9.1 | 0.09 | -0.62 | -0.06 | 127.3 | 0 |
| 28 | 0.0 | 0.0 | 1.0 | 104.4 | 104.0 | 104.0 | 104.7 | 104.3 | 110.6 | 103.9 | 9.1 | 0.09 | -0.24 | -0.56 | 45.4 | 0 |
| 29 | 0.0 | 0.0 | 1.0 | 104.5 | 104.0 | 104.1 | 104.9 | 104.5 | 111.4 | 105.0 | 9.0 | 0.09 | 0.13 | -0.53 | 28.6 | 0 |
| 30 | 0.17 | 0.83 | 0.0 | 107.1 | 106.5 | 106.9 | 107.4 | 107.4 | 112.2 | 108.5 | 7.4 | 0.07 | -0.05 | 0.94 | 74.3 | 0 |
| 31 | 0.22 | 0.78 | 0.0 | 109.6 | 109.1 | 109.6 | 110.0 | 109.9 | 112.4 | 110.4 | 6.6 | 0.06 | 0.04 | 2.02 | 340.7 | 0 |
| 32 | 0.0 | 0.0 | 1.0 | 112.3 | 111.8 | 111.8 | 112.7 | 112.1 | 119.9 | 113.2 | 7.0 | 0.06 | -0.36 | 0.86 | 105.5 | 0 |
| 33 | 0.0 | 0.0 | 1.0 | 117.1 | 116.5 | 116.6 | 117.5 | 117.0 | 119.9 | 116.2 | 7.6 | 0.07 | -0.7 | 0.43 | 177.8 | 0 |
| 34 | 0.0 | 0.0 | 1.0 | 119.1 | 118.5 | 118.7 | 119.6 | 119.1 | 124.8 | 118.1 | 7.8 | 0.07 | -0.6 | -0.17 | 122.3 | 0 |
| 35 | 0.0 | 0.0 | 1.0 | 119.2 | 118.5 | 118.8 | 119.6 | 119.1 | 124.8 | 118.9 | 7.9 | 0.07 | -0.37 | -0.71 | 87.9 | 0 |
| 36 | 0.0 | 0.01 | 0.99 | 122.5 | 121.8 | 122.2 | 123.1 | 122.2 | 126.5 | 120.8 | 7.3 | 0.06 | -0.59 | -0.19 | 117.9 | 0 |
| 37 | 1.0 | 0.0 | 0.0 | 123.2 | 121.8 | 123.0 | 123.3 | 122.4 | 126.7 | 124.9 | 3.6 | 0.03 | 0.34 | 1.15 | 148.9 | 0 |
| 38 | 0.0 | 0.99 | 0.01 | 123.5 | 123.0 | 124.1 | 124.0 | 124.3 | 126.8 | 126.1 | 3.5 | 0.03 | 0.74 | 1.81 | 456.6 | 0 |
| 39 | 0.64 | 0.36 | 0.0 | 125.1 | 124.4 | 125.0 | 125.3 | 124.6 | 128.9 | 127.3 | 3.8 | 0.03 | 0.89 | 1.89 | 560.8 | 0 |
| 40 | 0.53 | 0.47 | 0.0 | 128.3 | 127.6 | 128.3 | 128.6 | 128.0 | 129.2 | 129.0 | 3.5 | 0.03 | 0.56 | 2.89 | 800.3 | 0 |
| 41 | 0.54 | 0.46 | 0.0 | 128.4 | 127.6 | 128.4 | 128.6 | 128.1 | 130.0 | 129.4 | 3.8 | 0.03 | 0.58 | 2.03 | 455.6 | 0 |
| 42 | 0.29 | 0.71 | 0.0 | 130.7 | 130.7 | 129.7 | 131.1 | 131.2 | 133.5 | 131.8 | 4.1 | 0.03 | 0.16 | 0.54 | 33.0 | 0 |
| 43 | 0.36 | 0.64 | 0.0 | 134.3 | 134.0 | 134.0 | 134.6 | 134.0 | 137.5 | 135.4 | 3.9 | 0.03 | -0.43 | 0.84 | 122.0 | 0 |
| 44 | 0.35 | 0.65 | 0.0 | 137.5 | 137.4 | 136.8 | 137.9 | 137.8 | 140.1 | 138.3 | 3.8 | 0.03 | -1.23 | 2.32 | 953.1 | 0 |
| 45 | 0.16 | 0.84 | 0.0 | 140.9 | 141.2 | 139.4 | 141.4 | 142.1 | 146.3 | 141.8 | 5.2 | 0.04 | -0.99 | 0.53 | 348.1 | 0 |
| 46 | 0.09 | 0.91 | 0.0 | 141.5 | 141.7 | 140.0 | 141.9 | 142.8 | 146.5 | 142.7 | 5.4 | 0.04 | -0.81 | -0.05 | 218.1 | 0 |
| 47 | 0.23 | 0.77 | 0.0 | 144.2 | 144.2 | 143.1 | 144.5 | 145.0 | 148.9 | 145.5 | 4.7 | 0.03 | -0.93 | 0.95 | 361.6 | 0 |
| 48 | 0.23 | 0.77 | 0.0 | 144.3 | 144.3 | 143.1 | 144.7 | 145.2 | 149.1 | 145.8 | 4.9 | 0.03 | -0.64 | 0.24 | 143.4 | 0 |
| 49 | 0.36 | 0.64 | 0.0 | 147.9 | 147.2 | 147.8 | 148.3 | 148.0 | 149.5 | 148.6 | 3.3 | 0.02 | -1.44 | 3.81 | 1903.0 | 0 |
| 50 | 0.27 | 0.73 | 0.0 | 152.5 | 152.4 | 151.6 | 153.0 | 153.1 | 155.1 | 153.3 | 3.3 | 0.02 | -1.24 | 2.76 | 1151.0 | 0 |
| 51 | 0.65 | 0.35 | 0.0 | 157.6 | 156.4 | 158.3 | 157.7 | 157.0 | 157.8 | 158.5 | 2.4 | 0.02 | -2.33 | 14.85 | 20179.5 | 0 |
| 52 | 0.0 | 1.0 | 0.0 | 165.5 | 163.6 | 163.7 | 166.1 | 163.6 | 160.4 | 162.9 | 2.7 | 0.02 | -0.5 | 0.01 | 81.9 | 0 |
| 53 | 0.27 | 0.73 | 0.0 | 165.8 | 165.3 | 165.2 | 166.2 | 166.0 | 163.6 | 165.4 | 2.1 | 0.01 | -0.32 | 0.3 | 41.7 | 0 |
| 54 | 0.91 | 0.09 | 0.0 | 166.2 | 165.7 | 169.0 | 166.3 | 167.1 | 171.6 | 171.0 | 2.9 | 0.02 | 0.61 | 0.56 | 152.0 | 0 |
| 55 | 0.68 | 0.32 | 0.0 | 167.8 | 166.4 | 169.7 | 167.9 | 167.4 | 174.5 | 172.9 | 3.9 | 0.02 | 0.8 | 0.23 | 215.8 | 0 |
| 56 | 0.51 | 0.49 | 0.0 | 175.3 | 174.2 | 175.8 | 175.5 | 174.4 | 175.8 | 177.1 | 2.5 | 0.01 | 1.67 | 4.3 | 2473.8 | 0 |
| 57 | 0.52 | 0.48 | 0.0 | 178.9 | 177.8 | 180.2 | 179.1 | 178.5 | 180.7 | 181.1 | 2.0 | 0.01 | 0.99 | 2.71 | 940.1 | 0 |
| 58 | 0.29 | 0.71 | 0.0 | 181.8 | 181.3 | 181.1 | 182.3 | 182.6 | 185.9 | 184.2 | 3.4 | 0.02 | 0.11 | -0.54 | 28.5 | 0 |
| 59 | 0.29 | 0.71 | 0.0 | 181.8 | 181.4 | 181.2 | 182.4 | 182.7 | 186.7 | 185.2 | 3.4 | 0.02 | 0.82 | 0.37 | 236.1 | 0 |
| 60 | 0.44 | 0.56 | 0.0 | 183.5 | 182.8 | 183.8 | 183.9 | 183.7 | 187.1 | 186.4 | 2.9 | 0.02 | 1.28 | 1.74 | 795.3 | 0 |
| 61 | 0.6 | 0.4 | 0.0 | 186.2 | 185.1 | 187.1 | 186.5 | 185.6 | 187.2 | 188.9 | 2.1 | 0.01 | 1.38 | 3.14 | 1457.4 | 0 |

| 62 | 0.69 | 0.31 | 0.0 | 194.8 | 193.3 | 196.1 | 195.0 | 193.9 | 193.3 | 196.3 | 1.8 | 0.01 | 0.79 | 1.33 | 355.6 | 0 |
|---|---|---|---|---|---|---|---|---|---|---|---|---|---|---|---|---|
| 63 | 0.71 | 0.29 | 0.0 | 197.0 | 195.4 | 198.2 | 197.2 | 195.9 | 195.6 | 198.3 | 1.6 | 0.01 | 0.73 | 1.16 | 290.1 | 0 |
| 64 | 0.71 | 0.29 | 0.0 | 197.1 | 195.4 | 198.2 | 197.3 | 196.0 | 195.7 | 198.3 | 1.6 | 0.01 | 0.72 | 1.16 | 286.4 | 0 |
| 65 | 0.71 | 0.29 | 0.0 | 200.1 | 198.2 | 201.8 | 200.2 | 199.0 | 197.7 | 201.3 | 2.3 | 0.01 | 0.72 | 0.73 | 219.2 | 0 |
| 66 | 1.0 | 0.0 | 0.0 | 383.4 | 382.7 | 382.2 | 384.3 | 382.0 | 380.1 | 387.6 | 7.2 | 0.02 | -0.08 | 0.08 | 2.5 | 0 |
| 67 | 1.0 | 0.0 | 0.0 | 383.7 | 382.8 | 382.5 | 384.8 | 382.4 | 380.8 | 388.1 | 7.1 | 0.02 | -0.08 | 0.08 | 2.6 | 0 |
| 68 | 1.0 | 0.0 | 0.0 | 384.4 | 383.9 | 383.1 | 385.4 | 383.1 | 381.2 | 388.7 | 7.3 | 0.02 | -0.08 | 0.07 | 2.6 | 0 |
| 69 | 1.0 | 0.0 | 0.0 | 384.8 | 383.9 | 383.4 | 385.8 | 383.3 | 381.9 | 389.1 | 7.2 | 0.02 | -0.08 | 0.08 | 2.6 | 0 |
| 70 | 1.0 | 0.0 | 0.0 | 385.0 | 384.0 | 383.8 | 386.0 | 383.6 | 381.9 | 389.3 | 7.2 | 0.02 | -0.08 | 0.08 | 2.6 | 0 |
| 71 | 1.0 | 0.0 | 0.0 | 385.6 | 385.0 | 384.2 | 386.6 | 384.3 | 382.3 | 389.8 | 7.3 | 0.02 | -0.08 | 0.07 | 2.5 | 0 |
| 72 | 1.0 | 0.0 | 0.0 | 386.6 | 385.7 | 385.2 | 387.6 | 385.1 | 383.6 | 391.0 | 7.3 | 0.02 | -0.09 | 0.08 | 3.0 | 0 |
| 73 | 1.0 | 0.0 | 0.0 | 386.7 | 385.9 | 385.5 | 387.7 | 385.3 | 383.7 | 391.1 | 7.3 | 0.02 | -0.08 | 0.08 | 2.9 | 0 |
| 74 | 1.0 | 0.0 | 0.0 | 386.9 | 386.4 | 385.5 | 387.9 | 385.6 | 383.7 | 391.4 | 7.4 | 0.02 | -0.08 | 0.07 | 2.4 | 0 |
| 75 | 1.0 | 0.0 | 0.0 | 387.8 | 387.0 | 386.4 | 388.7 | 386.3 | 384.7 | 392.2 | 7.3 | 0.02 | -0.08 | 0.08 | 2.7 | 0 |
| 76 | 1.0 | 0.0 | 0.0 | 393.1 | 392.5 | 391.6 | 393.9 | 391.7 | 389.9 | 397.3 | 7.1 | 0.02 | -0.08 | 0.07 | 2.5 | 0 |
| 77 | 1.0 | 0.0 | 0.0 | 395.3 | 394.8 | 393.9 | 396.2 | 394.0 | 392.1 | 399.5 | 7.2 | 0.02 | -0.08 | 0.07 | 2.5 | 0 |
| 78 | 1.0 | 0.0 | 0.0 | 440.2 | 443.7 | 434.8 | 440.8 | 441.0 | 445.5 | 450.7 | 8.6 | 0.02 | -0.08 | -0.0 | 1.9 | 0 |

Table S20: Uracil dimer h-bonded harmonic vibrational frequencies.

| | S | B | T | PBE | RPBE | PBEsol | PW91 | optPBE-vdW | BEEF-vdW | $\mu$ | $\sigma$ | COV | Skew | Kurt. | JB | Imag. |
|---|---|---|---|---|---|---|---|---|---|---|---|---|---|---|---|---|
| 1 | | | | 2.8 | 2.6 | 3.2 | 2.7 | 1.3 | 9.9 | 8.6 | 4.0 | 0.47 | -0.45 | -0.53 | 89.7 | 944 |
| 2 | | | | 5.9 | 5.5 | 6.3 | 6.0 | 5.3 | 10.3 | 8.2 | 5.2 | 0.63 | -0.35 | -1.02 | 127.0 | 674 |
| 3 | | | | 8.3 | 7.7 | 9.2 | 8.3 | 7.4 | 11.7 | 8.7 | 4.9 | 0.56 | 0.07 | -0.68 | 40.5 | 368 |
| 4 | | | | 10.5 | 9.4 | 11.6 | 10.7 | 10.3 | 11.7 | 9.3 | 5.0 | 0.54 | -0.13 | -0.65 | 40.1 | 265 |
| 5 | | | | 11.4 | 10.8 | 12.2 | 11.7 | 11.0 | 12.3 | 10.8 | 4.2 | 0.39 | -0.43 | 0.84 | 121.9 | 189 |
| 6 | | | | 17.6 | 16.0 | 19.1 | 17.8 | 17.5 | 16.7 | 14.6 | 4.7 | 0.32 | -1.4 | 1.64 | 881.8 | 118 |
| 7 | 0.0 | 0.0 | 1.0 | 19.2 | 18.8 | 19.8 | 19.3 | 18.3 | 23.6 | 18.3 | 7.4 | 0.41 | -0.53 | -0.0 | 94.4 | 62 |
| 8 | 0.0 | 0.0 | 1.0 | 20.7 | 20.2 | 21.5 | 20.8 | 19.8 | 24.7 | 19.0 | 7.6 | 0.4 | -0.56 | -0.14 | 106.4 | 22 |
| 9 | 0.0 | 0.0 | 1.0 | 22.1 | 21.4 | 22.5 | 22.1 | 21.4 | 26.2 | 20.4 | 7.5 | 0.37 | -0.26 | -0.63 | 55.9 | 11 |
| 10 | 0.0 | 0.0 | 1.0 | 23.2 | 22.4 | 23.8 | 23.3 | 22.3 | 27.2 | 22.3 | 6.2 | 0.28 | 0.11 | -0.19 | 7.4 | 5 |
| 11 | 0.19 | 0.81 | 0.0 | 47.8 | 47.3 | 48.1 | 47.9 | 47.8 | 49.1 | 44.9 | 7.7 | 0.17 | -2.45 | 7.24 | 6363.2 | 0 |
| 12 | 0.2 | 0.8 | 0.0 | 49.6 | 48.8 | 50.2 | 49.7 | 49.6 | 50.2 | 45.7 | 7.6 | 0.17 | -1.95 | 4.14 | 2695.3 | 0 |
| 13 | 0.0 | 0.0 | 1.0 | 50.1 | 49.6 | 50.4 | 50.3 | 49.6 | 54.9 | 50.5 | 6.2 | 0.12 | -0.41 | 4.19 | 1520.8 | 0 |
| 14 | 0.0 | 0.0 | 1.0 | 50.2 | 49.6 | 50.4 | 50.3 | 49.6 | 55.0 | 51.2 | 5.6 | 0.11 | -0.43 | 4.76 | 1952.6 | 0 |
| 15 | 0.06 | 0.94 | 0.0 | 62.7 | 62.5 | 62.4 | 62.9 | 63.1 | 64.2 | 62.4 | 4.8 | 0.08 | -3.16 | 13.07 | 17548.6 | 0 |
| 16 | 0.15 | 0.85 | 0.0 | 63.8 | 63.2 | 64.0 | 64.1 | 64.0 | 64.6 | 63.1 | 4.7 | 0.07 | -3.11 | 10.61 | 12590.5 | 0 |
| 17 | 0.19 | 0.81 | 0.0 | 66.3 | 65.6 | 66.5 | 66.5 | 66.3 | 67.8 | 65.9 | 3.5 | 0.05 | -2.14 | 8.55 | 7611.8 | 0 |
| 18 | 0.15 | 0.85 | 0.0 | 66.6 | 65.9 | 67.1 | 66.9 | 66.8 | 68.0 | 66.4 | 3.1 | 0.05 | -2.19 | 11.87 | 13347.7 | 0 |
| 19 | 0.25 | 0.75 | 0.0 | 68.9 | 68.3 | 69.2 | 69.1 | 68.7 | 69.6 | 68.5 | 3.0 | 0.04 | -1.78 | 5.98 | 4038.4 | 0 |
| 20 | 0.2 | 0.8 | 0.0 | 70.4 | 69.6 | 71.1 | 70.6 | 70.0 | 71.2 | 69.6 | 3.2 | 0.05 | -1.33 | 3.18 | 1426.1 | 0 |
| 21 | 0.0 | 0.0 | 1.0 | 82.7 | 81.4 | 83.7 | 82.9 | 81.6 | 89.3 | 80.4 | 9.5 | 0.12 | -0.22 | -1.3 | 157.1 | 0 |
| 22 | 0.0 | 0.0 | 1.0 | 82.8 | 81.7 | 83.7 | 82.9 | 81.7 | 89.4 | 80.7 | 9.3 | 0.12 | -0.19 | -1.33 | 159.0 | 0 |
| 23 | 0.0 | 0.0 | 1.0 | 87.2 | 86.6 | 87.4 | 87.4 | 86.6 | 92.1 | 85.7 | 7.8 | 0.09 | -0.71 | 0.14 | 168.1 | 0 |
| 24 | 0.0 | 0.0 | 1.0 | 87.2 | 86.7 | 87.5 | 87.4 | 86.6 | 92.1 | 85.8 | 7.7 | 0.09 | -0.64 | -0.07 | 135.3 | 0 |
| 25 | 0.0 | 0.0 | 1.0 | 92.2 | 90.8 | 93.4 | 92.5 | 90.9 | 92.9 | 90.4 | 5.7 | 0.06 | -1.25 | 2.82 | 1179.8 | 0 |
| 26 | 0.0 | 0.0 | 1.0 | 92.3 | 91.0 | 93.5 | 92.6 | 91.0 | 93.1 | 90.9 | 5.1 | 0.06 | -1.03 | 3.08 | 1145.0 | 0 |
| 27 | 0.73 | 0.27 | 0.0 | 94.4 | 93.2 | 95.4 | 94.5 | 93.7 | 96.0 | 94.3 | 6.3 | 0.07 | 0.46 | 1.92 | 379.7 | 0 |
| 28 | 0.73 | 0.27 | 0.0 | 94.6 | 93.5 | 95.6 | 94.8 | 94.0 | 96.0 | 94.9 | 6.0 | 0.06 | 0.54 | 2.64 | 678.4 | 0 |
| 29 | 0.0 | 0.0 | 1.0 | 98.5 | 97.9 | 98.6 | 98.9 | 98.3 | 104.8 | 99.5 | 8.4 | 0.08 | 0.29 | -0.34 | 37.6 | 0 |
| 30 | 0.0 | 0.0 | 1.0 | 98.6 | 97.9 | 98.7 | 98.9 | 98.4 | 104.8 | 100.0 | 8.1 | 0.08 | 0.29 | -0.13 | 28.5 | 0 |
| 31 | 0.0 | 0.0 | 1.0 | 108.9 | 103.5 | 114.2 | 109.8 | 104.9 | 107.3 | 103.1 | 7.9 | 0.08 | 0.54 | -0.54 | 121.9 | 0 |
| 32 | 0.0 | 0.0 | 1.0 | 112.0 | 106.9 | 115.6 | 112.9 | 108.5 | 110.7 | 105.3 | 8.3 | 0.08 | 0.17 | -0.93 | 81.4 | 0 |
| 33 | 0.55 | 0.45 | 0.0 | 117.7 | 116.6 | 118.2 | 117.8 | 117.3 | 117.0 | 114.2 | 6.3 | 0.06 | -1.26 | 0.82 | 581.6 | 0 |
| 34 | 0.44 | 0.56 | 0.0 | 117.8 | 116.8 | 118.3 | 118.0 | 117.5 | 117.5 | 114.5 | 6.5 | 0.06 | -1.19 | 0.65 | 504.7 | 0 |
| 35 | 0.0 | 0.0 | 1.0 | 117.9 | 117.0 | 119.3 | 118.4 | 117.6 | 121.0 | 119.4 | 2.5 | 0.02 | 0.05 | 0.56 | 27.0 | 0 |
| 36 | 0.0 | 0.0 | 1.0 | 118.2 | 117.1 | 120.7 | 118.8 | 117.7 | 121.1 | 119.6 | 2.6 | 0.02 | 0.34 | 1.0 | 121.8 | 0 |
| 37 | 0.3 | 0.7 | 0.0 | 120.3 | 119.3 | 121.1 | 120.6 | 120.2 | 125.5 | 122.9 | 3.7 | 0.03 | 1.0 | 0.06 | 335.9 | 0 |
| 38 | 0.43 | 0.57 | 0.0 | 120.6 | 119.5 | 121.4 | 120.8 | 120.4 | 125.5 | 123.1 | 3.7 | 0.03 | 1.04 | 0.24 | 367.0 | 0 |
| 39 | 0.49 | 0.51 | 0.0 | 132.3 | 131.5 | 132.4 | 132.7 | 132.1 | 131.4 | 132.3 | 2.5 | 0.02 | -0.31 | 4.63 | 1817.7 | 0 |
| 40 | 0.48 | 0.52 | 0.0 | 132.5 | 131.7 | 132.6 | 132.8 | 132.3 | 131.6 | 132.4 | 2.6 | 0.02 | -0.35 | 4.33 | 1605.5 | 0 |
| 41 | 0.64 | 0.36 | 0.0 | 145.6 | 143.5 | 147.6 | 145.7 | 144.9 | 142.9 | 145.2 | 1.8 | 0.01 | -1.27 | 6.43 | 3978.9 | 0 |
| 42 | 0.62 | 0.38 | 0.0 | 145.8 | 143.7 | 147.7 | 145.9 | 145.1 | 143.3 | 145.4 | 1.7 | 0.01 | -1.48 | 8.53 | 6793.7 | 0 |
| 43 | 0.37 | 0.63 | 0.0 | 151.6 | 150.9 | 152.1 | 152.0 | 151.7 | 154.4 | 153.0 | 2.8 | 0.02 | 0.24 | -0.62 | 51.1 | 0 |
| 44 | 0.37 | 0.63 | 0.0 | 151.7 | 151.1 | 152.2 | 152.1 | 151.9 | 154.6 | 153.2 | 2.7 | 0.02 | 0.34 | -0.62 | 69.3 | 0 |
| 45 | 0.33 | 0.67 | 0.0 | 165.9 | 165.7 | 165.3 | 166.2 | 166.6 | 168.7 | 167.3 | 3.1 | 0.02 | -1.02 | 0.54 | 369.4 | 0 |
| 46 | 0.32 | 0.68 | 0.0 | 166.0 | 165.9 | 165.3 | 166.3 | 166.8 | 169.1 | 167.5 | 3.1 | 0.02 | -0.97 | 0.36 | 323.9 | 0 |
| 47 | 0.32 | 0.68 | 0.0 | 170.0 | 169.2 | 169.8 | 170.5 | 170.4 | 172.0 | 171.8 | 3.1 | 0.02 | 0.03 | -0.2 | 3.5 | 0 |
| 48 | 0.33 | 0.67 | 0.0 | 170.1 | 169.3 | 170.1 | 170.5 | 170.5 | 172.3 | 172.0 | 3.2 | 0.02 | 0.14 | -0.13 | 7.7 | 0 |
| 49 | 0.6 | 0.4 | 0.0 | 174.3 | 172.4 | 176.7 | 174.5 | 173.5 | 176.0 | 176.4 | 2.4 | 0.01 | 1.68 | 3.26 | 1828.3 | 0 |
| 50 | 0.62 | 0.38 | 0.0 | 174.4 | 172.5 | 177.1 | 174.6 | 173.5 | 176.3 | 176.7 | 2.4 | 0.01 | 1.57 | 3.06 | 1602.4 | 0 |
| 51 | 0.47 | 0.53 | 0.0 | 187.5 | 186.4 | 188.7 | 187.8 | 187.3 | 188.1 | 188.3 | 2.5 | 0.01 | -0.3 | -0.53 | 52.9 | 0 |
| 52 | 0.33 | 0.67 | 0.0 | 188.0 | 186.6 | 189.4 | 188.5 | 187.6 | 188.9 | 189.1 | 2.7 | 0.01 | 0.04 | -0.46 | 18.1 | 0 |
| 53 | 0.75 | 0.25 | 0.0 | 200.9 | 199.8 | 202.0 | 201.2 | 200.2 | 199.6 | 202.8 | 2.3 | 0.01 | 0.58 | 0.61 | 143.1 | 0 |
| 54 | 0.71 | 0.29 | 0.0 | 201.2 | 200.0 | 202.3 | 201.5 | 200.4 | 200.1 | 203.1 | 2.1 | 0.01 | 0.73 | 1.12 | 280.7 | 0 |
| 55 | 0.56 | 0.44 | 0.0 | 209.2 | 207.7 | 210.2 | 209.3 | 207.7 | 207.0 | 210.3 | 3.5 | 0.02 | 0.2 | 0.08 | 13.8 | 0 |
| 56 | 0.73 | 0.27 | 0.0 | 209.5 | 208.0 | 211.2 | 209.5 | 208.2 | 208.6 | 211.3 | 3.6 | 0.02 | 0.4 | 0.58 | 81.4 | 0 |
| 57 | 0.85 | 0.15 | 0.0 | 212.5 | 210.6 | 214.9 | 212.6 | 210.8 | 209.2 | 213.9 | 3.7 | 0.02 | 0.87 | 0.68 | 293.4 | 0 |
| 58 | 0.84 | 0.16 | 0.0 | 213.3 | 211.4 | 215.6 | 213.3 | 211.5 | 210.4 | 214.9 | 3.6 | 0.02 | 0.79 | 0.63 | 243.0 | 0 |
| 59 | 0.99 | 0.01 | 0.0 | 370.2 | 384.6 | 346.8 | 369.3 | 378.0 | 386.1 | 393.1 | 7.1 | 0.02 | -0.11 | 0.12 | 4.9 | 0 |
| 60 | 0.99 | 0.01 | 0.0 | 377.6 | 387.6 | 356.9 | 376.9 | 384.7 | 386.1 | 393.2 | 7.0 | 0.02 | -0.09 | 0.08 | 3.0 | 0 |
| 61 | 1.0 | 0.0 | 0.0 | 388.2 | 387.6 | 386.1 | 389.1 | 387.1 | 391.1 | 397.7 | 7.6 | 0.02 | -0.15 | 0.08 | 8.2 | 0 |

| | | | | | | | | | | | | | | | |
|---|---|---|---|---|---|---|---|---|---|---|---|---|---|---|---|
| 62 | 1.0 | 0.0 | 0.0 | 388.2 | 390.5 | 386.4 | 389.1 | 387.2 | 391.1 | 398.4 | 7.1 | 0.02 | -0.09 | 0.1 | 3.5 | 0 |
| 63 | 1.0 | 0.0 | 0.0 | 393.6 | 393.2 | 392.3 | 394.3 | 392.3 | 391.8 | 401.4 | 8.4 | 0.02 | 0.08 | -0.04 | 2.4 | 0 |
| 64 | 1.0 | 0.0 | 0.0 | 393.7 | 393.3 | 392.3 | 394.3 | 392.3 | 397.3 | 405.6 | 8.8 | 0.02 | -0.03 | -0.08 | 0.8 | 0 |
| 65 | 1.0 | 0.0 | 0.0 | 435.0 | 435.2 | 433.1 | 435.7 | 433.1 | 436.9 | 441.4 | 7.7 | 0.02 | -0.08 | 0.05 | 2.2 | 0 |
| 66 | 1.0 | 0.0 | 0.0 | 435.0 | 435.5 | 433.1 | 435.7 | 433.1 | 436.9 | 441.5 | 7.7 | 0.02 | -0.08 | 0.05 | 2.2 | 0 |

Table S21: Uracil dimer stack harmonic vibrational frequencies.

| | S | B | T | PBE | RPBE | PBEsol | PW91 | optPBE-vdW | BEEF-vdW | $\mu$ | $\sigma$ | COV | Skew | Kurt. | JB | Imag. |
|---|---|---|---|---|---|---|---|---|---|---|---|---|---|---|---|---|
| 1 | | | | 1.0 | 1.2 | 1.3 | 0.5 | 2.9 | 8.8 | 7.6 | 4.3 | 0.57 | -0.11 | -0.87 | 66.6 | 864 |
| 2 | | | | 2.4 | 1.9 | 2.0 | 1.1 | 3.4 | 8.8 | 7.2 | 4.5 | 0.63 | -0.05 | -0.97 | 79.5 | 712 |
| 3 | | | | 3.3 | 3.0 | 4.1 | 2.9 | 5.5 | 9.2 | 6.8 | 5.1 | 0.75 | 0.19 | -1.07 | 107.9 | 470 |
| 4 | | | | 4.1 | 3.4 | 4.5 | 3.5 | 6.0 | 10.5 | 7.2 | 5.0 | 0.7 | 0.33 | -1.04 | 126.7 | 287 |
| 5 | | | | 5.4 | 3.9 | 6.7 | 4.8 | 7.9 | 10.7 | 7.7 | 5.3 | 0.68 | 0.38 | -0.87 | 112.8 | 182 |
| 6 | | | | 6.1 | 5.2 | 8.3 | 5.1 | 9.6 | 11.8 | 9.2 | 4.8 | 0.53 | -0.0 | -0.51 | 21.6 | 95 |
| 7 | 0.01 | 0.02 | 0.97 | 18.3 | 17.5 | 19.7 | 18.2 | 19.5 | 23.0 | 17.0 | 8.7 | 0.51 | -0.49 | -0.7 | 120.8 | 41 |
| 8 | 0.02 | 0.02 | 0.96 | 20.0 | 18.0 | 20.8 | 19.4 | 20.5 | 24.7 | 18.6 | 8.3 | 0.45 | -0.5 | -0.55 | 109.9 | 20 |
| 9 | 0.03 | 0.0 | 0.97 | 20.4 | 20.3 | 20.9 | 20.5 | 21.2 | 25.1 | 19.5 | 8.2 | 0.42 | -0.47 | -0.46 | 90.8 | 11 |
| 10 | 0.04 | 0.01 | 0.95 | 20.9 | 20.6 | 22.7 | 20.9 | 21.8 | 25.6 | 21.2 | 6.7 | 0.32 | -0.29 | -0.12 | 29.0 | 2 |
| 11 | 0.18 | 0.76 | 0.06 | 46.2 | 46.0 | 46.1 | 46.2 | 46.1 | 48.4 | 42.6 | 9.3 | 0.22 | -2.11 | 4.82 | 3421.9 | 0 |
| 12 | 0.18 | 0.76 | 0.06 | 46.5 | 46.2 | 46.2 | 46.3 | 46.1 | 48.4 | 43.0 | 8.5 | 0.2 | -2.0 | 4.37 | 2923.0 | 0 |
| 13 | 0.03 | 0.04 | 0.94 | 48.3 | 47.6 | 48.8 | 48.5 | 48.4 | 53.9 | 48.4 | 8.4 | 0.17 | -1.64 | 6.35 | 4256.5 | 0 |
| 14 | 0.03 | 0.04 | 0.93 | 48.8 | 48.0 | 49.4 | 49.0 | 49.0 | 54.4 | 48.8 | 8.1 | 0.17 | -1.26 | 4.03 | 1877.6 | 0 |
| 15 | 0.08 | 0.9 | 0.03 | 62.6 | 62.3 | 62.3 | 62.6 | 62.5 | 64.2 | 59.2 | 8.2 | 0.14 | -1.62 | 2.72 | 1489.2 | 0 |
| 16 | 0.07 | 0.89 | 0.04 | 62.7 | 62.4 | 62.3 | 62.7 | 62.5 | 64.2 | 59.4 | 8.0 | 0.13 | -1.51 | 1.74 | 1008.5 | 0 |
| 17 | 0.19 | 0.79 | 0.01 | 65.0 | 64.7 | 64.9 | 65.1 | 64.9 | 66.8 | 64.1 | 5.0 | 0.08 | -2.53 | 8.8 | 8593.8 | 0 |
| 18 | 0.2 | 0.78 | 0.02 | 65.1 | 64.7 | 65.0 | 65.2 | 65.0 | 66.9 | 64.3 | 4.8 | 0.08 | -2.48 | 9.17 | 9059.7 | 0 |
| 19 | 0.21 | 0.78 | 0.01 | 66.9 | 66.6 | 67.2 | 67.2 | 67.1 | 68.6 | 66.4 | 3.9 | 0.06 | -1.03 | 2.19 | 755.1 | 0 |
| 20 | 0.22 | 0.76 | 0.01 | 67.2 | 66.8 | 67.2 | 67.3 | 67.2 | 68.6 | 66.5 | 3.8 | 0.06 | -1.05 | 2.2 | 770.1 | 0 |
| 21 | 0.01 | 0.04 | 0.95 | 70.3 | 68.8 | 71.0 | 70.1 | 69.7 | 77.7 | 73.3 | 9.1 | 0.12 | 0.65 | -0.75 | 187.5 | 0 |
| 22 | 0.01 | 0.04 | 0.95 | 70.6 | 68.9 | 71.4 | 70.7 | 70.1 | 78.2 | 73.5 | 9.2 | 0.12 | 0.63 | -0.82 | 187.6 | 0 |
| 23 | 0.0 | 0.01 | 0.98 | 84.0 | 82.2 | 83.8 | 84.1 | 83.9 | 91.2 | 82.0 | 9.8 | 0.12 | -0.45 | -1.0 | 149.2 | 0 |
| 24 | 0.0 | 0.01 | 0.98 | 84.2 | 82.7 | 84.5 | 84.2 | 84.4 | 91.7 | 82.5 | 9.7 | 0.12 | -0.46 | -0.95 | 143.9 | 0 |
| 25 | 0.0 | 0.0 | 0.99 | 87.5 | 87.3 | 87.7 | 87.9 | 87.1 | 91.8 | 86.3 | 7.9 | 0.09 | -0.66 | 0.17 | 146.5 | 0 |
| 26 | 0.01 | 0.0 | 0.99 | 87.7 | 87.4 | 87.7 | 87.9 | 87.2 | 92.0 | 86.7 | 7.8 | 0.09 | -0.66 | 0.24 | 152.3 | 0 |
| 27 | 0.01 | 0.0 | 0.98 | 90.7 | 89.5 | 91.5 | 91.0 | 89.4 | 92.6 | 90.4 | 5.3 | 0.06 | 0.25 | 1.56 | 224.9 | 0 |
| 28 | 0.1 | 0.03 | 0.87 | 92.1 | 90.3 | 93.6 | 92.4 | 91.4 | 92.7 | 91.2 | 5.2 | 0.06 | 0.1 | 1.8 | 272.8 | 0 |
| 29 | 0.72 | 0.23 | 0.05 | 93.5 | 92.2 | 94.6 | 93.7 | 93.0 | 96.1 | 94.9 | 6.1 | 0.06 | 1.02 | 2.3 | 787.4 | 0 |
| 30 | 0.66 | 0.22 | 0.12 | 93.6 | 92.3 | 94.8 | 93.8 | 93.2 | 98.3 | 95.7 | 6.5 | 0.07 | 0.83 | 1.41 | 395.2 | 0 |
| 31 | 0.01 | 0.0 | 0.99 | 97.7 | 97.1 | 97.9 | 98.1 | 97.2 | 104.6 | 100.2 | 6.7 | 0.07 | 0.81 | 0.26 | 226.2 | 0 |
| 32 | 0.01 | 0.0 | 0.99 | 98.1 | 97.5 | 98.3 | 98.5 | 97.7 | 105.1 | 100.6 | 6.9 | 0.07 | 0.78 | 0.15 | 202.3 | 0 |
| 33 | 0.01 | 0.01 | 0.98 | 115.8 | 115.3 | 115.3 | 116.4 | 115.7 | 116.1 | 112.7 | 6.3 | 0.06 | -1.23 | 0.59 | 536.8 | 0 |
| 34 | 0.02 | 0.01 | 0.96 | 116.0 | 115.3 | 115.5 | 116.4 | 115.8 | 116.1 | 112.8 | 6.3 | 0.06 | -1.23 | 0.58 | 529.5 | 0 |
| 35 | 0.59 | 0.38 | 0.03 | 116.6 | 115.6 | 117.4 | 116.6 | 116.2 | 120.9 | 118.7 | 2.6 | 0.02 | -0.03 | -0.64 | 34.3 | 0 |
| 36 | 0.61 | 0.38 | 0.01 | 116.7 | 115.7 | 117.4 | 116.7 | 116.3 | 121.0 | 118.8 | 2.6 | 0.02 | -0.01 | -0.62 | 31.8 | 0 |
| 37 | 0.22 | 0.77 | 0.01 | 119.6 | 118.7 | 120.1 | 120.0 | 119.7 | 124.7 | 122.0 | 3.5 | 0.03 | 1.01 | 0.25 | 345.8 | 0 |
| 38 | 0.22 | 0.77 | 0.01 | 119.8 | 118.8 | 120.3 | 120.1 | 119.9 | 124.7 | 122.1 | 3.5 | 0.03 | 1.01 | 0.27 | 349.3 | 0 |
| 39 | 0.51 | 0.49 | 0.0 | 131.6 | 130.7 | 131.8 | 131.9 | 131.4 | 130.9 | 131.7 | 2.4 | 0.02 | -0.53 | 3.97 | 1407.3 | 0 |
| 40 | 0.5 | 0.5 | 0.0 | 131.9 | 130.8 | 131.8 | 132.0 | 131.4 | 131.0 | 131.7 | 2.4 | 0.02 | -0.5 | 4.05 | 1452.4 | 0 |
| 41 | 0.63 | 0.37 | 0.0 | 143.2 | 141.2 | 144.1 | 143.0 | 141.7 | 140.4 | 142.3 | 2.1 | 0.01 | -1.94 | 9.66 | 9037.7 | 0 |
| 42 | 0.64 | 0.36 | 0.0 | 143.4 | 141.3 | 144.5 | 143.3 | 142.0 | 140.6 | 142.5 | 2.1 | 0.01 | -1.96 | 9.74 | 9181.8 | 0 |
| 43 | 0.33 | 0.66 | 0.0 | 148.1 | 147.7 | 148.5 | 148.4 | 148.8 | 152.6 | 150.9 | 3.1 | 0.02 | 0.32 | 0.57 | 61.1 | 0 |
| 44 | 0.35 | 0.65 | 0.0 | 148.1 | 147.8 | 148.7 | 148.5 | 148.9 | 152.8 | 151.1 | 3.1 | 0.02 | 0.35 | 0.53 | 63.1 | 0 |
| 45 | 0.32 | 0.68 | 0.01 | 165.1 | 164.2 | 164.4 | 165.3 | 165.0 | 165.7 | 165.3 | 2.1 | 0.01 | -1.29 | 1.63 | 779.4 | 0 |
| 46 | 0.28 | 0.72 | 0.01 | 165.3 | 164.5 | 164.5 | 165.5 | 165.7 | 167.1 | 166.1 | 2.4 | 0.01 | -1.18 | 0.96 | 543.2 | 0 |
| 47 | 0.36 | 0.64 | 0.01 | 168.3 | 167.3 | 168.5 | 168.4 | 168.6 | 173.2 | 171.4 | 4.4 | 0.03 | 0.24 | -0.33 | 28.8 | 0 |
| 48 | 0.32 | 0.67 | 0.01 | 169.1 | 167.8 | 168.7 | 169.1 | 169.4 | 173.5 | 171.9 | 4.4 | 0.03 | 0.09 | -0.24 | 7.2 | 0 |
| 49 | 0.32 | 0.68 | 0.0 | 170.3 | 169.9 | 170.2 | 170.6 | 171.2 | 175.7 | 173.8 | 4.0 | 0.02 | 0.7 | -0.03 | 164.5 | 0 |
| 50 | 0.37 | 0.63 | 0.01 | 170.6 | 170.0 | 171.0 | 170.8 | 171.2 | 175.7 | 174.1 | 3.8 | 0.02 | 0.81 | 0.18 | 221.4 | 0 |
| 51 | 0.47 | 0.52 | 0.01 | 180.6 | 179.6 | 181.4 | 180.8 | 180.5 | 182.8 | 183.2 | 2.7 | 0.01 | 0.79 | 0.38 | 222.4 | 0 |
| 52 | 0.48 | 0.52 | 0.01 | 180.8 | 179.8 | 181.8 | 181.0 | 180.6 | 182.9 | 183.4 | 2.6 | 0.01 | 0.88 | 0.61 | 291.6 | 0 |
| 53 | 0.74 | 0.26 | 0.0 | 200.7 | 199.2 | 202.0 | 201.2 | 199.7 | 199.4 | 202.4 | 2.2 | 0.01 | 0.8 | 1.15 | 324.4 | 0 |
| 54 | 0.74 | 0.26 | 0.0 | 200.9 | 199.4 | 202.0 | 201.2 | 199.8 | 199.4 | 202.5 | 2.1 | 0.01 | 0.83 | 1.26 | 360.9 | 0 |
| 55 | 0.84 | 0.16 | 0.0 | 210.1 | 208.2 | 211.9 | 210.2 | 207.5 | 206.6 | 210.5 | 4.5 | 0.02 | 0.38 | -0.09 | 47.6 | 0 |
| 56 | 0.83 | 0.17 | 0.0 | 211.1 | 209.1 | 213.0 | 211.2 | 208.6 | 207.7 | 211.7 | 4.4 | 0.02 | 0.37 | -0.07 | 46.3 | 0 |
| 57 | 0.83 | 0.16 | 0.0 | 215.8 | 214.2 | 217.9 | 216.2 | 213.9 | 213.0 | 217.0 | 4.5 | 0.02 | 0.39 | -0.08 | 50.3 | 0 |
| 58 | 0.83 | 0.17 | 0.0 | 216.2 | 214.7 | 218.5 | 216.7 | 214.5 | 213.5 | 217.4 | 4.4 | 0.02 | 0.39 | -0.06 | 51.4 | 0 |
| 59 | 1.0 | 0.0 | 0.0 | 388.4 | 387.7 | 386.8 | 389.5 | 387.3 | 386.3 | 393.3 | 7.0 | 0.02 | -0.08 | 0.07 | 2.6 | 0 |
| 60 | 1.0 | 0.0 | 0.0 | 388.4 | 387.7 | 387.1 | 389.5 | 387.5 | 386.3 | 393.3 | 7.0 | 0.02 | -0.08 | 0.07 | 2.6 | 0 |
| 61 | 1.0 | 0.0 | 0.0 | 393.9 | 393.5 | 392.4 | 394.8 | 392.4 | 391.2 | 398.6 | 7.2 | 0.02 | -0.08 | 0.07 | 2.4 | 0 |

| 62 | 1.0 | 0.0 | 0.0 | 394.3 | 393.8 | 392.5 | 394.9 | 392.6 | 391.2 | 398.6 | 7.2 | 0.02 | -0.08 | 0.07 | 2.4 | 0 |
| 63 | 1.0 | 0.0 | 0.0 | 432.2 | 434.6 | 429.0 | 433.8 | 429.8 | 434.5 | 438.8 | 7.6 | 0.02 | -0.08 | 0.06 | 2.3 | 0 |
| 64 | 1.0 | 0.0 | 0.0 | 432.4 | 434.6 | 429.2 | 434.0 | 429.9 | 434.6 | 438.9 | 7.6 | 0.02 | -0.08 | 0.06 | 2.3 | 0 |
| 65 | 1.0 | 0.0 | 0.0 | 439.9 | 440.6 | 438.1 | 440.8 | 437.3 | 441.3 | 445.9 | 7.7 | 0.02 | -0.07 | 0.05 | 1.9 | 0 |
| 66 | 1.0 | 0.0 | 0.0 | 440.0 | 440.7 | 438.2 | 441.2 | 437.3 | 441.6 | 446.2 | 7.7 | 0.02 | -0.07 | 0.05 | 1.9 | 0 |

Table S22: Adenine-thymine stack harmonic vibrational frequencies.

| | S | B | T | PBE | RPBE | PBEsol | PW91 | optPBE-vdW | BEEF-vdW | $\mu$ | $\sigma$ | COV | Skew | Kurt. | JB | Imag. |
|---|---|---|---|---|---|---|---|---|---|---|---|---|---|---|---|---|
| 1 | | | | 1.3 | 0.6 | 1.7 | 0.5 | 3.0 | 7.9 | 7.1 | 4.1 | 0.57 | -0.19 | -0.89 | 78.3 | 896 |
| 2 | | | | 2.4 | 1.7 | 3.5 | 2.2 | 3.9 | 8.1 | 7.0 | 4.4 | 0.62 | -0.05 | -0.98 | 80.8 | 775 |
| 3 | | | | 3.1 | 2.1 | 4.6 | 3.1 | 5.6 | 8.9 | 7.0 | 4.6 | 0.65 | 0.08 | -1.04 | 92.3 | 622 |
| 4 | | | | 4.0 | 2.9 | 5.5 | 3.8 | 6.9 | 10.0 | 7.4 | 4.8 | 0.65 | 0.08 | -1.1 | 103.8 | 483 |
| 5 | | | | 4.8 | 4.0 | 7.1 | 4.4 | 9.1 | 10.8 | 8.2 | 4.9 | 0.59 | 0.04 | -1.05 | 91.7 | 376 |
| 6 | | | | 6.4 | 4.8 | 8.0 | 6.1 | 10.8 | 11.1 | 9.2 | 4.9 | 0.53 | -0.12 | -0.84 | 63.7 | 299 |
| 7 | 0.05 | 0.02 | 0.94 | 13.4 | 13.2 | 13.9 | 13.7 | 15.0 | 18.6 | 13.4 | 7.2 | 0.54 | -0.23 | -1.01 | 101.3 | 211 |
| 8 | 0.06 | 0.01 | 0.93 | 16.4 | 16.9 | 16.6 | 16.5 | 16.4 | 23.3 | 16.4 | 8.7 | 0.53 | -0.27 | -0.97 | 103.9 | 117 |
| 9 | 0.02 | 0.01 | 0.97 | 19.1 | 18.2 | 19.3 | 19.2 | 19.5 | 23.9 | 18.1 | 8.6 | 0.47 | -0.46 | -0.72 | 113.0 | 63 |
| 10 | 0.02 | 0.01 | 0.98 | 19.5 | 19.0 | 20.6 | 19.5 | 21.4 | 31.1 | 21.7 | 10.8 | 0.5 | -0.23 | -0.98 | 97.8 | 35 |
| 11 | 0.01 | 0.01 | 0.99 | 25.8 | 25.5 | 26.0 | 25.7 | 26.5 | 35.9 | 25.6 | 11.0 | 0.43 | -0.37 | -0.94 | 119.2 | 22 |
| 12 | 0.12 | 0.85 | 0.03 | 33.2 | 33.5 | 32.8 | 33.2 | 33.9 | 36.4 | 30.5 | 9.9 | 0.33 | -1.28 | 0.7 | 590.7 | 16 |
| 13 | 0.06 | 0.92 | 0.02 | 33.3 | 33.6 | 33.0 | 33.4 | 34.2 | 38.7 | 33.3 | 8.8 | 0.27 | -1.27 | 1.96 | 857.6 | 10 |
| 14 | 0.0 | 0.01 | 0.99 | 36.1 | 35.5 | 36.4 | 36.1 | 36.3 | 39.6 | 35.0 | 8.5 | 0.24 | -1.11 | 2.15 | 798.7 | 5 |
| 15 | 0.0 | 0.02 | 0.97 | 36.7 | 37.0 | 37.3 | 36.9 | 37.2 | 41.5 | 36.9 | 8.9 | 0.24 | -0.69 | 1.26 | 292.5 | 4 |
| 16 | 0.21 | 0.76 | 0.04 | 46.5 | 46.3 | 46.5 | 46.5 | 46.5 | 47.7 | 42.5 | 10.5 | 0.25 | -0.96 | 0.78 | 355.6 | 2 |
| 17 | 0.02 | 0.02 | 0.96 | 48.3 | 47.8 | 48.5 | 48.5 | 48.3 | 53.6 | 45.8 | 10.9 | 0.24 | -0.9 | 0.82 | 325.0 | 0 |
| 18 | 0.05 | 0.13 | 0.82 | 49.4 | 48.5 | 52.0 | 48.7 | 55.6 | 56.3 | 48.7 | 11.2 | 0.23 | -1.02 | 0.72 | 387.4 | 0 |
| 19 | 0.19 | 0.8 | 0.01 | 55.7 | 55.3 | 55.6 | 55.7 | 56.6 | 63.1 | 53.0 | 12.1 | 0.23 | -1.03 | 0.5 | 372.8 | 0 |
| 20 | 0.06 | 0.21 | 0.73 | 60.1 | 59.8 | 59.9 | 59.9 | 59.7 | 64.4 | 56.1 | 11.2 | 0.2 | -1.17 | 1.15 | 565.1 | 0 |
| 21 | 0.17 | 0.64 | 0.19 | 63.5 | 62.2 | 63.4 | 63.5 | 63.1 | 66.2 | 59.2 | 9.0 | 0.15 | -0.83 | 0.12 | 228.3 | 0 |
| 22 | 0.11 | 0.47 | 0.42 | 63.9 | 63.3 | 63.9 | 63.9 | 63.7 | 67.0 | 63.2 | 8.1 | 0.13 | -0.32 | -0.04 | 33.7 | 0 |
| 23 | 0.06 | 0.32 | 0.62 | 64.4 | 63.6 | 65.2 | 64.4 | 64.4 | 69.9 | 65.7 | 6.8 | 0.1 | -0.01 | -0.0 | 0.0 | 0 |
| 24 | 0.22 | 0.71 | 0.06 | 65.3 | 64.9 | 66.0 | 65.7 | 65.4 | 70.1 | 67.3 | 6.4 | 0.1 | 0.07 | -0.44 | 17.5 | 0 |
| 25 | 0.01 | 0.02 | 0.97 | 69.5 | 68.3 | 69.3 | 68.9 | 68.2 | 74.1 | 69.6 | 7.3 | 0.11 | 0.3 | -0.14 | 32.5 | 0 |
| 26 | 0.02 | 0.08 | 0.9 | 70.7 | 69.4 | 70.7 | 70.2 | 69.8 | 74.6 | 71.2 | 7.0 | 0.1 | 0.55 | -0.37 | 113.0 | 0 |
| 27 | 0.14 | 0.76 | 0.1 | 73.0 | 72.5 | 72.7 | 73.0 | 72.9 | 75.4 | 73.9 | 6.7 | 0.09 | 0.28 | -0.28 | 32.4 | 0 |
| 28 | 0.32 | 0.67 | 0.02 | 74.4 | 73.8 | 74.5 | 74.6 | 74.5 | 76.0 | 76.5 | 5.9 | 0.08 | 0.73 | 0.12 | 178.3 | 0 |
| 29 | 0.0 | 0.0 | 0.99 | 80.2 | 79.7 | 80.3 | 80.4 | 79.7 | 83.5 | 79.5 | 6.1 | 0.08 | 0.29 | -0.46 | 46.1 | 0 |
| 30 | 0.01 | 0.01 | 0.98 | 81.5 | 81.1 | 81.3 | 81.6 | 80.8 | 85.2 | 81.3 | 6.1 | 0.07 | 0.38 | -0.0 | 48.3 | 0 |
| 31 | 0.0 | 0.01 | 0.99 | 82.4 | 81.6 | 82.8 | 82.6 | 81.8 | 86.9 | 83.4 | 5.7 | 0.07 | 0.37 | 0.27 | 51.1 | 0 |
| 32 | 0.71 | 0.28 | 0.01 | 88.0 | 87.0 | 88.8 | 88.1 | 87.4 | 87.1 | 87.5 | 4.1 | 0.05 | -0.02 | 4.61 | 1770.7 | 0 |
| 33 | 0.8 | 0.19 | 0.01 | 89.6 | 88.2 | 90.6 | 89.7 | 89.1 | 88.8 | 89.4 | 4.2 | 0.05 | -0.07 | 3.58 | 1069.1 | 0 |
| 34 | 0.0 | 0.0 | 0.99 | 90.0 | 88.8 | 90.8 | 90.4 | 89.1 | 93.5 | 91.2 | 4.6 | 0.05 | 0.02 | 1.83 | 278.7 | 0 |
| 35 | 0.0 | 0.04 | 0.96 | 92.8 | 91.9 | 93.2 | 93.1 | 91.8 | 96.3 | 93.3 | 4.4 | 0.05 | 0.4 | 1.48 | 237.1 | 0 |
| 36 | 0.43 | 0.56 | 0.01 | 97.4 | 96.3 | 97.9 | 97.5 | 96.7 | 96.9 | 95.9 | 4.7 | 0.05 | 0.28 | 0.68 | 64.3 | 0 |
| 37 | 0.0 | 0.0 | 0.99 | 97.6 | 96.4 | 98.0 | 97.9 | 96.9 | 100.8 | 98.2 | 4.6 | 0.05 | 0.32 | 0.99 | 114.6 | 0 |
| 38 | 0.0 | 0.0 | 0.99 | 101.3 | 100.4 | 101.3 | 101.6 | 101.1 | 108.6 | 102.4 | 6.1 | 0.06 | -0.07 | -0.98 | 80.8 | 0 |
| 39 | 0.03 | 0.13 | 0.84 | 107.7 | 107.1 | 106.8 | 108.1 | 107.5 | 110.2 | 106.3 | 6.1 | 0.06 | -0.31 | -0.73 | 76.5 | 0 |
| 40 | 0.19 | 0.68 | 0.13 | 107.8 | 108.0 | 108.1 | 108.3 | 107.7 | 114.6 | 110.0 | 5.4 | 0.05 | -0.48 | -0.0 | 77.3 | 0 |
| 41 | 0.21 | 0.78 | 0.0 | 113.2 | 112.4 | 113.5 | 113.6 | 113.3 | 115.7 | 113.2 | 4.7 | 0.04 | 0.04 | -0.06 | 0.9 | 0 |
| 42 | 0.0 | 0.0 | 1.0 | 115.5 | 115.2 | 115.1 | 115.7 | 115.3 | 117.2 | 116.3 | 4.7 | 0.04 | 0.47 | 0.76 | 122.5 | 0 |
| 43 | 0.47 | 0.46 | 0.06 | 116.4 | 115.7 | 116.5 | 116.7 | 116.5 | 123.5 | 119.1 | 5.4 | 0.05 | 0.29 | -0.3 | 35.4 | 0 |
| 44 | 0.18 | 0.68 | 0.14 | 122.7 | 122.7 | 122.0 | 123.2 | 123.8 | 125.6 | 124.0 | 4.6 | 0.04 | -0.93 | 1.69 | 525.4 | 0 |
| 45 | 0.44 | 0.56 | 0.0 | 123.8 | 122.8 | 123.4 | 123.4 | 124.2 | 128.3 | 125.3 | 4.7 | 0.04 | -0.49 | 0.22 | 83.8 | 0 |
| 46 | 0.0 | 0.63 | 0.37 | 127.4 | 127.7 | 125.9 | 127.9 | 128.9 | 131.3 | 128.4 | 5.4 | 0.04 | -0.64 | -0.63 | 169.2 | 0 |
| 47 | 0.49 | 0.51 | 0.0 | 131.4 | 130.6 | 132.0 | 131.5 | 130.9 | 135.2 | 132.3 | 3.4 | 0.03 | -0.72 | 0.74 | 220.6 | 0 |
| 48 | 0.53 | 0.47 | 0.0 | 138.1 | 136.7 | 139.1 | 138.2 | 137.3 | 135.9 | 137.8 | 1.4 | 0.01 | -3.33 | 23.92 | 51382.3 | 0 |
| 49 | 0.57 | 0.4 | 0.03 | 139.1 | 136.9 | 140.7 | 139.4 | 137.8 | 137.6 | 139.3 | 1.8 | 0.01 | -0.85 | 9.67 | 8036.1 | 0 |
| 50 | 0.46 | 0.54 | 0.01 | 143.9 | 143.3 | 144.7 | 144.6 | 144.3 | 145.2 | 145.3 | 1.7 | 0.01 | -1.02 | 4.12 | 1758.2 | 0 |
| 51 | 0.78 | 0.22 | 0.0 | 149.5 | 147.4 | 151.0 | 149.4 | 147.9 | 146.6 | 148.8 | 1.6 | 0.01 | -1.41 | 7.55 | 5416.9 | 0 |
| 52 | 0.45 | 0.55 | 0.0 | 150.6 | 149.4 | 151.7 | 150.9 | 150.5 | 150.8 | 151.5 | 1.3 | 0.01 | -2.67 | 21.95 | 42530.7 | 0 |
| 53 | 0.46 | 0.54 | 0.0 | 152.4 | 151.6 | 153.2 | 152.8 | 152.3 | 153.7 | 153.6 | 1.2 | 0.01 | 0.04 | 0.19 | 3.5 | 0 |
| 54 | 0.72 | 0.28 | 0.0 | 161.5 | 159.5 | 162.9 | 161.7 | 159.7 | 156.0 | 160.0 | 2.8 | 0.02 | -0.34 | -0.85 | 98.6 | 0 |
| 55 | 0.29 | 0.66 | 0.04 | 164.0 | 162.4 | 163.6 | 164.5 | 162.8 | 160.8 | 162.8 | 2.3 | 0.01 | -0.77 | 0.98 | 276.9 | 0 |
| 56 | 0.64 | 0.36 | 0.0 | 164.4 | 164.0 | 166.1 | 164.6 | 164.9 | 165.9 | 165.8 | 2.2 | 0.01 | -1.84 | 5.24 | 3419.7 | 0 |
| 57 | 0.64 | 0.36 | 0.0 | 166.0 | 164.3 | 166.3 | 166.0 | 165.0 | 167.4 | 167.0 | 2.1 | 0.01 | -1.58 | 4.88 | 2816.1 | 0 |
| 58 | 0.22 | 0.76 | 0.02 | 167.6 | 167.6 | 166.9 | 168.2 | 168.9 | 168.7 | 169.0 | 2.3 | 0.01 | -0.83 | 1.43 | 401.2 | 0 |
| 59 | 0.05 | 0.95 | 0.0 | 169.3 | 169.0 | 168.5 | 170.0 | 169.8 | 170.8 | 170.9 | 2.5 | 0.01 | -0.41 | -0.21 | 58.7 | 0 |
| 60 | 0.49 | 0.51 | 0.0 | 170.5 | 169.4 | 171.9 | 170.6 | 170.4 | 172.8 | 173.0 | 3.1 | 0.02 | 0.27 | -0.22 | 28.2 | 0 |
| 61 | 0.47 | 0.52 | 0.0 | 170.7 | 170.2 | 172.2 | 171.2 | 171.7 | 174.1 | 174.0 | 2.9 | 0.02 | 0.39 | 0.01 | 51.2 | 0 |

| | | | | | | | | | | | | | | | | |
|---|---|---|---|---|---|---|---|---|---|---|---|---|---|---|---|---|
| 62 | 0.6 | 0.4 | 0.0 | 172.7 | 171.2 | 172.8 | 172.8 | 172.0 | 179.4 | 176.4 | 4.0 | 0.02 | 0.68 | -0.6 | 182.5 | 0 |
| 63 | 0.0 | 0.71 | 0.28 | 175.5 | 176.6 | 174.4 | 176.2 | 177.8 | 180.7 | 179.5 | 3.8 | 0.02 | 0.11 | -0.87 | 66.4 | 0 |
| 64 | 0.12 | 0.68 | 0.21 | 177.7 | 177.6 | 175.4 | 178.5 | 179.1 | 181.8 | 181.0 | 3.5 | 0.02 | 0.32 | -0.31 | 42.3 | 0 |
| 65 | 0.44 | 0.53 | 0.04 | 179.5 | 178.9 | 180.9 | 180.1 | 179.4 | 182.9 | 182.8 | 2.9 | 0.02 | 1.47 | 3.19 | 1570.8 | 0 |
| 66 | 0.58 | 0.42 | 0.01 | 180.5 | 179.5 | 181.9 | 180.6 | 180.2 | 185.7 | 184.4 | 3.7 | 0.02 | 1.3 | 0.98 | 642.1 | 0 |
| 67 | 0.55 | 0.45 | 0.0 | 182.0 | 180.7 | 182.9 | 182.1 | 181.2 | 187.9 | 186.3 | 3.6 | 0.02 | 1.06 | 0.54 | 399.5 | 0 |
| 68 | 0.36 | 0.64 | 0.0 | 193.1 | 192.7 | 192.5 | 193.5 | 193.5 | 193.0 | 194.4 | 2.1 | 0.01 | -0.25 | 0.35 | 30.5 | 0 |
| 69 | 0.68 | 0.32 | 0.0 | 196.0 | 194.2 | 198.4 | 196.1 | 194.7 | 193.4 | 197.6 | 2.7 | 0.01 | 0.78 | 0.73 | 247.1 | 0 |
| 70 | 0.58 | 0.4 | 0.02 | 199.7 | 198.5 | 201.4 | 199.8 | 199.4 | 201.6 | 202.6 | 1.8 | 0.01 | 0.38 | 1.42 | 217.1 | 0 |
| 71 | 0.75 | 0.25 | 0.0 | 203.7 | 202.3 | 205.2 | 204.0 | 202.7 | 203.4 | 205.7 | 2.2 | 0.01 | 0.8 | 1.33 | 360.4 | 0 |
| 72 | 0.84 | 0.15 | 0.0 | 209.3 | 207.2 | 211.2 | 209.3 | 206.8 | 205.7 | 210.7 | 3.5 | 0.02 | 0.83 | 0.92 | 299.7 | 0 |
| 73 | 0.83 | 0.17 | 0.0 | 214.4 | 212.9 | 216.5 | 214.5 | 212.2 | 211.6 | 215.9 | 4.0 | 0.02 | 0.57 | 0.19 | 110.0 | 0 |
| 74 | 1.0 | 0.0 | 0.0 | 368.4 | 368.1 | 365.9 | 369.3 | 367.2 | 365.5 | 372.7 | 6.7 | 0.02 | -0.07 | 0.09 | 2.4 | 0 |
| 75 | 1.0 | 0.0 | 0.0 | 375.5 | 374.8 | 374.1 | 376.3 | 373.5 | 371.3 | 379.5 | 7.9 | 0.02 | -0.08 | 0.07 | 2.8 | 0 |
| 76 | 1.0 | 0.0 | 0.0 | 379.1 | 378.3 | 378.6 | 379.7 | 376.5 | 374.2 | 382.4 | 7.9 | 0.02 | -0.08 | 0.07 | 2.7 | 0 |
| 77 | 1.0 | 0.0 | 0.0 | 382.3 | 381.8 | 380.4 | 383.5 | 381.3 | 380.2 | 387.4 | 7.0 | 0.02 | -0.08 | 0.07 | 2.6 | 0 |
| 78 | 1.0 | 0.0 | 0.0 | 387.7 | 387.1 | 386.1 | 388.3 | 386.4 | 385.1 | 392.2 | 7.1 | 0.02 | -0.08 | 0.05 | 2.2 | 0 |
| 79 | 1.0 | 0.0 | 0.0 | 393.6 | 392.9 | 391.8 | 394.5 | 392.8 | 391.1 | 398.1 | 7.0 | 0.02 | -0.08 | 0.07 | 2.6 | 0 |
| 80 | 0.99 | 0.0 | 0.0 | 434.8 | 434.9 | 432.9 | 435.7 | 431.2 | 435.8 | 440.0 | 7.5 | 0.02 | -0.07 | 0.05 | 1.7 | 0 |
| 81 | 1.0 | 0.0 | 0.0 | 435.2 | 435.6 | 433.3 | 436.0 | 432.6 | 436.5 | 441.0 | 7.6 | 0.02 | -0.07 | 0.05 | 1.8 | 0 |
| 82 | 1.0 | 0.0 | 0.0 | 440.7 | 441.0 | 439.2 | 441.4 | 438.1 | 441.9 | 446.6 | 7.6 | 0.02 | -0.07 | 0.05 | 1.9 | 0 |
| 83 | 0.99 | 0.0 | 0.0 | 441.4 | 441.5 | 439.9 | 442.6 | 438.7 | 442.3 | 446.9 | 7.6 | 0.02 | -0.07 | 0.05 | 1.7 | 0 |
| 84 | 0.99 | 0.0 | 0.01 | 450.0 | 450.1 | 448.2 | 450.8 | 445.6 | 450.1 | 454.9 | 7.9 | 0.02 | -0.07 | 0.06 | 2.1 | 0 |

Table S23: Adenine-thymine Watson-Crick harmonic vibrational frequencies.

| | S | B | T | PBE | RPBE | PBE sol | PW91 | optPBE -vdW | BEEF -vdW | $\mu$ | $\sigma$ | COV | Skew | Kurt. | JB | Imag. |
|---|---|---|---|---|---|---|---|---|---|---|---|---|---|---|---|---|
| 1 | | | | 3.2 | 2.7 | 3.1 | 2.9 | 2.2 | 8.5 | 7.6 | 3.7 | 0.49 | -0.38 | -0.57 | 74.4 | 950 |
| 2 | | | | 3.8 | 3.3 | 4.2 | 3.9 | 3.1 | 9.7 | 7.7 | 4.3 | 0.56 | -0.41 | -0.83 | 113.7 | 819 |
| 3 | | | | 7.6 | 6.9 | 8.6 | 7.8 | 7.3 | 10.4 | 8.3 | 4.9 | 0.59 | -0.23 | -0.85 | 78.5 | 638 |
| 4 | | | | 8.3 | 7.4 | 8.9 | 8.3 | 7.5 | 11.8 | 8.8 | 4.9 | 0.56 | -0.24 | -0.78 | 70.3 | 506 |
| 5 | | | | 12.6 | 11.2 | 13.0 | 12.7 | 12.0 | 13.0 | 9.7 | 5.2 | 0.53 | -0.48 | -0.76 | 126.7 | 368 |
| 6 | | | | 13.3 | 12.3 | 13.9 | 13.6 | 12.7 | 14.2 | 11.4 | 4.9 | 0.43 | -0.91 | 0.23 | 282.0 | 273 |
| 7 | 0.0 | 0.0 | 0.99 | 13.8 | 13.2 | 15.6 | 13.8 | 13.5 | 17.4 | 13.6 | 6.2 | 0.46 | -0.34 | -0.33 | 47.4 | 193 |
| 8 | 0.0 | 0.0 | 1.0 | 17.2 | 17.8 | 16.8 | 17.5 | 18.0 | 23.2 | 16.7 | 8.2 | 0.49 | -0.25 | -0.77 | 69.9 | 110 |
| 9 | 0.0 | 0.0 | 1.0 | 19.6 | 19.2 | 19.8 | 19.7 | 19.2 | 24.7 | 18.8 | 7.9 | 0.42 | -0.27 | -0.6 | 54.7 | 63 |
| 10 | 0.0 | 0.0 | 1.0 | 21.2 | 20.5 | 21.8 | 21.2 | 20.1 | 31.5 | 22.4 | 10.1 | 0.45 | -0.07 | -1.07 | 96.6 | 38 |
| 11 | 0.0 | 0.0 | 1.0 | 27.4 | 27.0 | 27.6 | 27.4 | 26.6 | 35.6 | 26.6 | 10.6 | 0.4 | -0.28 | -1.03 | 114.6 | 24 |
| 12 | 0.07 | 0.92 | 0.02 | 34.7 | 34.7 | 34.4 | 34.8 | 35.1 | 38.2 | 31.3 | 9.9 | 0.32 | -0.96 | -0.09 | 309.1 | 12 |
| 13 | 0.0 | 0.0 | 1.0 | 35.9 | 35.5 | 36.0 | 36.0 | 35.3 | 39.2 | 33.9 | 8.2 | 0.24 | -1.18 | 1.76 | 722.1 | 8 |
| 14 | 0.0 | 0.02 | 0.97 | 36.2 | 35.7 | 36.3 | 36.3 | 35.5 | 39.8 | 35.2 | 7.8 | 0.22 | -1.08 | 2.16 | 779.1 | 4 |
| 15 | 0.14 | 0.86 | 0.0 | 37.9 | 36.9 | 39.0 | 38.0 | 37.5 | 40.5 | 37.7 | 8.7 | 0.23 | -0.43 | 1.05 | 153.9 | 2 |
| 16 | 0.2 | 0.78 | 0.02 | 48.8 | 46.6 | 49.6 | 48.9 | 48.4 | 49.8 | 42.8 | 10.1 | 0.24 | -0.48 | -0.28 | 82.1 | 1 |
| 17 | 0.0 | 0.02 | 0.98 | 49.5 | 48.1 | 50.0 | 49.7 | 48.7 | 54.4 | 46.8 | 10.2 | 0.22 | -0.74 | 0.4 | 197.7 | 0 |
| 18 | 0.01 | 0.0 | 0.99 | 53.3 | 48.9 | 57.4 | 53.9 | 50.1 | 57.3 | 49.7 | 9.7 | 0.2 | -0.9 | 0.75 | 316.8 | 0 |
| 19 | 0.21 | 0.79 | 0.0 | 57.0 | 56.4 | 58.8 | 57.2 | 56.8 | 58.4 | 55.8 | 8.3 | 0.15 | -0.85 | 1.5 | 430.8 | 0 |
| 20 | 0.17 | 0.83 | 0.0 | 64.2 | 62.7 | 64.4 | 64.4 | 62.0 | 64.9 | 59.9 | 7.8 | 0.13 | -1.28 | 2.57 | 1095.4 | 0 |
| 21 | 0.2 | 0.74 | 0.06 | 64.6 | 63.6 | 65.0 | 64.8 | 64.0 | 65.8 | 62.7 | 5.7 | 0.09 | -1.11 | 2.2 | 812.4 | 0 |
| 22 | 0.01 | 0.04 | 0.94 | 64.9 | 64.0 | 66.4 | 64.9 | 64.6 | 66.8 | 65.3 | 5.4 | 0.08 | 0.01 | 1.46 | 178.3 | 0 |
| 23 | 0.24 | 0.76 | 0.0 | 66.2 | 65.6 | 66.5 | 66.4 | 66.3 | 68.7 | 67.2 | 5.4 | 0.08 | 0.18 | 0.36 | 22.0 | 0 |
| 24 | 0.0 | 0.0 | 1.0 | 70.4 | 69.5 | 71.0 | 70.5 | 69.3 | 73.7 | 69.8 | 6.1 | 0.09 | -0.17 | -0.61 | 40.3 | 0 |
| 25 | 0.0 | 0.0 | 1.0 | 70.5 | 69.7 | 72.1 | 70.8 | 69.4 | 75.4 | 71.4 | 6.2 | 0.09 | 0.32 | -0.63 | 67.3 | 0 |
| 26 | 0.17 | 0.82 | 0.01 | 73.9 | 73.4 | 74.3 | 74.1 | 73.9 | 77.2 | 74.7 | 5.8 | 0.08 | -0.06 | -0.04 | 1.3 | 0 |
| 27 | 0.29 | 0.71 | 0.0 | 76.5 | 75.5 | 77.5 | 76.8 | 76.4 | 77.8 | 77.5 | 5.4 | 0.07 | 0.33 | -0.17 | 38.0 | 0 |
| 28 | 0.0 | 0.0 | 1.0 | 80.4 | 79.6 | 80.9 | 80.6 | 79.6 | 83.0 | 79.9 | 5.3 | 0.07 | 0.08 | -0.65 | 37.5 | 0 |
| 29 | 0.0 | 0.0 | 1.0 | 82.7 | 82.0 | 83.2 | 82.9 | 81.7 | 85.5 | 81.9 | 5.3 | 0.06 | -0.02 | -0.39 | 13.0 | 0 |
| 30 | 0.74 | 0.26 | 0.0 | 88.4 | 86.5 | 89.3 | 88.5 | 87.5 | 87.1 | 84.2 | 5.7 | 0.07 | 0.04 | -0.23 | 4.8 | 0 |
| 31 | 0.26 | 0.05 | 0.69 | 90.7 | 87.3 | 91.8 | 90.9 | 87.9 | 89.1 | 88.0 | 4.2 | 0.05 | -0.42 | 3.47 | 1063.8 | 0 |
| 32 | 0.56 | 0.11 | 0.33 | 90.8 | 89.4 | 92.2 | 91.0 | 89.5 | 91.0 | 90.3 | 4.5 | 0.05 | -0.21 | 2.3 | 455.9 | 0 |
| 33 | 0.0 | 0.02 | 0.98 | 93.1 | 89.6 | 94.1 | 93.6 | 90.2 | 93.4 | 91.8 | 4.6 | 0.05 | -0.02 | 1.66 | 229.3 | 0 |
| 34 | 0.0 | 0.02 | 0.98 | 93.9 | 92.5 | 97.6 | 94.6 | 92.5 | 96.6 | 93.7 | 4.4 | 0.05 | 0.38 | 1.64 | 270.8 | 0 |
| 35 | 0.0 | 0.0 | 1.0 | 97.5 | 96.4 | 98.2 | 97.8 | 96.1 | 97.3 | 95.9 | 4.4 | 0.05 | 0.29 | 1.24 | 156.3 | 0 |
| 36 | 0.4 | 0.6 | 0.0 | 97.6 | 96.8 | 100.7 | 97.9 | 97.2 | 100.9 | 98.3 | 4.6 | 0.05 | 0.42 | 1.33 | 204.4 | 0 |
| 37 | 0.0 | 0.0 | 1.0 | 100.6 | 100.0 | 103.0 | 100.9 | 99.2 | 109.0 | 102.4 | 6.2 | 0.06 | 0.14 | -0.87 | 70.0 | 0 |
| 38 | 0.0 | 0.04 | 0.96 | 108.6 | 108.3 | 107.9 | 108.8 | 107.8 | 110.2 | 106.8 | 6.1 | 0.06 | -0.51 | -0.57 | 113.6 | 0 |
| 39 | 0.2 | 0.8 | 0.0 | 109.4 | 108.4 | 110.4 | 109.7 | 109.2 | 114.7 | 110.4 | 5.5 | 0.05 | -0.83 | 0.79 | 282.1 | 0 |
| 40 | 0.22 | 0.78 | 0.0 | 113.5 | 112.5 | 113.8 | 113.8 | 113.5 | 116.9 | 113.1 | 5.7 | 0.05 | -0.65 | 0.78 | 191.6 | 0 |
| 41 | 0.0 | 0.0 | 1.0 | 117.9 | 116.9 | 118.0 | 118.3 | 117.6 | 118.9 | 116.1 | 5.8 | 0.05 | 0.11 | -0.26 | 9.7 | 0 |
| 42 | 0.43 | 0.48 | 0.08 | 118.4 | 117.5 | 118.5 | 118.7 | 118.2 | 121.0 | 118.7 | 6.2 | 0.05 | 0.23 | -0.67 | 55.1 | 0 |
| 43 | 0.24 | 0.64 | 0.12 | 122.7 | 118.1 | 122.8 | 123.2 | 120.3 | 125.2 | 120.8 | 5.9 | 0.05 | 0.09 | -0.41 | 16.4 | 0 |
| 44 | 0.14 | 0.18 | 0.69 | 125.6 | 122.8 | 125.8 | 125.9 | 123.6 | 126.1 | 124.6 | 5.1 | 0.04 | -0.65 | 1.18 | 255.9 | 0 |
| 45 | 0.35 | 0.42 | 0.24 | 125.7 | 124.1 | 126.9 | 126.7 | 125.3 | 128.1 | 125.9 | 4.9 | 0.04 | -0.25 | 0.44 | 36.2 | 0 |
| 46 | 0.0 | 0.6 | 0.4 | 127.3 | 127.6 | 130.9 | 128.0 | 128.7 | 130.6 | 128.5 | 4.9 | 0.04 | -0.33 | -0.97 | 114.5 | 0 |
| 47 | 0.49 | 0.51 | 0.0 | 130.8 | 130.1 | 135.7 | 131.0 | 130.0 | 136.1 | 132.2 | 4.1 | 0.03 | -0.17 | -0.08 | 10.5 | 0 |
| 48 | 0.52 | 0.48 | 0.0 | 139.0 | 137.5 | 140.2 | 139.4 | 138.2 | 138.3 | 138.7 | 1.6 | 0.01 | -3.28 | 22.72 | 46593.0 | 0 |
| 49 | 0.38 | 0.59 | 0.03 | 141.8 | 140.5 | 141.7 | 142.1 | 141.6 | 138.9 | 141.2 | 2.1 | 0.02 | -1.92 | 8.47 | 7212.2 | 0 |
| 50 | 0.77 | 0.23 | 0.0 | 147.9 | 145.0 | 150.7 | 148.3 | 146.6 | 145.7 | 147.5 | 1.5 | 0.01 | -0.91 | 4.6 | 2039.8 | 0 |
| 51 | 0.77 | 0.23 | 0.0 | 149.2 | 147.6 | 152.1 | 149.3 | 147.9 | 146.7 | 149.3 | 1.4 | 0.01 | -0.79 | 4.86 | 2179.5 | 0 |
| 52 | 0.34 | 0.66 | 0.0 | 152.2 | 150.5 | 152.4 | 152.6 | 151.6 | 151.2 | 152.1 | 1.3 | 0.01 | -0.53 | 1.83 | 372.3 | 0 |
| 53 | 0.56 | 0.44 | 0.0 | 152.6 | 151.7 | 154.3 | 152.9 | 152.3 | 153.8 | 154.0 | 1.3 | 0.01 | 0.12 | 0.02 | 5.1 | 0 |
| 54 | 0.63 | 0.37 | 0.0 | 162.4 | 160.6 | 163.1 | 162.6 | 161.0 | 157.8 | 161.1 | 2.7 | 0.02 | -0.48 | -0.64 | 111.7 | 0 |
| 55 | 0.67 | 0.32 | 0.01 | 164.0 | 161.9 | 163.9 | 164.1 | 162.5 | 160.4 | 163.0 | 2.2 | 0.01 | -0.5 | 0.28 | 88.7 | 0 |
| 56 | 0.38 | 0.58 | 0.04 | 164.4 | 164.1 | 166.6 | 164.8 | 165.0 | 166.1 | 166.3 | 1.8 | 0.01 | -1.63 | 5.78 | 3662.3 | 0 |
| 57 | 0.65 | 0.35 | 0.0 | 166.9 | 165.0 | 167.3 | 167.1 | 165.6 | 167.7 | 167.7 | 1.6 | 0.01 | -1.68 | 7.75 | 5937.9 | 0 |
| 58 | 0.05 | 0.95 | 0.0 | 169.4 | 169.1 | 168.8 | 170.1 | 169.7 | 169.4 | 170.2 | 2.0 | 0.01 | -0.49 | -0.17 | 82.7 | 0 |
| 59 | 0.48 | 0.52 | 0.0 | 170.3 | 169.5 | 172.0 | 170.5 | 170.6 | 171.3 | 172.0 | 1.9 | 0.01 | -0.71 | 1.03 | 256.0 | 0 |

| | | | | | | | | | | | | | | | | |
|---|---|---|---|---|---|---|---|---|---|---|---|---|---|---|---|---|
| 60 | 0.45 | 0.55 | 0.0 | 170.9 | 170.1 | 172.4 | 171.2 | 171.7 | 174.7 | 174.1 | 2.8 | 0.02 | 0.01 | -0.72 | 43.6 | 0 |
| 61 | 0.56 | 0.44 | 0.0 | 174.0 | 172.6 | 172.7 | 174.2 | 173.3 | 179.4 | 176.7 | 3.8 | 0.02 | 0.48 | -0.54 | 100.4 | 0 |
| 62 | 0.0 | 0.72 | 0.28 | 175.5 | 176.7 | 175.5 | 176.1 | 177.5 | 180.9 | 179.6 | 3.5 | 0.02 | -0.01 | -1.06 | 94.1 | 0 |
| 63 | 0.08 | 0.69 | 0.23 | 178.1 | 178.5 | 175.8 | 178.7 | 179.4 | 181.8 | 181.2 | 3.3 | 0.02 | 0.12 | -0.85 | 65.1 | 0 |
| 64 | 0.46 | 0.52 | 0.03 | 180.3 | 179.6 | 181.9 | 180.6 | 180.0 | 184.0 | 182.8 | 2.8 | 0.02 | 0.29 | -0.26 | 33.3 | 0 |
| 65 | 0.6 | 0.4 | 0.0 | 181.2 | 180.0 | 182.6 | 181.3 | 181.0 | 184.4 | 184.4 | 2.8 | 0.02 | 1.14 | 2.04 | 776.4 | 0 |
| 66 | 0.54 | 0.46 | 0.0 | 183.3 | 181.9 | 184.5 | 183.5 | 182.4 | 185.4 | 185.9 | 3.2 | 0.02 | 1.32 | 1.69 | 816.5 | 0 |
| 67 | 0.43 | 0.57 | 0.0 | 185.5 | 183.1 | 188.9 | 186.1 | 184.6 | 188.0 | 187.7 | 3.3 | 0.02 | 1.05 | 0.7 | 408.3 | 0 |
| 68 | 0.6 | 0.4 | 0.0 | 194.9 | 193.4 | 195.7 | 195.0 | 193.8 | 192.9 | 195.6 | 1.8 | 0.01 | 0.24 | 0.29 | 25.9 | 0 |
| 69 | 0.56 | 0.44 | 0.0 | 196.8 | 195.0 | 198.7 | 197.1 | 196.1 | 194.6 | 198.1 | 2.3 | 0.01 | 0.74 | 1.24 | 307.3 | 0 |
| 70 | 0.42 | 0.58 | 0.0 | 202.4 | 201.5 | 203.7 | 202.6 | 202.1 | 201.2 | 203.6 | 1.9 | 0.01 | 0.6 | 1.53 | 314.1 | 0 |
| 71 | 0.76 | 0.24 | 0.0 | 203.2 | 201.8 | 205.0 | 203.5 | 202.4 | 203.4 | 205.8 | 2.0 | 0.01 | 0.79 | 1.64 | 429.6 | 0 |
| 72 | 0.7 | 0.3 | 0.0 | 205.8 | 204.3 | 207.6 | 206.1 | 204.5 | 206.5 | 209.6 | 2.9 | 0.01 | 0.46 | 0.37 | 82.6 | 0 |
| 73 | 0.76 | 0.24 | 0.0 | 214.8 | 213.3 | 216.8 | 215.0 | 213.2 | 212.6 | 217.0 | 3.6 | 0.02 | 0.67 | 0.51 | 172.0 | 0 |
| 74 | 0.99 | 0.01 | 0.0 | 337.1 | 363.3 | 292.5 | 335.1 | 351.2 | 364.9 | 371.6 | 7.2 | 0.02 | -0.21 | 0.18 | 16.7 | 0 |
| 75 | 1.0 | 0.0 | 0.0 | 368.4 | 367.8 | 366.9 | 369.1 | 366.9 | 369.7 | 375.5 | 8.0 | 0.02 | 0.05 | 0.05 | 0.9 | 0 |
| 76 | 1.0 | 0.0 | 0.0 | 374.8 | 374.1 | 373.9 | 375.3 | 372.4 | 371.6 | 378.9 | 8.0 | 0.02 | -0.07 | 0.06 | 2.0 | 0 |
| 77 | 1.0 | 0.0 | 0.0 | 379.4 | 378.6 | 376.4 | 379.9 | 376.9 | 374.7 | 383.3 | 7.9 | 0.02 | -0.09 | 0.05 | 2.9 | 0 |
| 78 | 1.0 | 0.0 | 0.0 | 383.6 | 384.1 | 378.5 | 384.5 | 383.5 | 382.4 | 389.3 | 7.0 | 0.02 | -0.08 | 0.06 | 2.5 | 0 |
| 79 | 1.0 | 0.0 | 0.0 | 387.0 | 387.0 | 381.2 | 388.0 | 385.5 | 384.9 | 392.2 | 7.1 | 0.02 | -0.08 | 0.07 | 2.7 | 0 |
| 80 | 1.0 | 0.0 | 0.0 | 393.9 | 393.1 | 385.2 | 394.9 | 392.8 | 391.6 | 398.8 | 7.0 | 0.02 | -0.08 | 0.07 | 2.7 | 0 |
| 81 | 1.0 | 0.0 | 0.0 | 398.7 | 412.3 | 391.8 | 398.6 | 407.5 | 418.8 | 422.8 | 8.2 | 0.02 | -0.06 | 0.01 | 1.2 | 0 |
| 82 | 1.0 | 0.0 | 0.0 | 440.7 | 440.9 | 438.8 | 441.4 | 438.2 | 441.9 | 446.5 | 7.6 | 0.02 | -0.07 | 0.05 | 1.7 | 0 |
| 83 | 1.0 | 0.0 | 0.0 | 441.4 | 441.3 | 440.0 | 442.1 | 439.3 | 442.4 | 446.9 | 7.6 | 0.02 | -0.07 | 0.04 | 1.7 | 0 |
| 84 | 0.99 | 0.0 | 0.0 | 446.3 | 447.9 | 443.4 | 446.8 | 444.7 | 449.2 | 453.9 | 7.8 | 0.02 | -0.07 | 0.06 | 1.9 | 0 |










\end{document}